%% file: journal_network_embedding.tex
\documentclass[journal]{IEEEtran}
\usepackage{graphicx}
\usepackage{color}
\usepackage{enumerate}
\usepackage{placeins}
\usepackage{float}
\usepackage{tabularx,colortbl}
\usepackage{graphics, subfigure, theorem, times, amsfonts, amsmath, amssymb, cite, verbatim}
\usepackage{multicol,multirow,booktabs}
\usepackage{tikz}
\usepackage{framed}
\usetikzlibrary{shapes,arrows}
\input{figures/tikz_styles}
\usepackage{tikz-3dplot} %for tikz-3dplot functionality
\usepackage{pgfplots}
\usepackage{color}
\input{my_sections.tex}

\usepackage{hyperref}
\input{mysymbol.sty}

\newtheorem{proposition}{\hspace{0pt}\bf Proposition}

\newtheorem{theorem}{\hspace{0pt}\bf Theorem}
\newtheorem{corollary}{\hspace{0pt}\bf Corollary}
\newtheorem{fact}{\hspace{0pt}\bf Fact}

\newtheorem{definition}{\hspace{0pt}\bf Definition}
\newcommand{\QED}{\hfill\ensuremath{\blacksquare}}

\newcommand{\C} {\text{\normalfont C} }
\newcommand{\PE} {\text{\normalfont PE} }
\newcommand{\PES} {\text{\normalfont PE, S} }
\newcommand{\PEQ} {\text{\normalfont PE, Q} }
\newcommand{\EE} {\text{\normalfont E} }

\newcommand   \bee     [1] {\bold{1}\left\{#1\right\}}
\newcommand   \beeInline     [1] {\bold{1}\{#1\}}

\newcommand{\hhatphi} {\ensuremath{\hat \phi} }
\newcommand{\hhatpsi} {\ensuremath{\hat \psi} }

\begin{document}

\title{Network Comparison: Embeddings and Interiors}

\author{Weiyu Huang and Alejandro Ribeiro
\thanks{Electrical and Systems Engineering, University of Pennsylvania, 200 S 33rd Street, Philadelphia, PA 19104. Email: {whuang, aribeiro}@seas.upenn.edu.}}

\maketitle

\input{sec_0_abstract.tex}

\IEEEpeerreviewmaketitle

\input{sec_1_introduction.tex}

\input{sec_2_preliminaries.tex}

\input{sec_3_embedding.tex}

\input{sec_4_interiors.tex}

\input{sec_4a_distance_equivalance_sample.tex}
\input{sec_5_application.tex}

\input{sec_6_conclusion.tex}
\input{sec_A_appendix.tex}

%%%%%%%%%%%%%%%%%%%%%%%%%%%%%%%%%%%%%%%%%%%%%%%%%%%%%%%%%%%%%%%%%%%%%
%%%   R   E   F   E   R   E   N   C   E  %%%%%%%%%%%%%%%%%%%%%%%%%%%%%%%
%%%%%%%%%%%%%%%%%%%%%%%%%%%%%%%%%%%%%%%%%%%%%%%%%%%%%%%%%%%%%%%%%%%%%
%

\urlstyle{same}
\bibliographystyle{IEEEtran}
\bibliography{embedding_biblio}

\end{document}

%% file: figures/tikz_styles.tex
\tikzstyle{phantom vertex} = [ ellipse, 
                               anchor = center, 
                               minimum height = 0.8*\unit, 
                               minimum width  = 0.8*\unit,
                               inner sep=0pt,
                               anchor=center]
\tikzstyle{red vertex}   = [black, fill = red!20,   phantom vertex, draw]
\tikzstyle{black vertex} = [black, fill = black!20, phantom vertex, draw]
\tikzstyle{blue vertex}  = [black, fill = blue!20,  phantom vertex, draw]
\tikzstyle{green vertex} = [black, fill = green!20,  phantom vertex, draw]
\tikzstyle{yellow vertex} = [black, fill = yellow!20,  phantom vertex, draw]
\tikzstyle{vertex}       = [draw, phantom vertex]

\tikzstyle{point} = [ellipse, inner sep=0pt, draw, fill=white, anchor = center,
                     minimum height = 0.05*\unit, minimum width  = 0.05*\unit]

%% file: my_sections.tex
\usepackage{needspace}

%% file: sec_0_abstract.tex
%!TEX root = journal_network_embedding.tex

\begin{abstract}
This paper presents methods to compare networks where relationships between pairs of nodes in a given network are defined. We define such network distance by searching for the optimal method to embed one network into another network, prove that such distance is a valid metric in the space of networks modulo permutation isomorphisms, and examine its relationship with other network metrics. The network distance defined can be approximated via multi-dimensional scaling, however, the lack of structure in networks results in poor approximations. To alleviate such problem, we consider methods to define the interiors of networks. We show that comparing interiors induced from a pair of networks yields the same result as the actual network distance between the original networks. Practical implications are explored by showing the ability to discriminate networks generated by different models. 
\end{abstract}

% keywords

%% file: sec_1_introduction.tex
%!TEX root = journal_network_embedding.tex

%%%%%%%%%%%%%%%%%%%%%%%%%%%%%%%%%%%%%%%%%%%%%%%%%%%%%%%%%%%%%%%%%%%%%
%%%   I   N   T   R   O   %%%%%%%%%%%%%%%%%%%%%%%%%%%%%%%%%%%
%%%%%%%%%%%%%%%%%%%%%%%%%%%%%%%%%%%%%%%%%%%%%%%%%%%%%%%%%%%%%%%%%%%%%
%
\section{Introduction}\label{sec_intro}

The field of network science is predicated on the empirical observation that network structure carries important information about phenomena of interest. Network structures have been observed to be fundamental in social organizations \cite{Wasserman1994} and differences in the structure of brain networks have been shown to have clinical value in neurology and psychology \cite{sporns2011}. The fundamental value of network structure has led to an extensive literature on network identification that is mostly concerned with the identification of network features that serve as valuable discriminators in different contexts. Examples of these features and application domains are clustering coefficients \cite{Wang2012} and motifs \cite {Choobdar2012} in social networks; neighborhood topology \cite{Singh2008}, betweenness \cite{Peng2010}, and wavelets \cite{Yong2010} in protein interaction networks; as well as graphlet degree distributions \cite{Shervashidze2009}, graph theoretic measures \cite{bullmore2009}, and single linkage dendrogram \cite{lee2012} and homology \cite{wilkerson2013} in different contexts. However valuable, it is material to recognize that it is possible for very different networks to be indistinguishable from the perspective of specific features, or, conversely, to have similar networks that differ substantially on the values of some features. One way to sidestep this limitation it to define and evaluate proper network distances. This is the objective of this paper.

Without getting into the details of how a valid distance might be defined, it is apparent that their computation is bound to be  combinatorial. Indeed, since permutations of unlabeled nodes result in identical networks, distances must rely on comparison between a combinatorial number of node correspondences -- evaluating distances is relatively simpler if nodes are labeled \cite{Onnela2007, Kossinets2006, Khmelev2001}. An important observation in this regard is that the space of finite metric spaces is a subset of the space of networks composed of those whose edges satisfy the triangle inequality. This observation is pertinent because there is a rich literature on the comparison of metric spaces that we can adopt as a basis for generalizations that apply to the comparison of networks. Of particular interest here are the Gromov-Hausdorff distance, which measures the size of the smallest modification that allows the spaces to be mapped onto each other \cite{Memoli2008, memoli2004}, and the partial embedding distance, which measures the size of the smallest modification that allows a space to be embedded in the other \cite{bronstein2006, bronstein2006a, bronstein2006b}. The computation of either of these distances is intractable. However, Gromov-Hausdorff distances can be tractably approximated using homological features \cite{Chazal2009a} and embedding distances can be approximated using multi-dimensional scaling (MDS) \cite{bronstein2006}.

In prior work we have defined generalizations of the Gromov-Hausdorff distance \cite{Carlsson2014, Huang2015} to networks, and utilized homological features for computationally tractable approximate evaluations \cite{Huang2016, Huang2016a}. Our starting point here is the partial embedding distance for metric spaces \cite{bronstein2006, bronstein2006a, bronstein2006b, memoli2004}. Our goal is to generalize embedding distances to arbitrary networks and utilize MDS techniques for their approximate computation. 

\subsection{Organization And Contributions}

Networks, metrics, embedding metrics, and the Gromov-Hausdorff distance for networks are defined in Section \ref{sec_preliminaries}. The issue of defining an embedding distance for networks is addressed in Section \ref{sec_embedding_distance}. The idea of an embedding distance $d(A,B)$ is to analyze how much we have to modify network $A$ to make it a subset of network $B$. This is an asymmetric relationship. In particular, having $d(A,B) = 0$ means that network $A$ can be embedded in network $B$ but the opposite need not be true. The first contribution of this paper is to:

\begin{itemize}
\item[(i)] Define network \textit{embeddings} and a corresponding notion of partial embedding distances. Partial embedding distances define an embedding metric $d$ such that $d(A,B) = 0$ if and only if $A$ can be embedded in $B$.
\end{itemize}

We attempt to use the MDS techniques in \cite{bronstein2006a} to approximate the computation of embedding distances but observe that the methodology yields poor results -- see Figure \ref{fig_mds_original} for an illustration of why this is not unexpected. To improve these results we observe that when edge dissimilarities satisfy a triangle inequality, an Euclidean interior is implicitly defined. In the case of arbitrary networks this is not true and motivates the definition of the interior of a network that we undertake in Section \ref{sec_space_induced}. The second contribution of this paper is to:

\begin{itemize}
\item[(ii)] Provide a definition of the \textit{interior} of a network. The interior of a set of nodes is the set of points that can be written as convex combinations of the nodes. When the network forms a metric space, the dissimilarity between a pair of points in the interior is the distance on the shortest path between the pair. When the dissimilarities in the network do not form a metric space, e.g. representing travel time between nodes, such construction would yield conflict. The problem can be solved by defining the dissimilarity between a pair of points as the travel time on the shortest path between the pair.
%The interior of a set of nodes is the set of points that can be written as convex combinations of the nodes. The dissimilarity between points in the interior is the difference between the convex combinations of the respective dissimilarities between the respective points. 
\end{itemize}

Having the ability to extend networks into their interiors, we extend different networks and compute partial embedding distances between their extensions. In principle, distances between two networks and their respective extensions need not be related. In Section \ref{sec_interior_distance} we show that a restriction in the embedding of the extended networks renders them identical. Our third and most important contribution is to:

\begin{itemize}
\item[(iii)] Define embeddings for extended networks such that points in one of the original networks -- prior to extension -- can only be embedded into original points of the other network. We show that the embedding distance that results from this restriction is the same embedding distance between the original networks.
\end{itemize}

The definition of interior is somewhat arbitrary, however because of (iii), the practical implication of interior definition is justified. We point out that a network extension is a dense set that includes all the convex combinations of sets of points. To make interior extensions practical we consider samplings of the interior in Section \ref{sec_partial_embedding_induced}. It is not difficult to show in light of Contribution (iii) that the embedding distance between a a pair of networks extended to samples of their interiors is also identical to the embedding distance between the original pair of networks -- if the restriction in the mapping of original nodes is retained.

We exploit Contributions (ii) and (iii) to approximate the computation of embedding distances using the MDS techniques in \cite{bronstein2006a} but applied to networks extended to their interiors. The definition of an interior markedly improves the quality of MDS distance approximations. We illustrate this fact in Section \ref{sec_application} with an artificial illustrative example and also demonstrate the ability to discriminate networks with different generative models. %Although interiors are defined for sets with an arbitrary number of points, %we don't use sets with more than three points in our numerical results. 
We only extend by adding points that are mid-points of original nodes in the networks, in order to make the process computationally tractable. The small number of points considered in interiors is sufficient to distinguish networks of different processes, despite that the original networks may possess different number of nodes.

%We further observe that it is sufficient to sample a small number of points in the interior of the network. In the numerical analyses only three points are sampled in the interior of each set of three points.

%% file: sec_2_preliminaries.tex
%!TEX root = journal_network_embedding.tex

%%%%%%%%%%%%%%%%%%%%%%%%%%%%%%%%%%%%%%%%%%%%%%%%%%%%%%%%%%%%%%%%%%%%%
%%%   S   E   C   T   I   O   N   %%%%%%%%%%%%%%%%%%%%%%%%%%%%%%%%%%%
%%%%%%%%%%%%%%%%%%%%%%%%%%%%%%%%%%%%%%%%%%%%%%%%%%%%%%%%%%%%%%%%%%%%%
%
\section{Preliminaries}\label{sec_preliminaries}

A network is defined as a pair $N_X=(X, r_X)$, where $X$ is a finite set of nodes and $r_X : X^2 = X \times X \rightarrow \reals_+$ is a function encoding dissimilarity between pairs. For $x,x'\in X$, values of this function are denoted as $r_X(x, x')$. We assume that $r_X(x, x') = 0$ if and only if $x = x'$ and we further restrict attention to symmetric networks where $r_X(x, x') = r_X(x', x)$ for all pairs of nodes $x, x' \in X$. The set of all such networks is denoted as $\ccalN$.

When defining a distance between networks we need to take into consideration that permutations of nodes amount to relabelling nodes and should be considered as same entities. We therefore say that two networks $N_X = (X, r_X)$ and $N_Y = (Y, r_Y)$ are isomorphic whenever there exists a bijection $\phi: X \rightarrow Y$ such that for all points $x, x' \in X$,
\begin{align} \label{eqn_network_order_2_isomorphism}
   r_X(x, x') = r_Y(\phi(x), \phi(x')).
\end{align}
Such a map is called an isometry. Since the map $\phi$ is bijective, (\ref{eqn_network_order_2_isomorphism}) can only be satisfied when $X$ is a permutation of $Y$. When networks are isomorphic we write $N_X \cong N_Y$. The space of networks where isomorphic networks $N_X \cong N_Y$ are represented by the same element is termed the set of networks modulo isomorphism and denoted by $\ccalN \mod \cong$. The space $\ccalN \mod \cong$ can be endowed with a valid metric \cite{Carlsson2014, Huang2015}. The definition of this distance requires introducing the prerequisite notion of correspondence \cite[Def. 7.3.17]{Burago01}.

%%%%%%%%%%%%%%%%%%%%%%%%%%%%%%%%%%%%%%%%%%%%%%%%%%%%%%%%%%%%%%%%%%%%%
%%%   D   E   F   I   N   I   T   I   O   N   %%%%%%%%%%%%%%%%%%%%%%%
%%%%%%%%%%%%%%%%%%%%%%%%%%%%%%%%%%%%%%%%%%%%%%%%%%%%%%%%%%%%%%%%%%%%%
%
\begin{definition}\label{dfn_correspondence}
A correspondence between two sets $X$ and $Y$ is a subset $C \subseteq X \times Y$ such that $\forall~x \in X$, there exists $y \in Y$ such that $(x,y) \in C$ and $\forall~ y \in Y$ there exists $x \in X$ such that $(x,y) \in C$. The set of all correspondences between $X$ and $Y$ is denoted as $\ccalC(X,Y)$.
\end{definition}

%%%%%%%%%%%%%%%%%%%%%%%%%%%%%%%%%%%%%%%%%%%%%%%%%%%%%%%%%%%%%%%%%%%%%
%%%   M   A   I   N       M   A   T   T   E   R   %%%%%%%%%%%%%%%%%%%
%%%%%%%%%%%%%%%%%%%%%%%%%%%%%%%%%%%%%%%%%%%%%%%%%%%%%%%%%%%%%%%%%%%%%
%
A correspondence in the sense of Definition \ref{dfn_correspondence} is a map between node sets $X$ and $Y$ so that every element of each set has at least one correspondent in the other set. Correspondences include permutations as particular cases but also allow mapping of a single point in $X$ to multiple correspondents in $Y$ or, vice versa. Most importantly, this allows definition of correspondences between networks with different numbers of elements. We can now define the distance between two networks by selecting the correspondence that makes them most similar as stated next.

%%%%%%%%%%%%%%%%%%%%%%%%%%%%%%%%%%%%%%%%%%%%%%%%%%%%%%%%%%%%%%%%%%%%%
%%%   D   E   F   I   N   I   T   I   O   N   %%%%%%%%%%%%%%%%%%%%%%%
%%%%%%%%%%%%%%%%%%%%%%%%%%%%%%%%%%%%%%%%%%%%%%%%%%%%%%%%%%%%%%%%%%%%%
%
\begin{definition}\label{dfn_conventional_network_distance}
Given two networks $N_X = (X, r_X)$ and $N_Y = (Y, r_Y)$ and a correspondence $C$ between the node spaces $X$ and $Y$ define the network difference with respect to $C$ as 
\begin{align}\label{eqn_conventional_network_distance_prelim}
    \Gamma_{X,Y} (C) \!
         := \! \max_{(x, y), (x', y') \in C} 
            \Big| r_X(x, x') - r_Y (y, y') \Big|.
\end{align}
The network distance between $N_X$ and $N_Y$ is then defined as
\begin{align}\label{eqn_conventional_network_distance}
   d_\C(N_X, N_Y) := \min_{C \in \ccalC(X,Y)} 
    \Big\{ \Gamma_{X,Y} (C) \Big\}.
\end{align} \end{definition}

%%%%%%%%%%%%%%%%%%%%%%%%%%%%%%%%%%%%%%%%%%%%%%%%%%%%%%%%%%%%%%%%%%%%%
%%%   M   A   I   N       M   A   T   T   E   R   %%%%%%%%%%%%%%%%%%%
%%%%%%%%%%%%%%%%%%%%%%%%%%%%%%%%%%%%%%%%%%%%%%%%%%%%%%%%%%%%%%%%%%%%%
%
For a given correspondence $C \in \ccalC(X,Y)$ the network difference $\Gamma_{X,Y}(C)$ selects the maximum distance difference $|r_X(x_1, x_2) - r_Y (y_1, y_2)|$ among all pairs of correspondents -- we compare $r_X(x_1, x_2)$ with $r_Y (y_1, y_2)$ when the points $x_1$ and $y_1$, as well as the points $x_2$ and $y_2$, are correspondents. The distance in \eqref{eqn_conventional_network_distance} is defined by selecting the correspondence that minimizes these maximal differences. The distance in Definition \ref{dfn_conventional_network_distance} is a proper metric in the space of networks modulo isomorphism. It is nonnegative, symmetric, satisfies the triangle inequality, and is null if and only if the networks are isomorphic \cite{Carlsson2014, Huang2015}. For future reference, the notion of metric is formally stated next. 

%%%%%%%%%%%%%%%%%%%%%%%%%%%%%%%%%%%%%%%%%%%%%%%%%%%%%%%%%%%%%%%%%%%%%
%%%   D   E   F   I   N   I   T   I   O   N   %%%%%%%%%%%%%%%%%%%%%%%
%%%%%%%%%%%%%%%%%%%%%%%%%%%%%%%%%%%%%%%%%%%%%%%%%%%%%%%%%%%%%%%%%%%%%
%
\begin{definition}\label{dfn_metric}
Given a space $\ccalS$ and an isomorphism $\cong$, a function $d : \ccalS \times \ccalS \rightarrow \reals$ is a metric in $\ccalS \mod \cong$ if for any $a, b, c \in \ccalS$ the function $d$ satisfies:
\begin{mylist}
\item [(i) {\bf Nonnegativity.}] $d(a,b) \ge 0$.
\item [(ii) {\bf Symmetry.}] $d(a,b) = d(b,a)$.
\item [(iii) {\bf Identity.}] $d(a,b) = 0$ if and only if $a \cong b$.
\item [(iv) {\bf Triangle inequality.}] $d(a,b) \le d(a,c) + d(c,b)$.
\end{mylist}\end{definition}

%%%%%%%%%%%%%%%%%%%%%%%%%%%%%%%%%%%%%%%%%%%%%%%%%%%%%%%%%%%%%%%%%%%%%
%%%   M   A   I   N       M   A   T   T   E   R   %%%%%%%%%%%%%%%%%%%
%%%%%%%%%%%%%%%%%%%%%%%%%%%%%%%%%%%%%%%%%%%%%%%%%%%%%%%%%%%%%%%%%%%%%
%
A metric $d$ in $\ccalS\mod\cong$ gives a proper notion of distance. Since zero distances imply elements being isomorphic, the distance between elements reflects how far they are from being isomorphic. The distance in Definition \ref{dfn_conventional_network_distance} is a metric in space $\ccalN \mod \cong$. Observe that since correspondences may be between networks with different number of elements, Definition \ref{dfn_conventional_network_distance} defines a distance $d_\C(N_X, N_Y)$ when the node cardinalities $|X|$ and $|Y|$ are different. In the particular case when the functions $r_X$ satisfy the triangle inequality, the set of networks $\ccalN$ reduces to the set of metric spaces $\ccalM$. In this case the metric in Definition \ref{dfn_conventional_network_distance} reduces to the Gromov-Hausdorff (GH) distance between metric spaces. The distances $d_\C(N_X, N_Y)$ in \eqref{eqn_conventional_network_distance} are valid metrics even if the triangle inequalities are violated by $r_X$ or $r_Y$ \cite{Carlsson2014, Huang2015}.

A related notion is that of an isometric embedding. We say that a map $\phi: X \rightarrow Y$ is an isometric embedding from $N_X = (X, r_X)$ to $N_Y = (Y, r_Y)$ if \eqref{eqn_network_order_2_isomorphism} holds for all points $x, x' \in X$. Since $r_X(x, x') = r_Y(\phi(x), \phi(x'))$ for any $x, x' \in X$, $r_X(x, x') > 0$ for $x \neq x'$ and $r_Y(y, y) = 0$, the map $\phi$ is injective. This implies that the condition can only be satisfied when $N_X = (X, r_X)$ is a sub-network of $N_Y = (Y, r_Y)$. Such a map is called an isometric embedding. When $N_X$ can be isometrically embedded into $N_Y$, we write $N_X \sqsubseteq N_Y$. Related to the notion of isometric embedding is the notion of an embedding metric that we state next.

%%%%%%%%%%%%%%%%%%%%%%%%%%%%%%%%%%%%%%%%%%%%%%%%%%%%%%%%%%%%%%%%%%%%%
%%%   D   E   F   I   N   I   T   I   O   N   %%%%%%%%%%%%%%%%%%%%%%%
%%%%%%%%%%%%%%%%%%%%%%%%%%%%%%%%%%%%%%%%%%%%%%%%%%%%%%%%%%%%%%%%%%%%%
%
\begin{definition}\label{dfn_embedding_metric}
Given a space $\ccalS$ and an isometric embedding $\sqsubseteq$, a function $d : \ccalS \times \ccalS \rightarrow \reals$ is an embedding metric in $\ccalS$ if for any $a, b, c \in \ccalS$ the function $d$ satisfies:
\begin{mylist}
\item [(i)   {\bf Nonnegativity.}] $d(a,b) \ge 0$.
\item [(ii)  {\bf Embedding identity.}] $d(a,b) = 0$ if and only if $a \sqsubseteq b$.
\item [(iii) {\bf Triangle inequality.}] $d(a,b) \le d(a,c) + d(c,b)$.
\end{mylist}\end{definition}

%%%%%%%%%%%%%%%%%%%%%%%%%%%%%%%%%%%%%%%%%%%%%%%%%%%%%%%%%%%%%%%%%%%%%
%%%   M   A   I   N       M   A   T   T   E   R   %%%%%%%%%%%%%%%%%%%
%%%%%%%%%%%%%%%%%%%%%%%%%%%%%%%%%%%%%%%%%%%%%%%%%%%%%%%%%%%%%%%%%%%%%
%
It is apparent that metrics are embedding metrics because bijections are injective, and that in general embedding metrics are not metrics because they are asymmetric. In this paper, we consider defining an embedding distance between networks and evaluate its relationship with the Gromov-Hausdorff distance (Section \ref{sec_embedding_distance}). We then consider the problem of augmenting the networks by adding points to fill their ``interior''. The interior is defined so that embedding metrics between the original networks and embedding metrics between these augmented spaces coincide (Section \ref{sec_space_induced}).

%% file: sec_3_embedding.tex
%!TEX root = journal_network_embedding.tex

%%%%%%%%%%%%%%%%%%%%%%%%%%%%%%%%%%%%%%%%%%%%%%%%%%%%%%%%%%%%%%%%%%%%%
%%%   S   E   C   T   I   O   N   %%%%%%%%%%%%%%%%%%%%%%%%%%%%%%%%%%%
%%%%%%%%%%%%%%%%%%%%%%%%%%%%%%%%%%%%%%%%%%%%%%%%%%%%%%%%%%%%%%%%%%%%%
%
\section{Embeddings}\label{sec_embedding_distance}

As is the case with correspondences, mappings also allow definition of associations between networks with different numbers of elements. We use this to define the distance from one network to another network by selecting the mapping that makes them most similar as we formally define next.

%%%%%%%%%%%%%%%%%%%%%%%%%%%%%%%%%%%%%%%%%%%%%%%%%%%%%%%%%%%%%%%%%%%%%
%%%   D   E   F   I   N   I   T   I   O   N   %%%%%%%%%%%%%%%%%%%%%%%
%%%%%%%%%%%%%%%%%%%%%%%%%%%%%%%%%%%%%%%%%%%%%%%%%%%%%%%%%%%%%%%%%%%%%
%
\begin{definition}\label{dfn_partial_embedding}
Given two networks $N_X = (X, r_X)$, $N_Y = (Y, r_Y)$, and a map $\phi: X \rightarrow Y$ from node space $X$ to the node space $Y$, define the network difference with respect to $\phi$ as 
\begin{align}\label{eqn_dfn_partial_embedding_prelim}
    \Delta_{X,Y} (\phi)
         :=  \max_{x, x' \in X} 
            \Big| r_X(x, x') - r_Y (\phi(x),\phi(x')) \Big|.
\end{align}
The partial embedding distance from $N_X$ to $N_Y$ is defined as
\begin{align}\label{eqn_dfn_partial_embedding}
   d_\PE(N_X, N_Y) := \min_{\phi: X \rightarrow Y} 
    \Big\{ \Delta_{X,Y} (\phi) \Big\}.
\end{align} \end{definition}

%%%%%%%%%%%%%%%%%%%%%%%%%%%%%%%%%%%%%%%%%%%%%%%%%%%%%%%%%%%%%%%%%%%%%
%%%   M   A   I   N       M   A   T   T   E   R   %%%%%%%%%%%%%%%%%%%
%%%%%%%%%%%%%%%%%%%%%%%%%%%%%%%%%%%%%%%%%%%%%%%%%%%%%%%%%%%%%%%%%%%%%
%
Both, Definition \ref{dfn_conventional_network_distance} and Definition \ref{dfn_partial_embedding} consider a mapping between the node space $X$ and the node space $Y$, compare dissimilarities, and set the network distance to the comparison that yields the smallest value in terms of maximum differences. The distinction between them is that in \eqref{eqn_conventional_network_distance_prelim} we consider correspondence, which requires each point in any node spaces ($X$ or $Y$) to have a correspondent in the other node space, whereas in \eqref{eqn_dfn_partial_embedding_prelim} we examine mappings, which only require all points in node space $X$ to have one correspondent in the node set $Y$. Moreover, in \eqref{eqn_conventional_network_distance_prelim}, a node $x\in X$ may have multiple correspondents, however, in \eqref{eqn_dfn_partial_embedding_prelim}, a node $x\in X$ can only have exactly one correspondent. Except for this distinction, Definition \ref{dfn_conventional_network_distance} and Definition \ref{dfn_partial_embedding} are analogous since $\Delta_{X,Y}(\phi)$ selects the difference $|r_X(x_1, x_2) - r_Y(y_1, y_2)|$ among all pairs. The distance $d_\PE(N_X, N_Y)$ is defined by selecting the mapping that minimizes these maximal differences. We show in the following proposition that the function $d_\PE : \ccalN \times \ccalN \rightarrow \reals_+$ is, indeed, an embedding metric in the space of networks.

%%%%%%%%%%%%%%%%%%%%%%%%%%%%%%%%%%%%%%%%%%%%%%%%%%%%%%%%%%%%%%%%%%%%%
%%%   P   R   O   P   O   S   I   T   I   O   N   %%%%%%%%%%%%%%%%%%%%%%%
%%%%%%%%%%%%%%%%%%%%%%%%%%%%%%%%%%%%%%%%%%%%%%%%%%%%%%%%%%%%%%%%%%%%%
%
\begin{proposition}\label{prop_partial_embedding_metric}
The function $d_\PE : \ccalN \times \ccalN \rightarrow \reals_+$ defined in \eqref{eqn_dfn_partial_embedding} is an embedding metric in the space $\ccalN$. \end{proposition}

%%%%%%%%%%%%%%%%%%%%%%%%%%%%%%%%%%%%%%%%%%%%%%%%%%%%%%%%%%%%%%%%%%%%%
%%%   P   R   O   O   F   %%%%%%%%%%%%%%%%%%%%%%%%%%%%%%%
%%%%%%%%%%%%%%%%%%%%%%%%%%%%%%%%%%%%%%%%%%%%%%%%%%%%%%%%%%%%%%%%%%%%%
%
\begin{myproof} See Appendix \ref{apx_proof_1} for proofs in Section \ref{sec_embedding_distance}. \end{myproof}

%%%%%%%%%%%%%%%%%%%%%%%%%%%%%%%%%%%%%%%%%%%%%%%%%%%%%%%%%%%%%%%%%%%%%
%%%   M   A   I   N       M   A   T   T   E   R   %%%%%%%%%%%%%%%%%%%
%%%%%%%%%%%%%%%%%%%%%%%%%%%%%%%%%%%%%%%%%%%%%%%%%%%%%%%%%%%%%%%%%%%%%
%
The embedding distance $d_\PE(N_X, N_Y)$ from one network $N_X$ to another network $N_Y$ is not a metric due to its asymmetry. We can construct a symmetric version from $d_\PE(N_X, N_Y)$ by taking the maximum from the embedding distance $d_\PE(N_X, N_Y)$ and $d_\PE(N_Y, N_X)$. This would give us a valid metric distance in $\ccalN \mod \cong$. A formal definition and theorem are shown next.

%%%%%%%%%%%%%%%%%%%%%%%%%%%%%%%%%%%%%%%%%%%%%%%%%%%%%%%%%%%%%%%%%%%%%
%%%   D   E   F   I   N   I   T   I   O   N   %%%%%%%%%%%%%%%%%%%%%%%
%%%%%%%%%%%%%%%%%%%%%%%%%%%%%%%%%%%%%%%%%%%%%%%%%%%%%%%%%%%%%%%%%%%%%
%
\begin{definition}\label{dfn_embedding}
Given two networks $N_X = (X, r_X)$, $N_Y = (Y, r_Y)$, define the embedding distance between the pair as
\begin{align}\label{eqn_dfn_embedding}
   d_\EE(N_X, N_Y) := \max\left\{ d_\PE(N_X, N_Y), d_\PE(N_Y, N_X)\right\}.
\end{align} 
where partial embedding distances $d_\PE(N_X, N_Y)$ and $d_\PE(N_X, N_Y)$ are defined in Definition \ref{dfn_partial_embedding}.
\end{definition}

%%%%%%%%%%%%%%%%%%%%%%%%%%%%%%%%%%%%%%%%%%%%%%%%%%%%%%%%%%%%%%%%%%%%%
%%%   T   H   E   O   R   E   M   %%%%%%%%%%%%%%%%%%%%%%%
%%%%%%%%%%%%%%%%%%%%%%%%%%%%%%%%%%%%%%%%%%%%%%%%%%%%%%%%%%%%%%%%%%%%%
%
\begin{theorem}\label{thm_embedding_distance_metric}
The function $d_\EE : \ccalN \times \ccalN \rightarrow \reals_+$ defined in \eqref{eqn_dfn_embedding} is a metric in the space $\ccalN \mod \cong$.
\end{theorem}

%%%%%%%%%%%%%%%%%%%%%%%%%%%%%%%%%%%%%%%%%%%%%%%%%%%%%%%%%%%%%%%%%%%%%
%%%   F   I   G   U   R   E   %%%%%%%%%%%%%%%%%%%%%%%%%%%%%%%%%%%%%%%
%%%%%%%%%%%%%%%%%%%%%%%%%%%%%%%%%%%%%%%%%%%%%%%%%%%%%%%%%%%%%%%%%%%%%
%
\begin{figure*}[t]
\centerline{\input{figures/embedding_example.tex}\vspace{-1mm} 
}
\begin{minipage}[h]{0.032\textwidth}
~
\end{minipage}
\begin{minipage}[h]{0.275\textwidth}
    	\centering
    	\includegraphics[trim=0.01cm 0.01cm 0.01cm 0.01cm, clip=true, width=1 \textwidth]
	{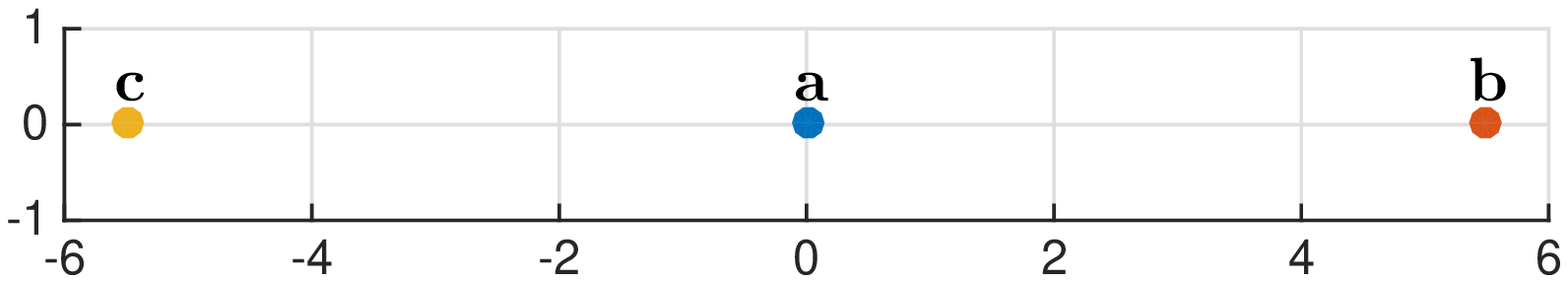}
\end{minipage}
\begin{minipage}[h]{0.054\textwidth}
~
\end{minipage}
\begin{minipage}[h]{0.275\textwidth}
    	\centering
    	\includegraphics[trim=0.01cm 0.01cm 0.01cm 0.01cm, clip=true, width=1 \textwidth]
	{figures/mds_original.eps}
\end{minipage}
\begin{minipage}[h]{0.054\textwidth}
~
\end{minipage}
\begin{minipage}[h]{0.275\textwidth}
    	\centering
    	\includegraphics[trim=0.01cm 0.01cm 0.01cm 0.01cm, clip=true, width=1 \textwidth]
	{figures/mds_original.eps}
\end{minipage}
\caption{An example where different networks result in identical multi-dimensional scaling results. We emphasize that the number of dimension used in multi-dimensional scaling would not distinguish networks since the triangle inequality property for relationships between nodes in the networks is violated. Such a caveat would be solved by inducing semimetrics in the space defined by the given networks, as we develop throughout Section \ref{sec_space_induced}. \vspace{-4mm}
}
\label{fig_mds_original}
\end{figure*}

%%%%%%%%%%%%%%%%%%%%%%%%%%%%%%%%%%%%%%%%%%%%%%%%%%%%%%%%%%%%%%%%%%%%%
%%%   M   A   I   N      M   A   T   T   E   R   %%%%%%%%%%%%%%%%%%%%%%%
%%%%%%%%%%%%%%%%%%%%%%%%%%%%%%%%%%%%%%%%%%%%%%%%%%%%%%%%%%%%%%%%%%%%%
%
Since embedding distances between two networks generate a well-defined metric, they provide a means to compare networks of arbitrary sizes. In comparing the embedding distance in \eqref{eqn_dfn_embedding} with the network distance in \eqref{eqn_conventional_network_distance} we see that both find the bottleneck that prevents the networks to be matched to each other. It is not there surprising to learn that they satisfy the relationship that we state in the following proposition. 

%%%%%%%%%%%%%%%%%%%%%%%%%%%%%%%%%%%%%%%%%%%%%%%%%%%%%%%%%%%%%%%%%%%%%
%%%   L   E   M   M   A   %%%%%%%%%%%%%%%%%%%%%%%
%%%%%%%%%%%%%%%%%%%%%%%%%%%%%%%%%%%%%%%%%%%%%%%%%%%%%%%%%%%%%%%%%%%%%
%
\begin{proposition}\label{lemma_GH}
The network distance $d_\C(N_X, N_Y)$ defined in \eqref{eqn_conventional_network_distance} can also be written as
\begin{align}\label{eqn_lemma_GH}
    d_\C(N_X, N_Y) \! = \!\!\! \min_{\substack{\phi:X \rightarrow Y\\ \psi:Y \rightarrow X}}\!\!\! 
        \max\left\{\! \Delta_{X, Y}(\phi),\! \Delta_{Y, X}(\psi),\! \delta_{X, Y}(\phi,\! \psi)\!\right\}\!,
\end{align}
where the network differences $\Delta_{X, Y}(\phi)$ and $\Delta_{Y, X}(\psi)$ with respect to mappings $\phi$ and $\psi$ are defined in \eqref{eqn_dfn_partial_embedding_prelim} and $\delta_{X, Y}(\phi, \psi)$ measures how far the mappings $\phi$ and $\psi$ are from being the inverse of each other, and is defined as
\begin{align}\label{eqn_lemma_GH_difference_mapping}
    \delta_{X, Y}(\phi, \psi) = \max_{x \in X, y \in Y} \left| r_X(x, \psi(y)) - r_Y(\phi(x), y) \right|.
\end{align}
\end{proposition}

%%%%%%%%%%%%%%%%%%%%%%%%%%%%%%%%%%%%%%%%%%%%%%%%%%%%%%%%%%%%%%%%%%%%%
%%%   M   A   I   N      M   A   T   T   E   R   %%%%%%%%%%%%%%%%%%%%%%%
%%%%%%%%%%%%%%%%%%%%%%%%%%%%%%%%%%%%%%%%%%%%%%%%%%%%%%%%%%%%%%%%%%%%%
%
A direct consequence of Lemma \ref{lemma_GH} is that the embedding distance  \eqref{eqn_dfn_embedding} is a lower bound of the network distance \eqref{eqn_conventional_network_distance}.

%%%%%%%%%%%%%%%%%%%%%%%%%%%%%%%%%%%%%%%%%%%%%%%%%%%%%%%%%%%%%%%%%%%%%
%%%   C   O   R   O   L   L   A   R   Y   %%%%%%%%%%%%%%%%%%%%%%%
%%%%%%%%%%%%%%%%%%%%%%%%%%%%%%%%%%%%%%%%%%%%%%%%%%%%%%%%%%%%%%%%%%%%%
%
\begin{corollary}\label{coro_lower_bound}
Function $d_\EE$ is a lower bound with $d_\C$ in \eqref{eqn_conventional_network_distance}, i.e. 
\begin{align}\label{eqn_equivalence}
    d_\EE(N_X, N_Y) \leq d_\C(N_X, N_Y),
\end{align}
for any networks $N_X$ and $N_Y$. 
\end{corollary}

%%%%%%%%%%%%%%%%%%%%%%%%%%%%%%%%%%%%%%%%%%%%%%%%%%%%%%%%%%%%%%%%%%%%%
%%%   F   I   G   U   R   E   %%%%%%%%%%%%%%%%%%%%%%%%%%%%%%%%%%%%%%%
%%%%%%%%%%%%%%%%%%%%%%%%%%%%%%%%%%%%%%%%%%%%%%%%%%%%%%%%%%%%%%%%%%%%%
%
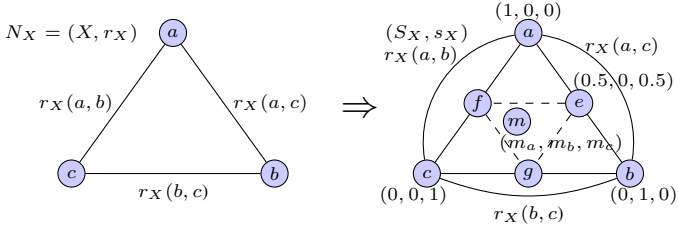
\begin{figure}[t]
\centerline{\input{figures/points_inside_network.tex}\vspace{-4mm} 
}
\caption{An example of induced space with points defined in the original network. We would like to induce a space $(S_X, s_X)$ such that $S_X$ includes infinite number of points formed by the convex of the original points $a$, $b$, and $c$, and $s_X$ is a semimetric for any pair of points in $S_X$. We want to induce the semimetric from the original network such that $s_X(a, b) = r_X(a, b)$, $s_X(a, c) = r_X(a, c)$, and $s_X(b, c) = r_X(b, c)$. Middle points, e.g. $e$, can be considered as the entity represents $50\%$ of $a$ and $50\%$ of $b$. An important observation is that any points in the space, e.g. $m$, can be written as a convex combination representing the proportion of their contents from original nodes -- $(m_a, m_b, m_c)$ with $m_a + m_b + m_c = 1$.\vspace{-4mm}
}
\label{fig_inside_example}
\end{figure}

%%%%%%%%%%%%%%%%%%%%%%%%%%%%%%%%%%%%%%%%%%%%%%%%%%%%%%%%%%%%%%%%%%%%%
%%%   M   A   I   N       M   A   T   T   E   R   %%%%%%%%%%%%%%%%%%%
%%%%%%%%%%%%%%%%%%%%%%%%%%%%%%%%%%%%%%%%%%%%%%%%%%%%%%%%%%%%%%%%%%%%%
%
The relationships in Lemma \ref{lemma_GH} and Corollary \ref{coro_lower_bound} are extensions of similar analyses that hold for the Gromov-Hausdorff distance between metrics spaces, \cite{kalton1997, linial1995}. As in the case of metric spaces, these results imply that the embedding distance $d_\EE(N_X, N_Y)$ can be used to lower bound the network distance $d_\C(N_X, N_Y)$ [cf. \eqref{eqn_equivalence}]. This value is in addition to the ability of the partial embedding distance $d_\PE(N_X, N_Y)$ of Definition \ref{dfn_partial_embedding} to measure how far the network $X$ is to being a subnetwork of network $Y$.

In the comparison of surfaces and shapes, the partial embedding distance $d_\PE(N_X, N_Y)$  has the attractive property of being approximable using multidimensional scaling techniques \cite{bronstein2006, bronstein2006b}. Our empirical analysis shows that the use of analogous techniques to estimate $d_\PE(N_X, N_Y)$ for arbitrary networks yields poor results and that this is related to how far the dissimilarities in $N_X$ and $N_Y$ are from satisfying the triangle inequality -- see the example in Figure \ref{fig_mds_original} and the numerical analysis in Section \ref{sec_application}. To improve the accuracy of multidimensional scaling estimates we propose to define the interior of a network by defining a space where dissimilarities between any pair of points represented by a convex combination of nodes in the given networks are defined (Section \ref{sec_space_induced}). We will further demonstrate that the proposed definition of the interior of a network is such that the partial embedding distances between networks with interiors are the same as the partial embedding distances between the corresponding original networks (Theorems \ref{thm_induced_space_distance_same} and \ref{thm_sampled_induced_space}). Empirical demonstrations will show that the comparison of networks with interiors using MDS techniques yields better results that are comparable to those obtained when comparing shapes and surfaces (Section \ref{sec_application}). 

%%%%%%%%%%%%%%%%%%%%%%%%%%%%%%%%%%%%%%%%%%%%%%%%%%%%%%%%%%%%%%%%%%%%%
%%%   R   E   M   A   R   K   %%%%%%%%%%%%%%%%%%%%%%%
%%%%%%%%%%%%%%%%%%%%%%%%%%%%%%%%%%%%%%%%%%%%%%%%%%%%%%%%%%%%%%%%%%%%%
%
%\begin{remark}\label{remark_lower_bound}\normalfont
%Another observation from Lemma \ref{lemma_GH} is that the two network distances are equivalent with $d_\EE \equiv d_\C$ if and only if the third term $\delta_{X, Y}(\phi, \psi)$ on the right hand side of \eqref{eqn_lemma_GH} is no greater than the maximum of the distances $\Delta_{X, Y}(\phi)$ and $\Delta_{Y, X}(\psi)$ measured by each of them, for the pair of mappings achieving the minimum in \eqref{eqn_lemma_GH}. In other words, the two networks can be mapped by a pair of mappings $\phi$ and $\psi$ such that the pair of maps are not that different from being inverse of each other.
%\end{remark}

%% file: figures/embedding_example.tex
%!TEX root = ../journal_network_embedding.tex

\def \thisplotscale {0.48}
\def \unit {\thisplotscale cm}

\def \length {3.9}

\pgfdeclarelayer{background}
\pgfdeclarelayer{foreground}
\pgfsetlayers{background,foreground}

\begin{tikzpicture}[-stealth, shorten >=0, x = 1.0*\unit, y=0.35*\unit, font=\scriptsize]

    \begin{pgfonlayer}{foreground}
         \node at (0, 0) (center1) {};
         \path (center1) ++ (135:\length * 1.5) node {$N_X$}; 
         \path (center1) ++ (90:\length) node (Xa) [fill = blue!20, vertex] {$a$};
         \path (center1) ++ (180:\length) node (Xb) [fill = blue!20, vertex] {$c$};
         \path (center1) ++ (0:\length) node (Xc) [fill = blue!20, vertex] {$b$};
         
         \node at (\length * 3.5, 0) (center2) {};
         \path (center2) ++ (135:\length * 1.5) node {$N_Y$}; 
         \path (center2) ++ (90:\length) node (Ya) [fill = blue!20, vertex] {$a$};
         \path (center2) ++ (180:\length) node (Yb) [fill = blue!20, vertex] {$c$};
         \path (center2) ++ (0:\length) node (Yc) [fill = blue!20, vertex] {$b$};
         
         \node at (\length * 7, 0) (center3) {};
         \path (center3) ++ (135:\length * 1.5) node {$N_Z$}; 
         \path (center3) ++ (90:\length) node (Za) [fill = blue!20, vertex] {$a$};
         \path (center3) ++ (180:\length) node (Zb) [fill = blue!20, vertex] {$c$};
         \path (center3) ++ (0:\length) node (Zc) [fill = blue!20, vertex] {$b$};
    \end{pgfonlayer}

    \begin{pgfonlayer}{background} 
        \path [-, left] (Xa) edge node {{$1$}} (Xb);
        \path [-, right] (Xa) edge node {{$1$}} (Xc);
        \path [-, below] (Xb) edge node {{$11$}} (Xc);
        
        \path [-, left] (Ya) edge node {{$3$}} (Yb);
        \path [-, right] (Ya) edge node {{$3$}} (Yc);
        \path [-, below] (Yb) edge node {{$11$}} (Yc);
        
        \path [-, left] (Za) edge node {{$5$}} (Zb);
        \path [-, right] (Za) edge node {{$5$}} (Zc);
        \path [-, below] (Zb) edge node {{$11$}} (Zc);
    \end{pgfonlayer}
    
\end{tikzpicture}

%% file: figures/points_inside_network.tex
%!TEX root = ../journal_network_embedding.tex

\def \thisplotscale {0.45}
\def \unit {\thisplotscale cm}

\def \length {3}

\pgfdeclarelayer{background}
\pgfdeclarelayer{foreground}
\pgfsetlayers{background,foreground}

\begin{tikzpicture}[-stealth, shorten >=0, x = 1.0*\unit, y=0.8*\unit, font=\scriptsize]

    \begin{pgfonlayer}{foreground}
         \node at (0, 0) (center1) {};
         \path (center1) ++ (120:\length * 2) node {$N_X = (X, r_X)$}; 
         \path (center1) ++ (90:\length * 1.7321) node (Xa) [fill = blue!20, vertex] {$a$};
         \path (center1) ++ (180:\length) node (Xc) [fill = blue!20, vertex] {$c$};
         \path (center1) ++ (0:\length) node (Xb) [fill = blue!20, vertex] {$b$};
         
         \path (center1) ++ (23:\length * 2) node {\Large $\Rightarrow$}; 
         
         \node at (\length * 3.5, 0) (center2) {};
         \path (center2) ++ (120:\length * 2) node {$(S_X, s_X)$}; 
         \path (center2) ++ (90:\length * 1.7321) node (Sa) [fill = blue!20, vertex] {$a$} ++ (0, 0.8) node {$(1, 0, 0)$};
         \path (center2) ++ (180:\length) node (Sc) [fill = blue!20, vertex] {$c$} ++ (-0.4, -0.8) node {$(0, 0, 1)$};
         \path (center2) ++ (0:\length) node (Sb) [fill = blue!20, vertex] {$b$} ++ (0.4, -0.8) node [] {$(0, 1, 0)$};
         
         \path (center2) ++ (60:\length * 1) node (Se) [fill = blue!20, vertex] {$e$} ++ (1.3, 0.8) node [] {$(0.5, 0, 0.5)$};
         \path (center2) ++ (120:\length * 1) node (Sf) [fill = blue!20, vertex] {$f$} ;
         \path (center2) ++ (0:\length * 0) node (Sg) [fill = blue!20, vertex] {$g$} ;
         
         \path (center2) ++ (100:\length * 0.65) node (Sm) [fill = blue!20, vertex] {$m$} ++ (1.3, -0.8) node [] {$(m_a, m_b, m_c)$};
    \end{pgfonlayer}
               
    \begin{pgfonlayer}{background} 
        \path [-, left] (Xa) edge node {{$r_X(a, b)$}} (Xc);
        \path [-, right] (Xa) edge node {{$r_X(a, c)$}} (Xb);
        \path [-, below] (Xc) edge node {{$r_X(b,c$)}} (Xb);
                
        \path [-, below] (Sa) edge node {} (Sc);
        \path [-, below] (Sa) edge node {} (Sb);
        \path [-, below] (Sc) edge node {} (Sb);
        
        \path [-, below, dashed] (Se) edge node {} (Sf);
        \path [-, below, dashed] (Se) edge node {} (Sg);
        \path [-, below, dashed] (Sf) edge node {} (Sg);
        
        \path [-, left, bend right = 45, pos = 0.3] (Sa) edge node {{$r_X(a, b)$}} (Sc);
        \path [-, right, bend left = 45, pos = 0.2] (Sa) edge node {{$r_X(a, c)$}} (Sb);
        \path [-, below, bend right = 20] (Sc) edge node {{$r_X(b,c$)}} (Sb);
    \end{pgfonlayer}
    
\end{tikzpicture}

%% file: sec_4_interiors.tex
%!TEX root = journal_network_embedding.tex

%%%%%%%%%%%%%%%%%%%%%%%%%%%%%%%%%%%%%%%%%%%%%%%%%%%%%%%%%%%%%%%%%%%%%
%%%   S   E   C   T   I   O   N   %%%%%%%%%%%%%%%%%%%%%%%%%%%%%%%%%%%
%%%%%%%%%%%%%%%%%%%%%%%%%%%%%%%%%%%%%%%%%%%%%%%%%%%%%%%%%%%%%%%%%%%%%
%
\section{Interiors}\label{sec_space_induced}

We provide a different perspective to think of networks as semimetric spaces where: (i) There are interior points defined by convex combinations of given nodes. (ii) Dissimilarities between these interior points are determined by the dissimilarities between the original points. To substantiate the formal definition below (Definition \ref{dfn_induced_space}) we discuss the problem of defining the interior of a network with three points. Such network is illustrated in Figure \ref{fig_inside_example} where nodes are denoted as $a$, $b$, and $c$ and dissimilarities are denoted as $r_X$. Our aim is to induce a space $(S_X, s_X)$ where the dissimilarities in the induced space are $s_X : S_X \times S_X \rightarrow \reals_+$. We require that $S_X$ preserve the distance of original points in $N_X$ such that $s_X(a, b) = r_X(a, b)$, $s_X(a, c) = r_X(a, c)$, and $s_X(b, c) = r_X(b,c)$. 

Points inside the network are represented in terms of convex combinations of the original points $a$, $b$, and $c$. Specifically, a point $m$ in the interior of the network is represented by the tuple $(m_a, m_b, m_c)$ which we interpret as indicating that $m$ contains an $m_a$ proportion of $a$, an $m_b$ proportion of $b$, and an $m_c$ proportion of $c$. Points $e$, $f$, and $g$ on Figure \ref{fig_inside_example} contain null proportions of some nodes and are interpreted as lying on the edges. Do notice that although we are thinking of $m$ as a point inside the triangle, a geometric representation does not hold. 

%%%%%%%%%%%%%%%%%%%%%%%%%%%%%%%%%%%%%%%%%%%%%%%%%%%%%%%%%%%%%%%%%%%%%
%%%   F   I   G   U   R   E   %%%%%%%%%%%%%%%%%%%%%%%%%%%%%%%%%%%%%%%
%%%%%%%%%%%%%%%%%%%%%%%%%%%%%%%%%%%%%%%%%%%%%%%%%%%%%%%%%%%%%%%%%%%%%
%
\begin{figure}[t]
\centerline{\input{figures/arbitrary_points}\vspace{-4mm}
 }
\caption{Comparing arbitrary points inside the space induced from networks of three nodes. Given a pair of nodes $p$ and $m$ in the induced space, we need to find paths from $p$ to $m$ that are consisted of vectors parallel to the direction of original nodes in the networks, e.g. $a$ to $b$, $a$ to $c$, and/or $b$ to $c$. We assume the direction of original nodes in the networks have unit amount of transformation. Potential choices of paths from $p$ to $m$ include: $p$ to $m$ via $q_1$, via $q_2$, or via $q_3$. Of them, the path $p$ to $m$ via $q_1$ has the smallest amount of transformation traversed along the path. There are paths in the form which involves vectors $p$ to $q_2$, $q_2$ to $q_4$, and $q_4$ to $m$; such paths would not give the optimal solution to \eqref{eqn_constraint_arbitraryNodes}.\vspace{-4mm}
}
\label{fig_arbitrary_points}
\end{figure}
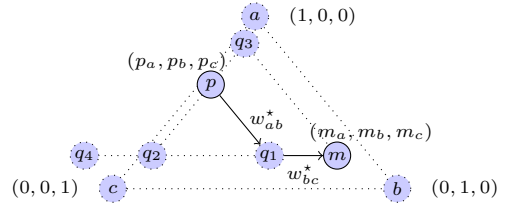

%%%%%%%%%%%%%%%%%%%%%%%%%%%%%%%%%%%%%%%%%%%%%%%%%%%%%%%%%%%%%%%%%%%%%
%%%   M   A   I   N      M   A   T   T   E   R   %%%%%%%%%%%%%%%%%%%%%%%%%%%%%%%%%%%
%%%%%%%%%%%%%%%%%%%%%%%%%%%%%%%%%%%%%%%%%%%%%%%%%%%%%%%%%%%%%%%%%%%%%
%
First we consider the case that the triangle inequality is satisfied by $r_X$. To evaluate the dissimilarities between $p$ represented by the tuple $(p_a, p_b, p_c)$ and $m$ represented by the tuple $(m_a, m_b, m_c)$ using dissimilarities in the original network, we need to find a path consisting of vectors parallel to the edges in the network that go to $m$ from $p$. Specifically, denote $w_{ab}$ as the proportion transversed in the direction from $a$ to $b$ in the path. For a positive value $w_{ab}$, compared to $p$, $m$ becomes more similar to $b$ by $w_{ab}$ units and less similar to $a$ by $-w_{ab}$ units; for a negative $w_{ab}$, compared to $p$, $m$ becomes more similar to $a$ and less similar to $b$. Proportion transversed in other directions, e.g. from $a$ to $c$ and from $b$ to $c$, are denoted as $w_{ac}$ and $w_{bc}$, respectively. For the path transversing $w_{ab}$ from $a$ to $b$, $w_{ac}$ from $a$ to $c$, and $w_{bc}$ from $b$ to $c$, the dissimilarity can be denoted as $|w_{ab}|r_X(a, b) + |w_{ac}|r_X(a, c) + |w_{bc}|r_X(b, c)$. There may be many different paths from $p$ to $m$, as illustrated in Figure \ref{fig_arbitrary_points}. Out of all paths, only the one yielding the smallest distance should be considered. This means the dissimilarity $s_X(p, m)$  between $p$ and $m$ can be defined by solving the following problem,
\begin{equation}\begin{aligned}\label{eqn_constraint_threeNodes_metric}
    \min \quad & \left| w_{ab} \right|r_X(a,b) + \left| w_{ac} \right|r_X(a,c) + \left| w_{bc} \right|r_X(b,c) \\
    \st \quad & m_a = p_a - w_{ab} - w_{ac}, \\
    & m_b = p_b + w_{ab} - w_{bc}, \\
    & m_c = p_c + w_{ac} + w_{bc}.
\end{aligned}\end{equation}

\noindent The constraints make sure that the path starts with tuple $(m_a, m_b, m_c)$ and ends with tuple $(p_a, p_b, p_c)$. This is like the definition of Manhattan distance. In fact, if Manhattan was a triangle with three endpoints and the roads in Manhattan were in a triangle grid, then the distance between any pair of points in Manhattan would be evaluated as in \eqref{eqn_constraint_threeNodes_metric}.

When relationships in $r_X$ do not satisfy triangle inequality, e.g. $r_X(a, b) + r_X(b, c) < r_X(a, c)$, however, the construction in \eqref{eqn_constraint_threeNodes_metric} is problematic since the optimal solution in \eqref{eqn_constraint_threeNodes_metric} would yield $s_X(a, c) = r_X(a, b) + r_X(b, c)$, which violates our requirement that $s_X(a, c)$ should be the same as $r_X(a, c)$. The problem arises because each segment in a given path contains two pieces of information -- the proportion of transformation, and the dissimilarity created of such transformation. E.g. for the path segment $pq_1$ in Figure \ref{fig_arbitrary_points}, it represents $w_{ab}^\star$ units of transformation from $a$ to $b$, and also denotes a dissimilarity between $p$ and $q_1$ as $|w_{ab}^\star|r_X(a, b)$. The two pieces of information unite when $r_X$ is a metric, however, create conflicts for dissimilarities in a general network. To resolve such issue, we could separate the amount of transformation from the dissimilarity incurred due to transformation. Firstly, we find the path with the smallest amount of transformation
\begin{equation}\begin{aligned}\label{eqn_constraint_threeNodes}
    \min \quad & \left| w_{ab} \right| + \left| w_{ac} \right| + \left| w_{bc} \right| \\
    \st \quad & m_a = p_a - w_{ab} - w_{ac}, \\
    & m_b = p_b + w_{ab} - w_{bc}, \\
    & m_c = p_c + w_{ac} + w_{bc}.
\end{aligned}\end{equation}
Then, for the optimal path $w_{ab}^\star$, $w_{ac}^\star$, and $w_{bc}^\star$ in \eqref{eqn_constraint_threeNodes}, define the dissimilarity as the distance transversed on the path, i.e.
\begin{align}\label{eqn_threeNode_pmDistance}
    s_X(p, m) \!=\! \left|w_{ab}^\star\right| r_X(a,b) \!+\! \left|w_{bc}^\star\right| r_X(b,c) \!+\! 
        \left| w_{ac}^\star\right| r_X(a,c).
\end{align}
The problem in \eqref{eqn_constraint_threeNodes} can always be solved since it is underdetermined due to the facts that $m_a + m_b + m_c = p_a + p_b + p_c = 1$. It traces back to \eqref{eqn_constraint_threeNodes_metric} when relationships in network are metrics. Moreover, it satisfy our requirement $s_X(a, b) = r_X(a, b)$, $s_X(a, c) = r_X(a, c)$, and $s_X(b, c) = r_X(b,c)$ for any networks. Regarding our previous example of a triangle-shaped Manhattan with three endpoints, suppose relationships in the network denote the amount of travel time between the endpoints. These relationship may not necessarily satisfy triangle inequalities. Suppose roads in Manhattan form a triangle grid, the problem in \eqref{eqn_constraint_threeNodes} is finding the shortest path between a pair of locations in Manhattan. The dissimilarity in \eqref{eqn_threeNode_pmDistance} describes the travel time between this pair of locations using the shortest path.

Given any network with arbitrary number of nodes, we define the induced space as a generalization to the case for nodes with three nodes we developed previously.

%%%%%%%%%%%%%%%%%%%%%%%%%%%%%%%%%%%%%%%%%%%%%%%%%%%%%%%%%%%%%%%%%%%%%
%%%   D   E   F   I   N   I   T   I   O   N   %%%%%%%%%%%%%%%%%%%%%%%%%%%%%%%%%%%
%%%%%%%%%%%%%%%%%%%%%%%%%%%%%%%%%%%%%%%%%%%%%%%%%%%%%%%%%%%%%%%%%%%%%
%
\begin{definition}\label{dfn_induced_space}
Given a network $N_X = (X, r_X)$ with $X = \{1, 2, \dots, n\}$, the induced space $(S_X, s_X)$ is defined such that the space $S_X$ is the convex hull of $X$ with $S_X = \{m = (m_1, m_2, \dots, m_n) \mid m_i \geq 0, \sum_{i\in X} m_i = 1\}$. Given a pair of nodes $m, p \in S_X$, the path yielding the smallest amount of transformation from $p$ to $m$ is obtained through the problem
\begin{equation}\begin{aligned}\label{eqn_constraint_arbitraryNodes}
    \left\{ w_{ij}^\star \right\} = \argmin ~ & \sum_{i, j \in X, i < j}\left| w_{ij} \right| \\
    \st ~ & m_i = p_i - \!\! \sum_{j \in X, j > i} w_{ij} + \!\!\sum_{j \in X, j < i} w_{ji}, ~ \forall i
\end{aligned}\end{equation}
The distance between $p$ and $m$ is then the distance traversed proportional to the original relationships weighted by the path,
\begin{align}\label{eqn_arbitrary_pmDistance}
    s_X(p, m) = \sum_{i, j \in X, i< j}\left|w_{ij}^\star\right| r_X(i, j).
\end{align}\end{definition}

%%%%%%%%%%%%%%%%%%%%%%%%%%%%%%%%%%%%%%%%%%%%%%%%%%%%%%%%%%%%%%%%%%%%%
%%%   M   A   I   N      M   A   T   T   E   R   %%%%%%%%%%%%%%%%%%%%%%%%%%%%%%%%%%%
%%%%%%%%%%%%%%%%%%%%%%%%%%%%%%%%%%%%%%%%%%%%%%%%%%%%%%%%%%%%%%%%%%%%%
%
The induced space $S_X$ is the convex hull constructed by all nodes $i \in X$. Each node in the induced space $m \in S_X$ can be represented as a tuple $(m_1, m_2, \dots, m_n)$ with $\sum_{i \in X}m_i = 1$ where $m_i$ represents the percentage of $m$ inheriting the property of node $i \in X$. To come up with distance between pairs of points $p, m \in S_X$ with the respective tuple representation $(p_1, p_2, \dots, p_n)$ and $(m_1, m_2, \dots, m_n)$, we consider each edge in the original space $X$, e.g. from $i$ to $j$, represents one unit of cost to transform $i$ into $j$. All edges are considered similarly with one unit of cost to transform the starting node into the ending node. We want to find the smallest amount of cost to transform $p$ into $m$. This is solved via \eqref{eqn_constraint_arbitraryNodes}, which is always solvable since the problem is underdetermined due to the facts that $\sum_{i \in X}m_i = \sum_{i \in X}p_i = 1$. This gives us the optimal path with weights $\{w_{ij}^\star\}$ meaning that the most cost-saving transformation from $p$ into $m$ is to undertaking $w_{ij}^\star$ unit of transformation along the direction of transforming $i$ into $j$. The distance in the induced space $s_X(p, m)$ is then the distance traversed proportional to the original relationships weighted by the path defined in \eqref{eqn_arbitrary_pmDistance}.

%%%%%%%%%%%%%%%%%%%%%%%%%%%%%%%%%%%%%%%%%%%%%%%%%%%%%%%%%%%%%%%%%%%%%
%%%   P   R   O   P   O   S   I   T   I   O   N   %%%%%%%%%%%%%%%%%%%%%%%%%%%%%%%%%%%
%%%%%%%%%%%%%%%%%%%%%%%%%%%%%%%%%%%%%%%%%%%%%%%%%%%%%%%%%%%%%%%%%%%%%
%
\begin{proposition}\label{prop_semimetric}
The space $(S_X, s_X)$ induced from $N_X = (X, r_X)$ defined in Definition \ref{dfn_induced_space} is a semimetric space in $S_X$. Moreover, the induced space preserves relationships: when $p, m \in X$, $s_X(p, m) = r_X(p, m)$.
\end{proposition}

%%%%%%%%%%%%%%%%%%%%%%%%%%%%%%%%%%%%%%%%%%%%%%%%%%%%%%%%%%%%%%%%%%%%%
%%%   P   R   O   O   F   %%%%%%%%%%%%%%%%%%%%%%%%%%%%%%%%%%%
%%%%%%%%%%%%%%%%%%%%%%%%%%%%%%%%%%%%%%%%%%%%%%%%%%%%%%%%%%%%%%%%%%%%%
%
\begin{myproof} See Appendix \ref{apx_proof_2} for proofs in Section \ref{sec_space_induced}. \end{myproof}

%%%%%%%%%%%%%%%%%%%%%%%%%%%%%%%%%%%%%%%%%%%%%%%%%%%%%%%%%%%%%%%%%%%%%
%%%   M   A   I   N      M   A   T   T   E   R   %%%%%%%%%%%%%%%%%%%%%%%%%%%%%%%%%%%
%%%%%%%%%%%%%%%%%%%%%%%%%%%%%%%%%%%%%%%%%%%%%%%%%%%%%%%%%%%%%%%%%%%%%
%
The semimetric established in Proposition \ref{prop_semimetric} guarantees that the points in the induced space with their dissimilarity $s_X(p, m)$ are well-behaved. We note that semimetric is the best property we can expect, since the triangle inequality may not be satisfied even for the dissimilarities in the original networks. Next we show that the embedding distance is preserved when interiors are considered.

%% file: figures/arbitrary_points.tex
%!TEX root = ../journal_network_embedding.tex

\def \thisplotscale {0.45}
\def \unit {\thisplotscale cm}

\def \length {4.2}

\pgfdeclarelayer{background}
\pgfdeclarelayer{foreground}
\pgfsetlayers{background,foreground}

\begin{tikzpicture}[-stealth, shorten >=0, x = 1.0*\unit, y=0.7*\unit, font=\scriptsize]

    \begin{pgfonlayer}{foreground}
         \node at (0, 0) (center1) {};
         \path (center1) ++ (90:\length * 1.7321) node (Xa) [fill = blue!20, dotted, vertex] {$a$} ++ (2, 0) node {$(1, 0, 0)$};
         \path (center1) ++ (180:\length) node (Xc) [fill = blue!20, dotted, vertex] {$c$} ++ (-2, 0) node {$(0, 0, 1)$};
         \path (center1) ++ (0:\length) node (Xb) [fill = blue!20, dotted, vertex] {$b$} ++ (2, 0) node {$(0, 1, 0)$};
         
         \path (center1) ++ (30:\length / 1.5) node (m) [fill = blue!20, vertex] {$m$} ++ (1, 1) node {$(m_a, m_b, m_c)$};
         \path (m) ++ (-2-1.732, 3) node (p) [fill = blue!20, vertex] {$p$} ++ (-1, 1) node {$(p_a, p_b, p_c)$};
         
         \path (m) ++ (-2, 0) node (q1) [fill = blue!20, vertex, dotted] {$q_1$};
         \path (m) ++ (-2 - 2 * 1.732, 0) node (q2) [fill = blue!20, vertex, dotted] {$q_2$};
         \path (p) ++ (1, 1.732) node (q3) [fill = blue!20, vertex, dotted] {$q_3$};
         \path (q2) ++ (-2, 0) node (q4) [fill = blue!20, vertex, dotted] {$q_4$};
         
    \end{pgfonlayer}
               
    \begin{pgfonlayer}{background} 
        \path [-, left, dotted] (Xa) edge node {{}} (Xc);
        \path [-, right, dotted] (Xa) edge node {{}} (Xb);
        \path [-, below, dotted] (Xc) edge node {{}} (Xb);
        
        \path [<-, below] (m) edge node {$w_{bc}^\star$} (q1);
        \path [<-, right] (q1) edge node {$w_{ab}^\star$} (p);
        
        \path [-, below, dotted] (q1) edge node {} (q2);
        \path [-, below, dotted] (q2) edge node {} (q3);
        \path [-, below, dotted] (q3) edge node {} (m);
        \path [-, below, dotted] (q2) edge node {} (q4);
    \end{pgfonlayer}
    
\end{tikzpicture}

%% file: sec_4a_distance_equivalance_sample.tex
%!TEX root = journal_network_embedding.tex

%%%%%%%%%%%%%%%%%%%%%%%%%%%%%%%%%%%%%%%%%%%%%%%%%%%%%%%%%%%%%%%%%%%%%
%%%   S   E   C   T   I   O   N   %%%%%%%%%%%%%%%%%%%%%%%%%%%%%%%%%%%
%%%%%%%%%%%%%%%%%%%%%%%%%%%%%%%%%%%%%%%%%%%%%%%%%%%%%%%%%%%%%%%%%%%%%
%
\subsection{Distances Between Networks Extended To Their Interiors}\label{sec_interior_distance}

Since semimetrics are induced purely from the relationships in the original network, a pair of networks $N_X$ and $N_Y$ can be compared by considering their induced space, as we state next.

%%%%%%%%%%%%%%%%%%%%%%%%%%%%%%%%%%%%%%%%%%%%%%%%%%%%%%%%%%%%%%%%%%%%%
%%%   D   E   F   I   N   I   T   I   O   N   %%%%%%%%%%%%%%%%%%%%%%%
%%%%%%%%%%%%%%%%%%%%%%%%%%%%%%%%%%%%%%%%%%%%%%%%%%%%%%%%%%%%%%%%%%%%%
%
\begin{definition}\label{dfn_induced_space_distance}
Given two networks $N_X = (X, r_X)$ and $N_Y = (Y, r_Y)$ with their respective induced space $(S_X, s_X)$ and $(S_Y, s_Y)$, for a map $\phi: S_X \rightarrow S_Y$ from the induced space $S_X$ to the induced space $S_Y$ such that $\phi(x) \in Y$ for any $x\in X$, define the network difference with respect to $\phi$ as 
\begin{align}\label{eqn_dfn_induced_space_distance_prelim}
    \Delta_{S_X, S_Y} (\phi)
         :=  \max_{x, x' \in S_X} 
            \Big| s_X(x, x') - s_Y (\phi(x),\phi(x')) \Big|.
\end{align}
The partial embedding distance from $N_X$ to $N_Y$ measured with respect to the induced spaces is then defined as
\begin{align}\label{eqn_dfn_induced_space_distance}
   d_\PES(N_X, N_Y) := \min_{\phi: S_X \rightarrow S_Y \mid \phi(x) \in Y, \forall x\in X } 
    \Big\{ \Delta_{S_X, S_Y} (\phi) \Big\}.
\end{align} \end{definition}

%%%%%%%%%%%%%%%%%%%%%%%%%%%%%%%%%%%%%%%%%%%%%%%%%%%%%%%%%%%%%%%%%%%%%
%%%   M   A   I   N      M   A   T   T   E   R   %%%%%%%%%%%%%%%%%%%%%%%%%%%%%%%%%%%
%%%%%%%%%%%%%%%%%%%%%%%%%%%%%%%%%%%%%%%%%%%%%%%%%%%%%%%%%%%%%%%%%%%%%
%
The partial embedding distance $d_\PES(N_X, N_Y)$ with respect to the induced space in \eqref{eqn_dfn_induced_space_distance} is defined similarly as the partial embedding distance $d_\PE(N_X, N_Y)$ in \eqref{eqn_dfn_partial_embedding} however considers the mapping between all elements in the induced spaces. %By the second part of Proposition \ref{prop_semimetric}, $s_X$ incorporates all information in $r_X$ and similarly for $s_Y$. Therefore, comparing the original networks via the induced spaces is reasonable. Simultaneously, since the induced space is richer, the problem where different networks result in identical multi-dimensional scaling results illustrated in Figure \ref{fig_mds_original} would be alleviated. 
Observe that we further require that the embedding satisfy $\phi(x) \in Y$ for any $x\in X$. This ensures the original nodes of network $X$ are mapped to original nodes of network $Y$. The restriction is incorporated because it makes the embedding distance $d_\PES(N_X, N_Y)$ with respect to the induced spaces identical to the original embedding distance $d_\PE(N_X, N_Y)$ as we state next.

%%%%%%%%%%%%%%%%%%%%%%%%%%%%%%%%%%%%%%%%%%%%%%%%%%%%%%%%%%%%%%%%%%%%%
%%%   F   I   G   U   R   E   %%%%%%%%%%%%%%%%%%%%%%%%%%%%%%%%%%%%%%%
%%%%%%%%%%%%%%%%%%%%%%%%%%%%%%%%%%%%%%%%%%%%%%%%%%%%%%%%%%%%%%%%%%%%%
%
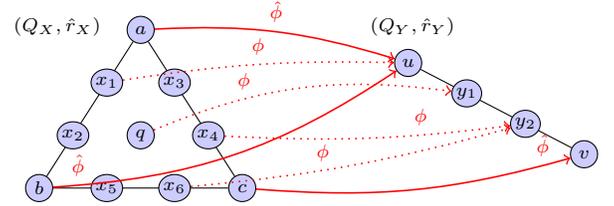
\begin{figure}[t]
\centerline{\input{figures/regular_sample_pair}\vspace{-3mm} 
}
\caption{An example of regular sample pair $(Q_X, \hhatr_X)$ and $(Q_Y, \hhatr_Y)$, where $Q_X = \{a, b, c, x_1, x_2, x_3, x_4, x_5, x_6, q\}$ collects original points and induced points that are combination of one-third of original nodes, and $Q_Y = \{u, v, y_1, y_2\}$. Here we illustrate a specific mapping $\hhatphi$ with $\hhatphi(a) = \hhatphi(b) = u$ and $\hhatphi(c) = v$; it is apparent that $\phi(x) \in Q_Y$ for any $x \in Q_X$. Note that $\tdQ_X = Q_X / \{q\}$ also form a regular sample pair with $Q_Y$.\vspace{-3mm}
}
\label{fig_regular_sample_pair}
\end{figure}

%%%%%%%%%%%%%%%%%%%%%%%%%%%%%%%%%%%%%%%%%%%%%%%%%%%%%%%%%%%%%%%%%%%%%
%%%   F   I   G   U   R   E   %%%%%%%%%%%%%%%%%%%%%%%%%%%%%%%%%%%%%%%
%%%%%%%%%%%%%%%%%%%%%%%%%%%%%%%%%%%%%%%%%%%%%%%%%%%%%%%%%%%%%%%%%%%%%
%
\begin{figure*}[t]
\centerline{\input{figures/embedding_sample_3.tex}\vspace{-1mm} 
}
\begin{minipage}[h]{0.032\textwidth}
~
\end{minipage}
\begin{minipage}[h]{0.275\textwidth}
    	\centering
    	\includegraphics[trim=0.01cm 0.01cm 0.01cm 0.01cm, clip=true, width=1 \textwidth]
	{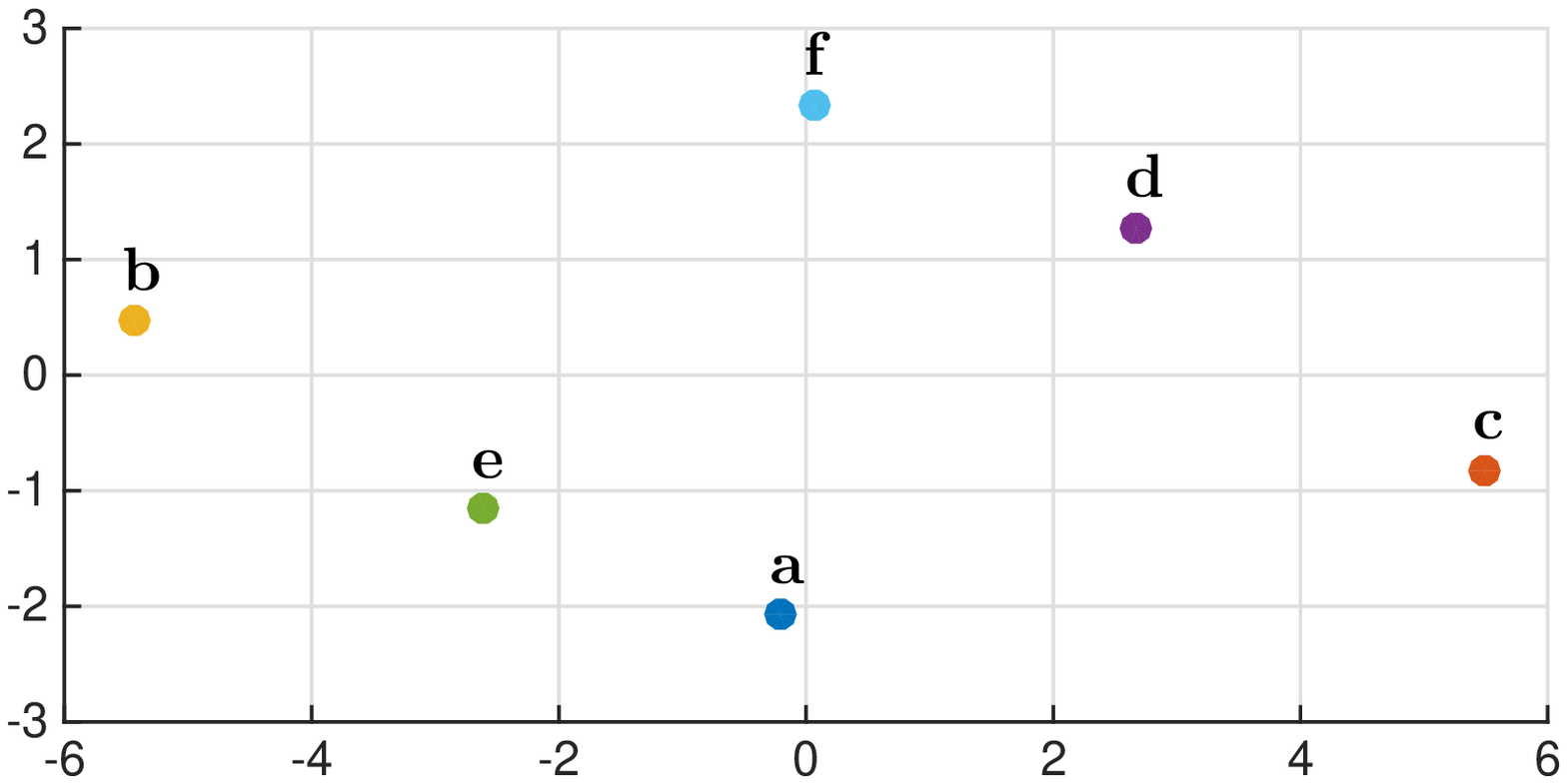}
\end{minipage}
\begin{minipage}[h]{0.054\textwidth}
~
\end{minipage}
\begin{minipage}[h]{0.275\textwidth}
    	\centering
    	\includegraphics[trim=0.01cm 0.01cm 0.01cm 0.01cm, clip=true, width=1 \textwidth]
	{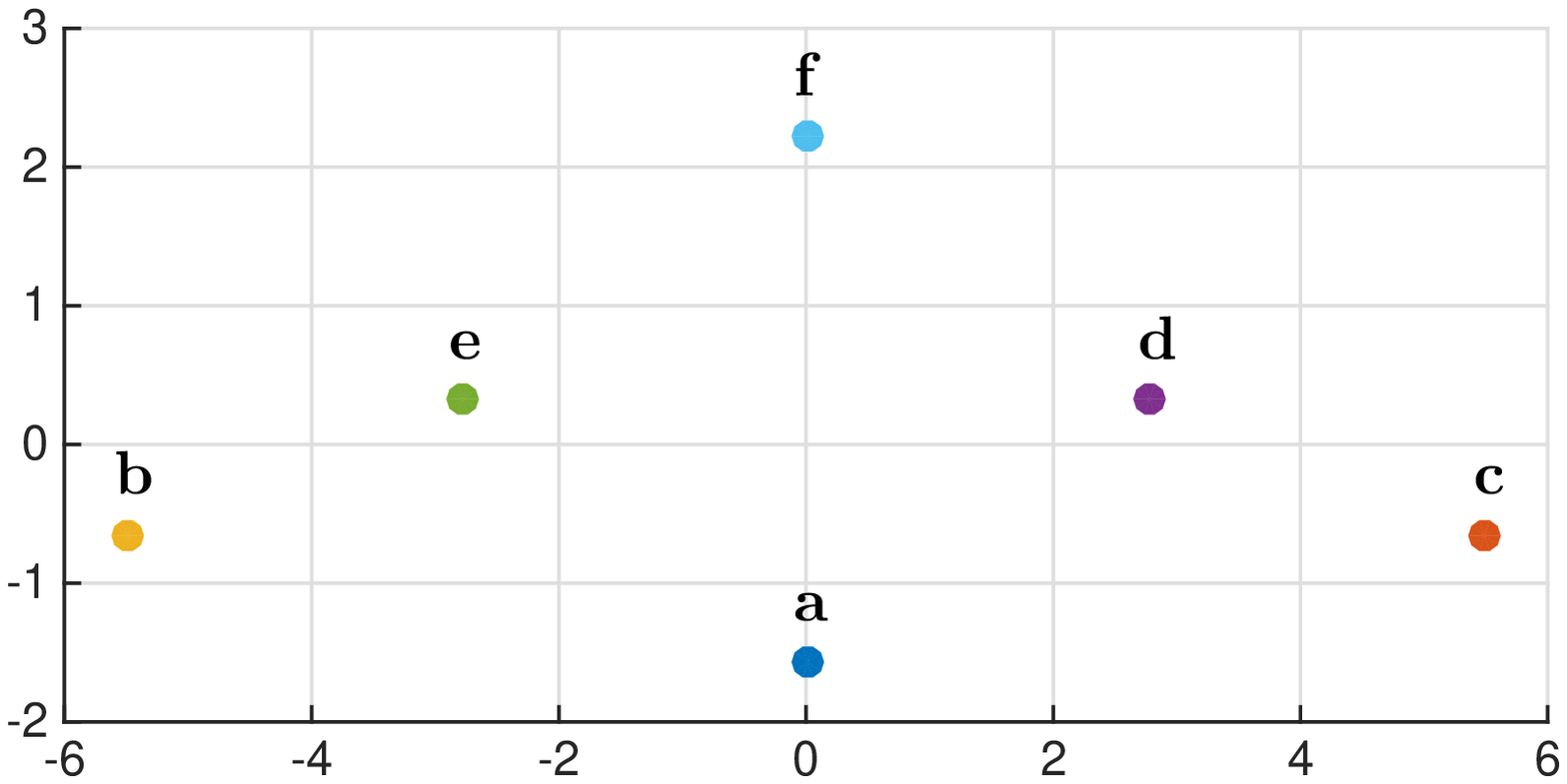}
\end{minipage}
\begin{minipage}[h]{0.054\textwidth}
~
\end{minipage}
\begin{minipage}[h]{0.275\textwidth}
    	\centering
    	\includegraphics[trim=0.01cm 0.01cm 0.01cm 0.01cm, clip=true, width=1 \textwidth]
	{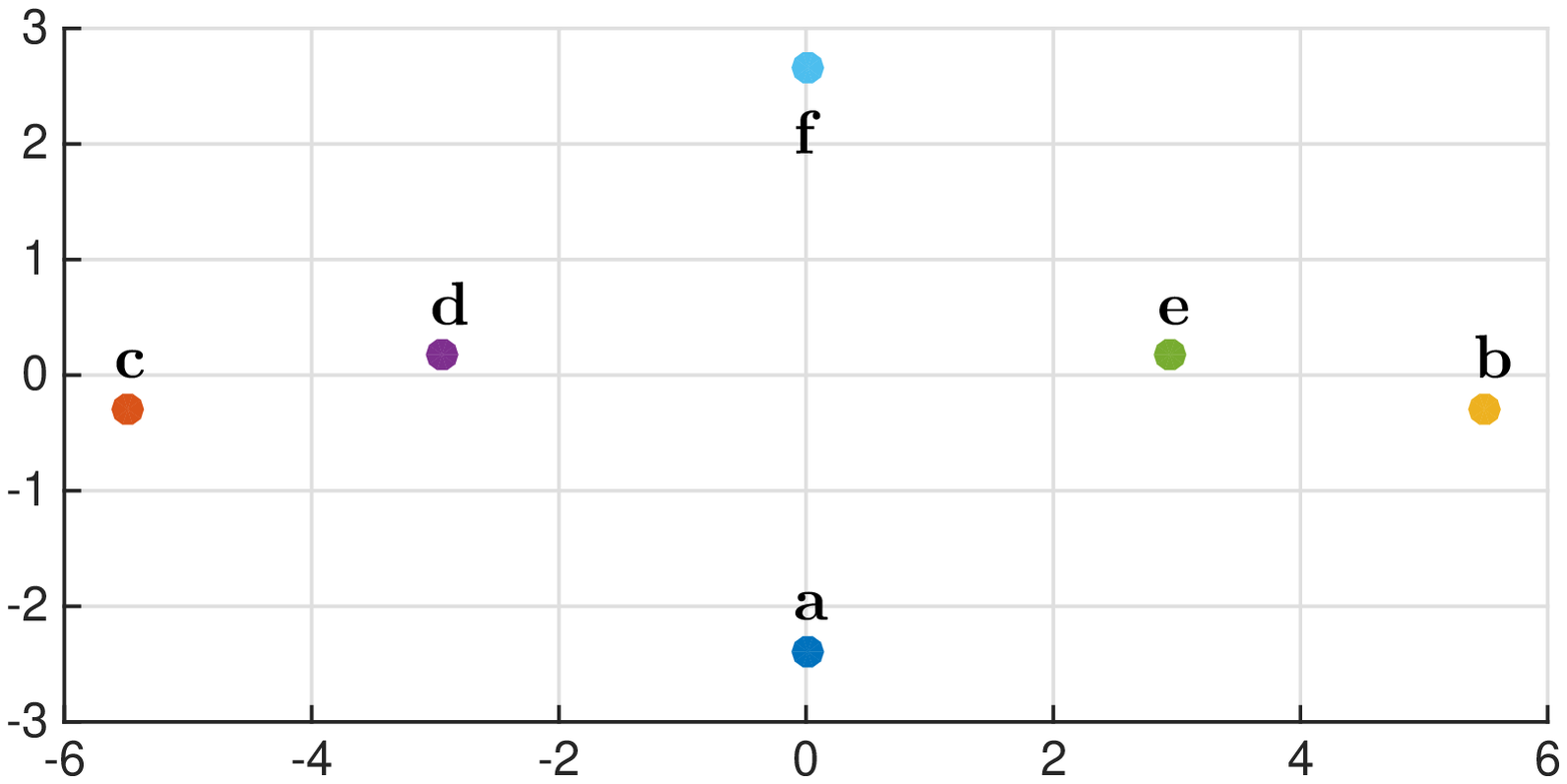}
\end{minipage}
\caption{The caveat illustrated in Figure \ref{fig_mds_original} where different networks results in identical multi-dimensional scaling results could be solved by considering the induced sample space where we utilize the same predetermined sampling strategy -- taking midpoints for all edges -- in the network. Multi-dimensional scaling by adding interiors would distinguish different networks. The dissimilarities between nodes $a$ and $f$ are not illustrated in the respective induced network due to space limit.\vspace{-3mm}
}
\label{fig_mds_midpoints}
\end{figure*}

%%%%%%%%%%%%%%%%%%%%%%%%%%%%%%%%%%%%%%%%%%%%%%%%%%%%%%%%%%%%%%%%%%%%%
%%%   T   H   E   O   R   E   M   %%%%%%%%%%%%%%%%%%%%%%%%%%%%%%%%%%%
%%%%%%%%%%%%%%%%%%%%%%%%%%%%%%%%%%%%%%%%%%%%%%%%%%%%%%%%%%%%%%%%%%%%%
%
\begin{theorem}\label{thm_induced_space_distance_same}
The function $d_\PES: \ccalN \times \ccalN \rightarrow \reals_+$ defined in \eqref{eqn_dfn_induced_space_distance} is an embedding metric in the space $\ccalN$ and  yields the same distance as the function $d_\PE$ defined in \eqref{eqn_dfn_partial_embedding}, 
\begin{align}\label{eqn_thm_induced_space_distance_same}
    d_\PES(N_X, N_Y) = d_\PE(N_X, N_Y),
    \quad \text {for all\ } N_X, N_Y.
\end{align} \end{theorem}

%%%%%%%%%%%%%%%%%%%%%%%%%%%%%%%%%%%%%%%%%%%%%%%%%%%%%%%%%%%%%%%%%%%%%
%%%   M   A   I   N      M   A   T   T   E   R   %%%%%%%%%%%%%%%%%%%%%%%%%%%%%%%%%%%
%%%%%%%%%%%%%%%%%%%%%%%%%%%%%%%%%%%%%%%%%%%%%%%%%%%%%%%%%%%%%%%%%%%%%
%
The statement in Theorem \ref{thm_induced_space_distance_same} justifies comparing networks via their respective induced space. Similar as in Definition \ref{dfn_embedding}, defining $\max\{d_\PES(N_X, N_Y), d_\PES(N_Y, N_X)\}$ would yield a metric in the space $\ccalN \mod \cong$ and this maximum is the same as $d_\EE$ defined in \eqref{eqn_dfn_embedding}. Since the induced spaces incorporate more information of the original networks while at the same time $d_\PES(N_X, N_Y) = d_\PE(N_X, N_Y)$, an approximation to $d_\PES(N_X, N_Y)$ via the induced space would be a better approximation to $d_\PE(N_X, N_Y)$. It may appear that the evaluation of the induced space is costly. However, we demonstrate in the next subsection that the partial embedding distances have a nice property that if we sample a number of points in the induced spaces respectively according to the same rule, the distance between the sampled induced space is the same as the original distance. Despite that the definition of interiors of networks is somewhat arbitrary, its practical usefulness can be justified from Theorem \ref{thm_induced_space_distance_same}.

%%%%%%%%%%%%%%%%%%%%%%%%%%%%%%%%%%%%%%%%%%%%%%%%%%%%%%%%%%%%%%%%%%%%%
%%%   S   E   C   T   I   O   N   %%%%%%%%%%%%%%%%%%%%%%%%%%%%%%%%%%%
%%%%%%%%%%%%%%%%%%%%%%%%%%%%%%%%%%%%%%%%%%%%%%%%%%%%%%%%%%%%%%%%%%%%%
%
\subsection{Sampling Of Interiors}\label{sec_partial_embedding_induced}

In this section, we consider a practical scenario where we only take several samples in the induced space. We show that comparing the combination of nodes in the respective original networks and sampled nodes in the induced space would yield the same result as comparing the original networks. Given a network $N_X = (X, r_X)$, our aim is to define a sampled induced space $(Q_X, \hhatr_X)$ where $Q_X \supset X$ includes more nodes compared to $X$. An example is in Figure \ref{fig_mds_midpoints}, where the original node space is given by $\{a, b, c\}$, and one version of sampled induced node space is $Q_X = \{a, b, c, e, f, g\}$, the union of the original nodes and the nodes in the midpoints of the edges in the original networks. The distance in the sampled induced space $\hhatr_X: Q_X \times Q_X \rightarrow \reals_+$ should preserve the distance of original points in $X$. A natural choice for $\hhatr_X$ is the restriction of the distance $s_X$ defined for the induced space $S_X$: i.e. given any pair of points $x, x' \in Q_X $, let $\hhatr_X(x, x') := s_X(x, x')$. Our key observation for such construction is that if the nodes in the induced spaces of a pair of networks are sampled according to the same strategy, then the distance between the sampled induced space is identical to the original distance. We start by formally describing what do we mean by a pair of networks sampled according to the same rule as next.

%%%%%%%%%%%%%%%%%%%%%%%%%%%%%%%%%%%%%%%%%%%%%%%%%%%%%%%%%%%%%%%%%%%%%
%%%   D   E   F   I   N   I   T   I   O   N   %%%%%%%%%%%%%%%%%%%%%%%%%%%%%%%%%%%
%%%%%%%%%%%%%%%%%%%%%%%%%%%%%%%%%%%%%%%%%%%%%%%%%%%%%%%%%%%%%%%%%%%%%
%
\begin{definition}\label{dfn_sample_same_rule}
Given a pair of networks $N_X = (X, r_X)$ and $N_Y = (Y, r_Y)$, their respective sampled space $(Q_X, \hhatr_X)$ and $(Q_Y, \hhatr_Y)$ form a regular sample pair, if for any mapping $\hhatphi: X \rightarrow Y$ in the original node set, we have $\phi(x) \in Q_Y$ for any $x \in Q_X$, where $\phi: S_X \rightarrow S_Y$ is the map induced from $\hhatphi$ such that $\phi: x \mapsto \phi(x)$ whose the $i$-th element in the tuple representation $[\phi(x)]_i$ is
\begin{align}\label{eqn_dfn_sample_same_rule_phi_mapping}
    [\phi(x)]_i = \sum_{j \in X} \bee{\hhatphi(j) = i} x_j,
\end{align}
and for any mapping $\hhatpsi: Y \rightarrow X$ in the original node set, we have $\psi(y) \in Q_X$ for any $x \in Q_X$ where $\psi: S_Y \rightarrow S_X$ is the map induced from $\hhatpsi$ such that the $j$-th element in the tuple representation of $\psi(y)$ is 
\begin{align}\label{eqn_dfn_sample_same_rule_psi_mapping}
    [\psi(y)]_j = \sum_{i \in Y} \bee{\hhatpsi(i) = j} y_i.
\end{align}
\end{definition}

%%%%%%%%%%%%%%%%%%%%%%%%%%%%%%%%%%%%%%%%%%%%%%%%%%%%%%%%%%%%%%%%%%%%%
%%%   M   A   I   N      M   A   T   T   E   R   %%%%%%%%%%%%%%%%%%%%%%%%%%%%%%%%%%%
%%%%%%%%%%%%%%%%%%%%%%%%%%%%%%%%%%%%%%%%%%%%%%%%%%%%%%%%%%%%%%%%%%%%%
%
\noindent In the definition, $\beeInline{\hhatphi(j) = i}$ is the indicator function such that it equals one if $\hhatphi$ maps $j\in X $ to $i\in Y$ and $\beeInline{\hhatphi(j) = i} = 0$ otherwise. The notation $[\phi(x)]_i$ denotes the proportion of $\phi(x)$ coming from $i$-th node in $Y$. It is easy to see that $\phi$ in \eqref{eqn_dfn_sample_same_rule_phi_mapping} is well-defined. Firstly, $[\phi(x)]_i \geq 0$ for any $i \in Y$, and
\begin{align}\label{eqn_dfn_sample_same_rule_phi_mapping_proof}
    \sum_{i \in Y}[\phi(x)]_i = \sum_{i \in Y}\sum_{j \in X} \bee{\hhatphi(j) = i} x_j 
        = \sum_{j \in X} x_j = 1,
\end{align}
ensuring $\phi(x)$ is in the induced convex hull space $S_Y$. Secondly, for any $j \in X$ in the original nodespace, its mapping $\phi(j)$ would have the tuple representation with $[\phi(j)]_i = \beeInline{\hhatphi(j) = i}$, a node in the original node space of $Y$. Consequently, for any $j \in X$, we have that $\phi(j) \in Y$. Combining these two observations imply that $\phi: S_X \rightarrow S_Y$ is well-defined. By symmetry, $\psi$ induced from $\hhatpsi$ is also well-defined from $S_Y$ to $S_X$. Definition \ref{dfn_sample_same_rule} states that for any point $x$ in $Q_X$, no matter how we relate points in $Q_X$ to points in $Q_Y$, the mapped node $\phi(x)$ should be in the induced sample space $Q_Y$. An example of regular sample pair is illustrated in Figure \ref{fig_regular_sample_pair}, where $Q_X = \{a, b, c, x_1, x_2, x_3, x_4, x_5, x_6, q\}$ is the collection of original node space and points that are combination of one-third of original nodes and $Q_Y = \{u, v, y_1, y_2\}$. Figure \ref{fig_regular_sample_pair} exemplifies the scenario for a specific mapping $\hhatphi$ with $\hhatphi(a) = \hhatphi(b) = u$ and $\hhatphi(c) = v$; it is apparent that $\phi(x) \in Q_Y$ for any $x \in Q_X$. We note that $\tdQ_X = Q_X / \{q\}$ also form a regular sample pair with $Q_Y$. A pair of networks $N_X$ and $N_Y$ can be compared by evaluating their difference in their respective sampled induced space as next.

%%%%%%%%%%%%%%%%%%%%%%%%%%%%%%%%%%%%%%%%%%%%%%%%%%%%%%%%%%%%%%%%%%%%%
%%%   D   E   F   I   N   I   T   I   O   N   %%%%%%%%%%%%%%%%%%%%%%%
%%%%%%%%%%%%%%%%%%%%%%%%%%%%%%%%%%%%%%%%%%%%%%%%%%%%%%%%%%%%%%%%%%%%%
%
\begin{definition}\label{dfn_sampled_space_distance}
Given two networks $N_X = (X, r_X)$ and $N_Y = (Y, r_Y)$ with their respective sampled induced space $(Q_X, \hhatr_X)$ and $(Q_Y, \hhatr_Y)$, for a map $\phi: Q_X \rightarrow Q_Y$ such that $\phi(x) \in Y$ for any $x\in X$, define the difference with respect to $\phi$ as 
\begin{align}\label{eqn_dfn_sampled_space_distance_prelim}
    \Delta_{Q_X, Q_Y} (\phi)
         :=  \max_{x, x' \in Q_X} 
            \Big| \hhatr_X(x, x') - \hhatr_Y (\phi(x),\phi(x')) \Big|.
\end{align}
The partial embedding distance from $N_X$ to $N_Y$ measured with respect to the sampled induced spaces is then defined as
\begin{align}\label{eqn_dfn_sampled_space_distance}
   d_\PEQ(N_X, N_Y) \!:=\! \min_{\phi: Q_X \rightarrow Q_Y \mid \phi(x) \in Y, \forall x\in X } 
    \Big\{ \Delta_{Q_X, Q_Y} (\phi) \Big\}.
\end{align} \end{definition}

%%%%%%%%%%%%%%%%%%%%%%%%%%%%%%%%%%%%%%%%%%%%%%%%%%%%%%%%%%%%%%%%%%%%%
%%%   M   A   I   N      M   A   T   T   E   R   %%%%%%%%%%%%%%%%%%%%%%%%%%%%%%%%%%%
%%%%%%%%%%%%%%%%%%%%%%%%%%%%%%%%%%%%%%%%%%%%%%%%%%%%%%%%%%%%%%%%%%%%%
%
Our key result is that $d_\PEQ(N_X, N_Y)$ is the same as the partial embedding distance $d_\PE(N_X, N_Y)$ defined in \eqref{eqn_dfn_partial_embedding} when the sampled space form a regular sample pair.

%%%%%%%%%%%%%%%%%%%%%%%%%%%%%%%%%%%%%%%%%%%%%%%%%%%%%%%%%%%%%%%%%%%%%
%%%   F   I   G   U   R   E   %%%%%%%%%%%%%%%%%%%%%%%
%%%%%%%%%%%%%%%%%%%%%%%%%%%%%%%%%%%%%%%%%%%%%%%%%%%%%%%%%%%%%%%%%%%%%
%
\begin{figure*}[t]
\begin{minipage}[h]{0.40\linewidth}
    	\centering
    	\input{figures/simple_points.tex}
\end{minipage}
\begin{minipage}[h]{0.045\linewidth}
        ~
\end{minipage}
\begin{minipage}[h]{0.22\linewidth}
    	\centering
    	\includegraphics[trim=1.5cm 0.9cm 1.5cm 0.7cm, clip=true, width=1 \textwidth]
	{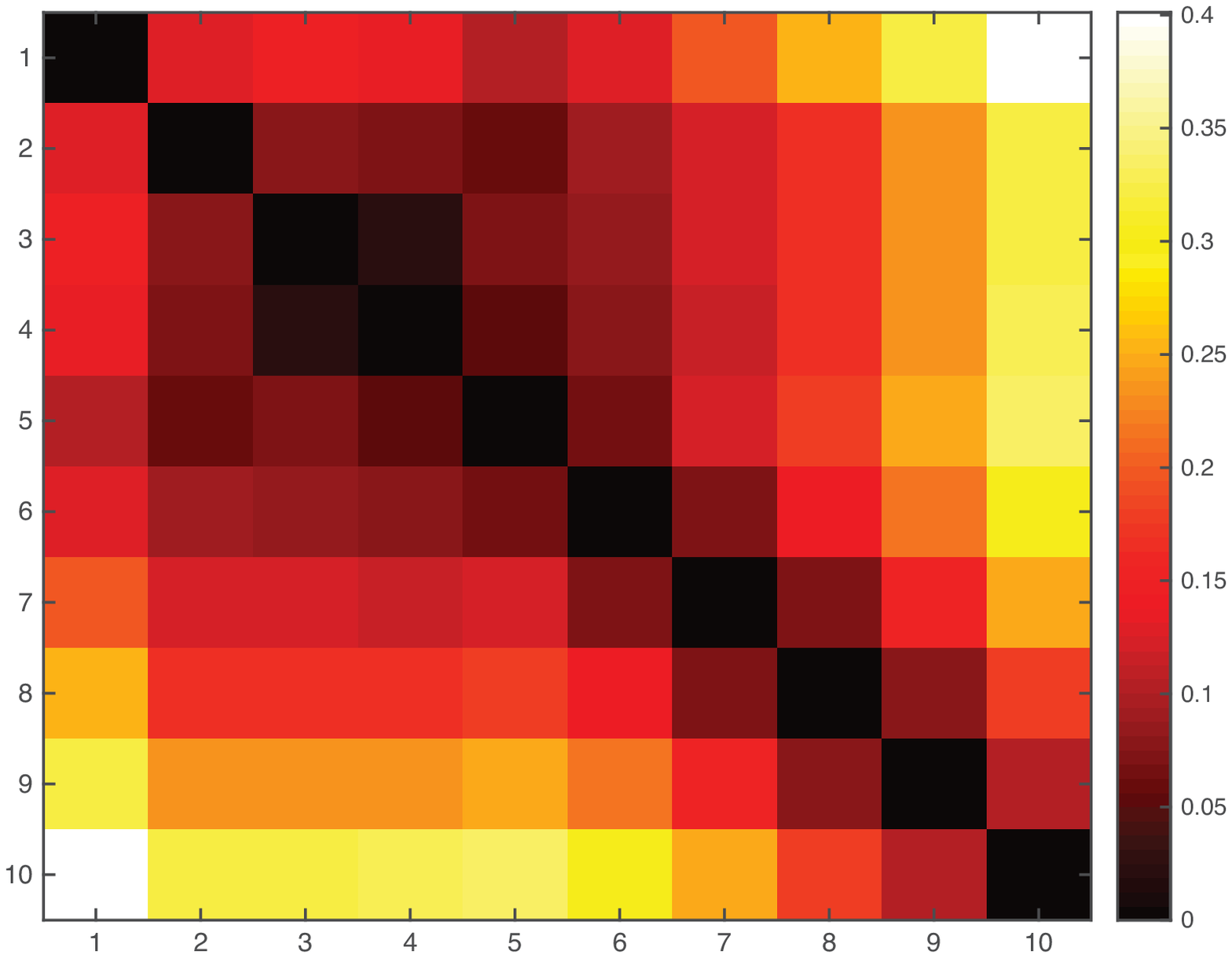}
	{\scriptsize (b) with interiors}
\end{minipage}
\begin{minipage}[h]{0.045\linewidth}
        ~
\end{minipage}
\begin{minipage}[h]{0.22\linewidth}
    	\centering
    	\includegraphics[trim=1.5cm 0.9cm 1.5cm 0.7cm, clip=true, width=1 \textwidth]
	{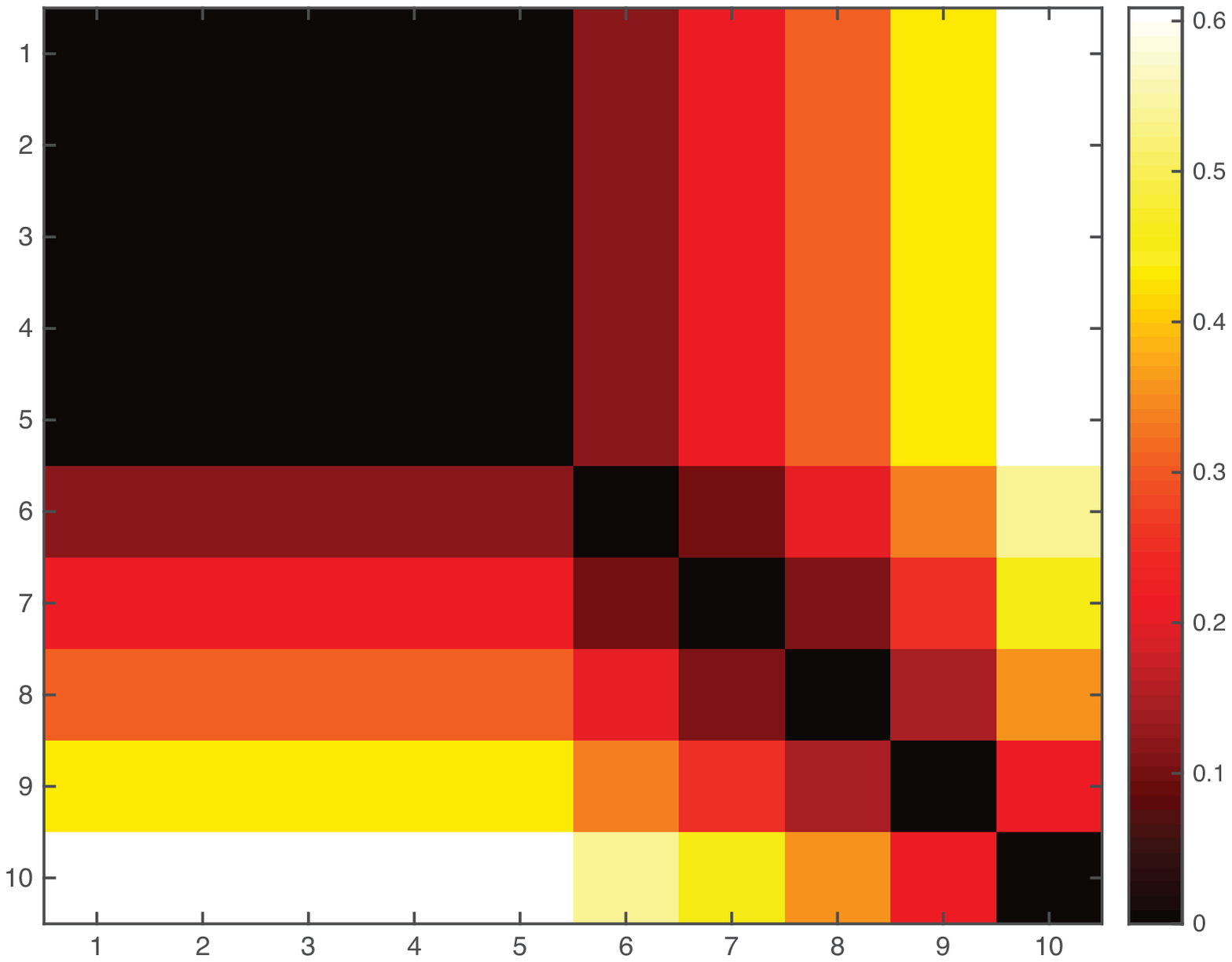}
	{\scriptsize (c) without interiors}
\end{minipage}
\caption{Example of the usefulness of considering the interior of networks. We consider 10 networks in the form (a) where $\gamma = 1, 2, \dots, 10$. Approximations of network embedding distances are evaluated. (b) and (c) illustrate the heat-maps of the distance approximations where the indices in both horizontal and vertical directions denote $\gamma$ in the networks. When interiors are considered by adding midpoints of edges, e.g. nodes $e$, $f$, and $g$ in (a), network distance approximations (b) yield more desired results, especially for $\gamma \leq 5$ where the relationships in the original networks fail to satisfy triangle inequality.}
\label{fig_application_0}
\end{figure*}

%%%%%%%%%%%%%%%%%%%%%%%%%%%%%%%%%%%%%%%%%%%%%%%%%%%%%%%%%%%%%%%%%%%%%
%%%   F   I   G   U   R   E   %%%%%%%%%%%%%%%%%%%%%%%%%%%%%%%%%%%%%%%
%%%%%%%%%%%%%%%%%%%%%%%%%%%%%%%%%%%%%%%%%%%%%%%%%%%%%%%%%%%%%%%%%%%%%
%
\begin{figure*}[t]
\begin{minipage}[h]{0.246\linewidth}
    	\centering
    	\includegraphics[trim=1.5cm 0.9cm 1.5cm 0.7cm, clip=true, width=1 \textwidth]
	{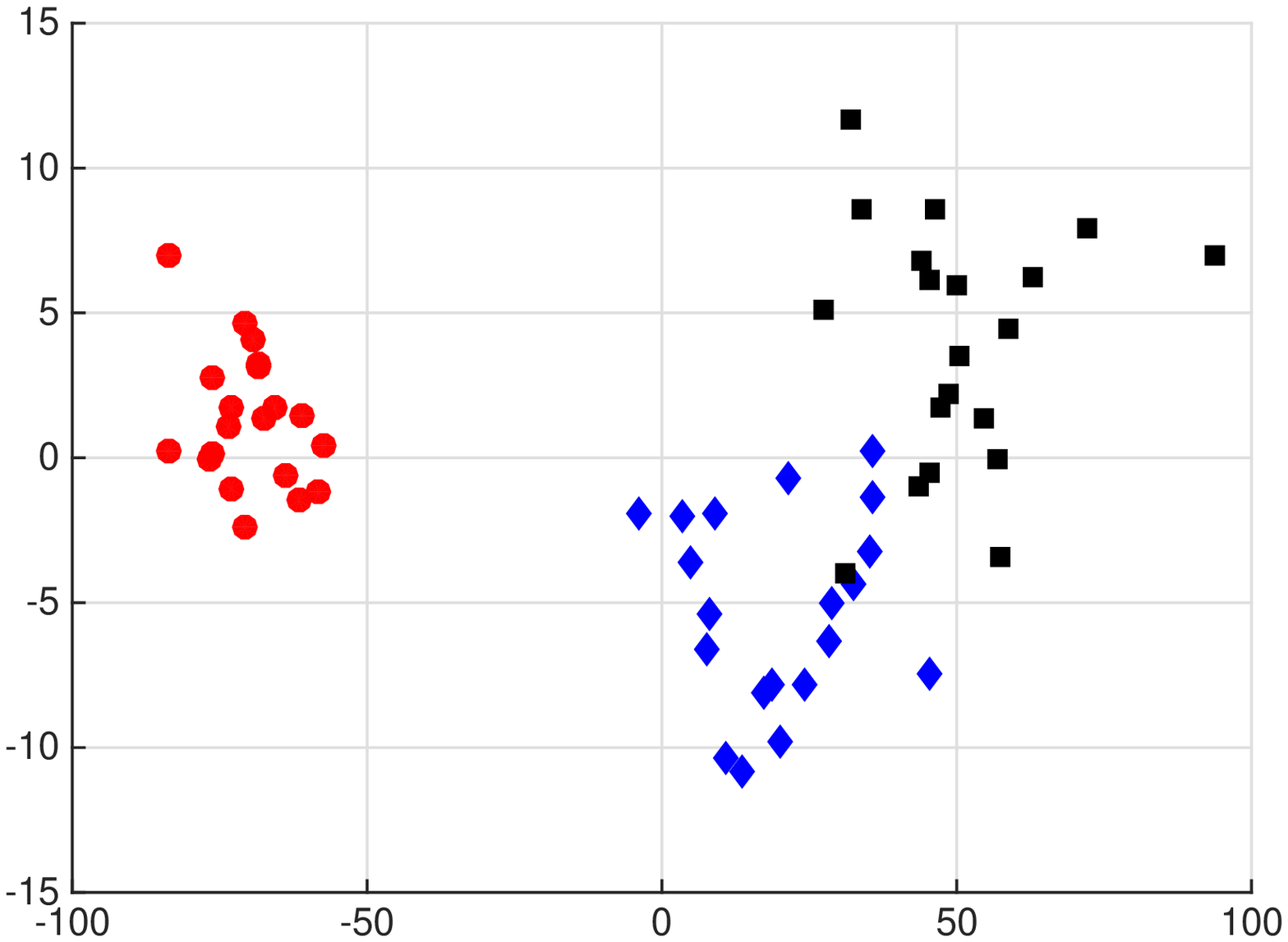}
	{\scriptsize (a) with interiors, $25$ nodes}
\end{minipage}
\begin{minipage}[h]{0.246\linewidth}
    	\centering
    	\includegraphics[trim=1.5cm 0.9cm 1.5cm 0.7cm, clip=true, width=1 \textwidth]
	{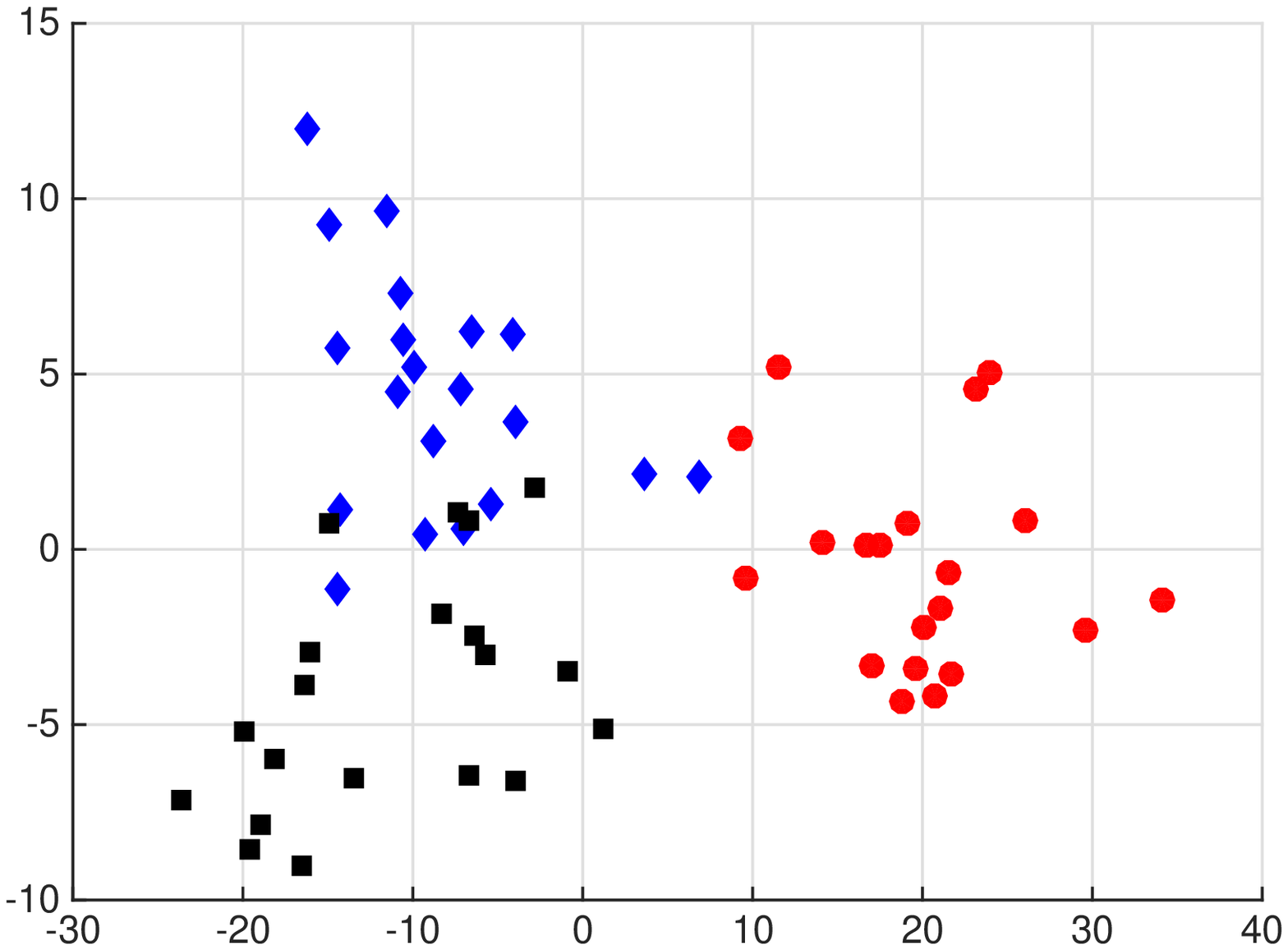}
	{\scriptsize (b) without interiors, $25$ nodes}
\end{minipage}
\begin{minipage}[h]{0.246\linewidth}
    	\centering
    	\includegraphics[trim=1.5cm 0.9cm 1.5cm 0.7cm, clip=true, width=1 \textwidth]
	{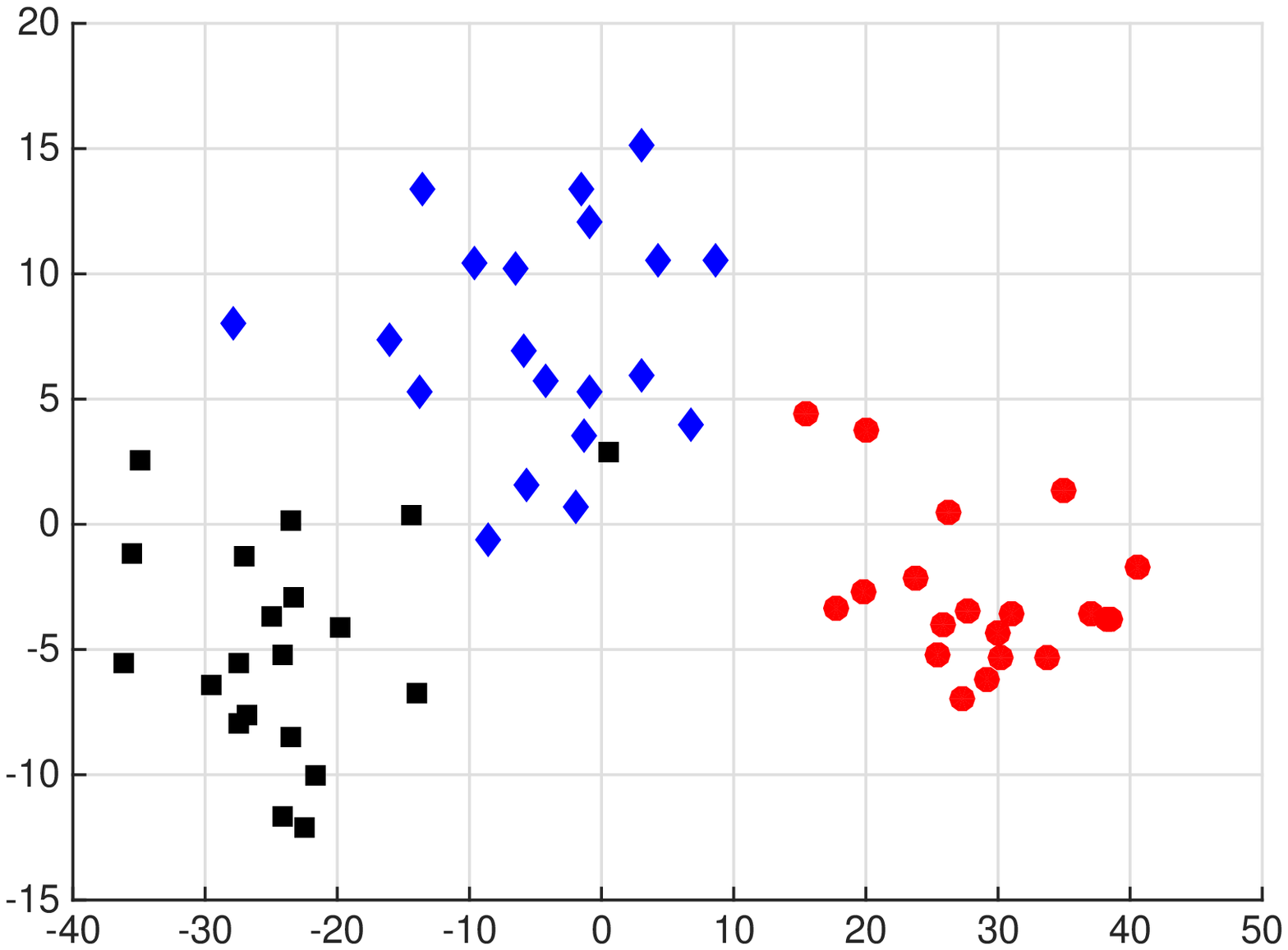}
	{\scriptsize (c) with interiors, $20$ to $25$ nodes}
\end{minipage}
\begin{minipage}[h]{0.246\linewidth}
    	\centering
    	\includegraphics[trim=1.5cm 0.9cm 1.5cm 0.7cm, clip=true, width=1 \textwidth]
	{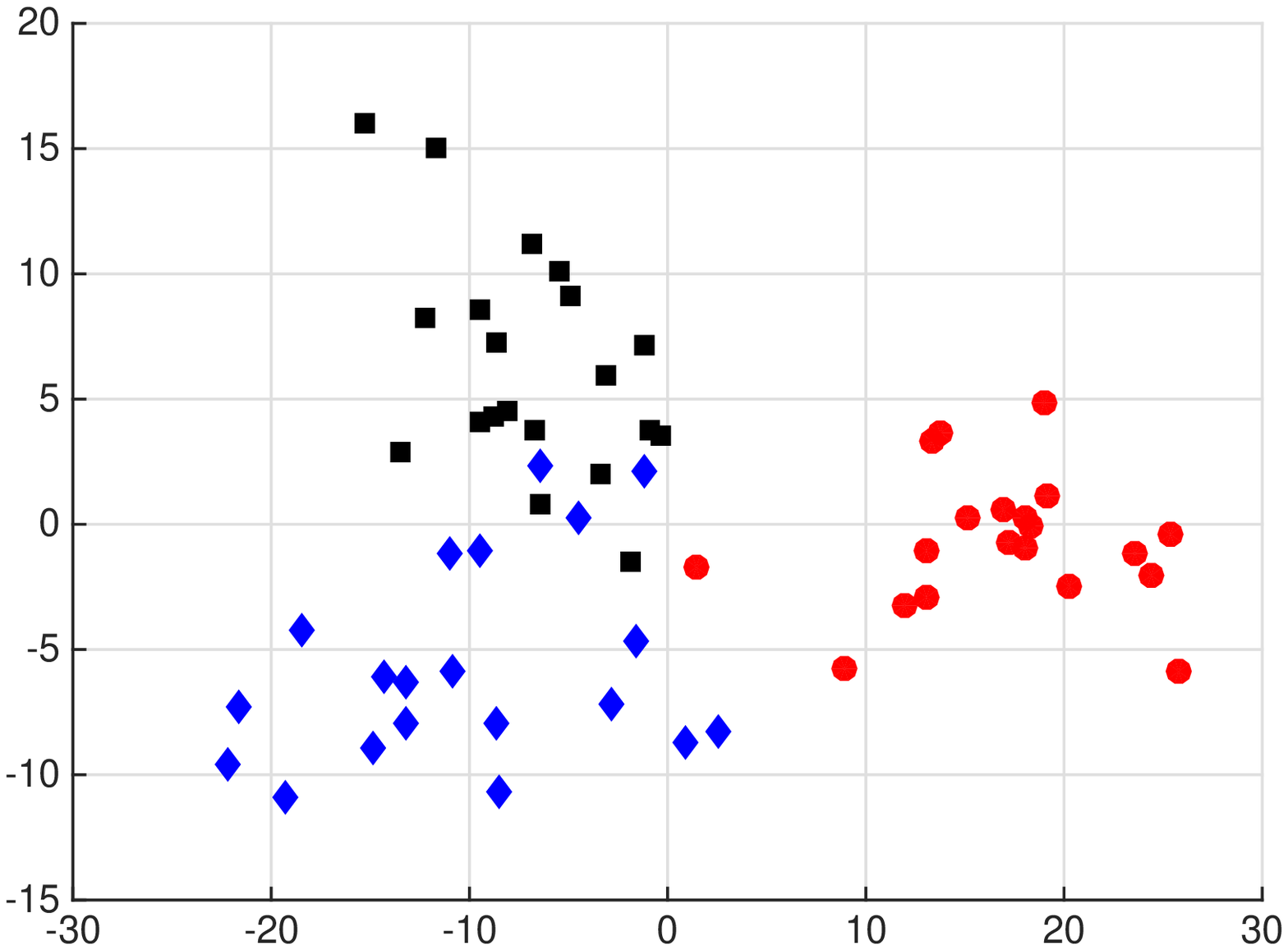}
	{\scriptsize (d) without interiors, $20$ to $25$ nodes}
\end{minipage}
\caption{Two dimensional Euclidean embeddings of the networks constructed from three different models with different number of nodes with respect to the approximation to the network embedding distance. In the embeddings, red circles denote networks constructed from the Erd\H os-R\' enyi model, blue diamonds represent networks constructed from the unit circle model, and black squares the networks from the correlation model.\vspace{-4mm}
}
\label{fig_application}
\end{figure*}

%%%%%%%%%%%%%%%%%%%%%%%%%%%%%%%%%%%%%%%%%%%%%%%%%%%%%%%%%%%%%%%%%%%%%
%%%   T   H   E   O   R   E   M   %%%%%%%%%%%%%%%%%%%%%%%%%%%%%%%%%%%
%%%%%%%%%%%%%%%%%%%%%%%%%%%%%%%%%%%%%%%%%%%%%%%%%%%%%%%%%%%%%%%%%%%%%
%
\begin{theorem}\label{thm_sampled_induced_space}
When the sampled spaces $Q_X$ and $Q_Y$ form a regular sample pair, the function $d_\PEQ: \ccalN \times \ccalN \rightarrow \reals_+$ defined in \eqref{eqn_dfn_sampled_space_distance_prelim} is an embedding metric in the space $\ccalN$. Moreover, it yields the same distance as the function $d_\PE$ defined in \eqref{eqn_dfn_partial_embedding}, i.e.
\begin{align}\label{eqn}
    d_\PEQ(N_X, N_Y) = d_\PE(N_X, N_Y),
\end{align}
for any networks $N_X$ and $N_Y$.
\end{theorem}

%%%%%%%%%%%%%%%%%%%%%%%%%%%%%%%%%%%%%%%%%%%%%%%%%%%%%%%%%%%%%%%%%%%%%
%%%   M   A   I   N      M   A   T   T   E   R   %%%%%%%%%%%%%%%%%%%%%%%%%%%%%%%%%%%
%%%%%%%%%%%%%%%%%%%%%%%%%%%%%%%%%%%%%%%%%%%%%%%%%%%%%%%%%%%%%%%%%%%%%
%
The statement in Theorem \ref{thm_sampled_induced_space} gives proper reasoning for differentiating networks via their sampled induced space. Similar as previous treatments, we could define $\max\{d_\PEQ(N_X, N_Y), d_\PEQ(N_Y, N_X)\}$ as a metric in the space $\ccalN \mod \cong$. Since the sampled induced spaces incorporate more information of the original networks, an approximation to $d_\PEQ(N_X, N_Y)$ via the sampled induced space would be a better approximation to $d_\PE(N_X, N_Y)$. Moreover, since we can construct the sampled induced space following some predetermined strategy -- taking midpoints for all edges in the networks, comparing networks via their sampled induced space is plausible in terms of complexity. Figure \ref{fig_mds_midpoints} illustrate the same network considered in Figure \ref{fig_mds_original} where the multi-dimensional scaling based on the sampled induced points would succeed in distinguishing networks that are different. We illustrate the practical usefulness of such methods in the next section.

%% file: figures/regular_sample_pair.tex
%!TEX root = ../journal_network_embedding.tex

\def \thisplotscale {0.45}
\def \unit {\thisplotscale cm}

\def \length {3}

\pgfdeclarelayer{background}
\pgfdeclarelayer{foreground}
\pgfdeclarelayer{midground}
\pgfsetlayers{background,midground,foreground}

\begin{tikzpicture}[-stealth, shorten >=0, x = 1.0*\unit, y=0.9*\unit, font=\scriptsize]

    \begin{pgfonlayer}{midground}
         \node at (0, 0) (center1) {};
         \path (center1) ++ (115:\length * 1.95) node (Xname) {$(Q_X, \hhatr_X)$}; 
         \path (center1) ++ (90:\length * 1.7321) node (Xa) [fill = blue!20, vertex] {$a$};
         \path (center1) ++ (180:\length) node (Xb) [fill = blue!20, vertex] {$b$};
         \path (center1) ++ (0:\length) node (Xc) [fill = blue!20, vertex] {$c$};
         
         \path (Xa) ++ (240: \length * 2 / 3) node (Xx1) [fill = blue!20, vertex] {$x_1$};
         \path (Xa) ++ (240: \length * 4 / 3) node (Xx2) [fill = blue!20, vertex] {$x_2$};
         \path (Xa) ++ (-60: \length * 2 / 3) node (Xx3) [fill = blue!20, vertex] {$x_3$};
         \path (Xa) ++ (-60: \length * 4 / 3) node (Xx4) [fill = blue!20, vertex] {$x_4$};
         \path (center1) ++ (180: \length / 3) node (Xx5) [fill = blue!20, vertex] {$x_5$};
         \path (center1) ++ (0: \length / 3) node (Xx6) [fill = blue!20, vertex] {$x_6$};
         \path (center1) ++ (90:\length * 1.7321 / 3) node (Xcenter) [fill = blue!20, vertex] {$q$};
         
         \node at (\length * 3.5, \length * 0.87) (center2) {};  
         \path (Xname) ++ (0:\length * 3.5) node {$(Q_Y, \hhatr_Y)$}; 
         \path (center2) ++ (150:\length) node (Yu) [fill = blue!20, vertex] {$u$};
         \path (center2) ++ (-30:\length) node (Yv) [fill = blue!20, vertex] {$v$};
         \path (Yu) ++ (-30: \length * 2 / 3) node (Yy1) [fill = blue!20, vertex] {$y_1$};
         \path (Yu) ++ (-30: \length * 4 / 3) node (Yy2) [fill = blue!20, vertex] {$y_2$};
     \end{pgfonlayer}
               
    \begin{pgfonlayer}{background} 
        \path [-, left] (Xa) edge node {} (Xc);
        \path [-, right] (Xa) edge node {} (Xb);
        \path [-, below] (Xc) edge node {} (Xb);
        
        \path [-] (Yu) edge node {} (Yv);
    \end{pgfonlayer}
        
    \begin{pgfonlayer}{foreground} % mappings 
        \path [->, above, red, semithick, bend left = 10] (Xa) edge node {$\hhatphi$} (Yu);
        \path [->, above, red, semithick, bend right = 15, pos = 0.06] (Xb) edge node {$\hhatphi$} (Yu);
        \path [->, above, red, semithick, bend right = 10, pos = 0.92] (Xc) edge node {$\hhatphi$} (Yv);
        
        \path [->, above, red, dotted, semithick, bend left = 5] (Xx1) edge node {$\phi$} (Yu);
        \path [->, above, red, dotted, semithick, bend right = 5, pos = 0.7] (Xx4) edge node {$\phi$} (Yy2);
        \path [->, above, red, dotted, semithick, bend right = 5, pos = 0.4] (Xx6) edge node {$\phi$} (Yy2);
        \path [->, above, red, dotted, semithick, bend left = 15, pos = 0.3] (Xcenter) edge node {$\phi$} (Yy1);
    \end{pgfonlayer}
    
\end{tikzpicture}

%% file: figures/embedding_sample_3.tex
%!TEX root = ../journal_network_embedding.tex

\def \thisplotscale {0.48}
\def \unit {\thisplotscale cm}

\def \length {3.9}

\pgfdeclarelayer{background}
\pgfdeclarelayer{foreground}
\pgfsetlayers{background,foreground}

\begin{tikzpicture}[-stealth, shorten >=0, x = 1.4*\unit, y=0.4*\unit, font=\scriptsize]

    \begin{pgfonlayer}{foreground}
         \node at (0, 0) (center1) {};
         \path (center1) ++ (110:\length * 1.7) node {$(Q_X, \hhatr_X)$}; 
         \path (center1) ++ (90:\length * 1.7321) node (Xa) [fill = blue!20, vertex] {$a$};
         \path (center1) ++ (0:\length) node (Xb) [fill = blue!20, vertex] {$b$};
         \path (center1) ++ (180:\length) node (Xc) [fill = blue!20, vertex] {$c$};
         \path (Xa) ++ (-60:\length) node (Xd) [fill = blue!20, vertex] {$d$};
         \path (Xa) ++ (240:\length) node (Xe) [fill = blue!20, vertex] {$e$};
         \path (center1) node (Xf) [fill = blue!20, vertex] {$f$};
         
         \node at (\length * 2.5, 0) (center2) {};
         \path (center2) ++ (110:\length * 1.7) node {$(Q_Y, \hhatr_Y)$}; 
         \path (center2) ++ (90:\length * 1.7321) node (Ya) [fill = blue!20, vertex] {$a$};
         \path (center2) ++ (0:\length) node (Yb) [fill = blue!20, vertex] {$b$};
         \path (center2) ++ (180:\length) node (Yc) [fill = blue!20, vertex] {$c$};
         \path (Ya) ++ (-60:\length) node (Yd) [fill = blue!20, vertex] {$d$};
         \path (Ya) ++ (240:\length) node (Ye) [fill = blue!20, vertex] {$e$};
         \path (center2) node (Yf) [fill = blue!20, vertex] {$f$};
         
         \node at (\length * 5, 0) (center3) {};
         \path (center3) ++ (110:\length * 1.7) node {$(Q_Z, \hhatr_Z)$}; 
         \path (center3) ++ (90:\length * 1.7321) node (Za) [fill = blue!20, vertex] {$a$};
         \path (center3) ++ (0:\length) node (Zb) [fill = blue!20, vertex] {$b$};
         \path (center3) ++ (180:\length) node (Zc) [fill = blue!20, vertex] {$c$};
         \path (Za) ++ (-60:\length) node (Zd) [fill = blue!20, vertex] {$d$};
         \path (Za) ++ (240:\length) node (Ze) [fill = blue!20, vertex] {$e$};
         \path (center3) node (Zf) [fill = blue!20, vertex] {$f$};
    \end{pgfonlayer}

    \begin{pgfonlayer}{background} 
        % X
        \path [-, right] (Xa) edge node {$0.5$} (Xd);
        \path [-, right] (Xd) edge node {$0.5$} (Xb);
        \path [-, left] (Xa) edge node {$0.5$} (Xe);
        \path [-, left] (Xe) edge node {$0.5$} (Xc);
        \path [-, below] (Xc) edge node {$5.5$} (Xf);
        \path [-, below] (Xf) edge node {$5.5$} (Xb);
        \path [-, above] (Xe) edge node {$5.5$} (Xd);
        \path [-, left] (Xe) edge node {$0.5$} (Xf);
        \path [-, right] (Xd) edge node {$0.5$} (Xf);
        
        \path [-, dotted, pos = 0.32] (Xe) edge node {$6$} (Xb);
        \path [-, dotted] (Xd) edge node {} (Xc);
        
        % Y
        \path [-, right] (Ya) edge node {$1.5$} (Yd);
        \path [-, right] (Yd) edge node {$1.5$} (Yb);
        \path [-, left] (Ya) edge node {$1.5$} (Ye);
        \path [-, left] (Ye) edge node {$1.5$} (Yc);
        \path [-, below] (Yc) edge node {$5.5$} (Yf);
        \path [-, below] (Yf) edge node {$5.5$} (Yb);
        \path [-, above] (Ye) edge node {$5.5$} (Yd);
        \path [-, left] (Ye) edge node {$1.5$} (Yf);
        \path [-, right] (Yd) edge node {$1.5$} (Yf);
        
        \path [-, dotted, pos = 0.32] (Ye) edge node {$7$} (Yb);
        \path [-, dotted] (Yd) edge node {} (Yc);
        
        % Z
        \path [-, right] (Za) edge node {$2.5$} (Zd);
        \path [-, right] (Zd) edge node {$2.5$} (Zb);
        \path [-, left] (Za) edge node {$2.5$} (Ze);
        \path [-, left] (Ze) edge node {$2.5$} (Zc);
        \path [-, below] (Zc) edge node {$5.5$} (Zf);
        \path [-, below] (Zf) edge node {$5.5$} (Zb);
        \path [-, above] (Ze) edge node {$5.5$} (Zd);
        \path [-, left] (Ze) edge node {$2.5$} (Zf);
        \path [-, right] (Zd) edge node {$2.5$} (Zf);
        
        \path [-, dotted, pos = 0.32] (Ze) edge node {$8$} (Zb);
        \path [-, dotted] (Zd) edge node {} (Zc);

    \end{pgfonlayer}
    
\end{tikzpicture}

%% file: figures/simple_points.tex
%!TEX root = ../journal_network_embedding.tex

\def \thisplotscale {0.53}
\def \unit {\thisplotscale cm}

\def \length {3}

\pgfdeclarelayer{background}
\pgfdeclarelayer{foreground}
\pgfsetlayers{background,foreground}

\begin{tikzpicture}[-stealth, shorten >=0, x = 1.3*\unit, y=1*\unit, font=\footnotesize]

    \begin{pgfonlayer}{foreground}
         \node at (0, 0) (center1) {};
         \path (center1) ++ (90:\length * 1.7321) node (Xa) [fill = blue!20, vertex] {$a$};
         \path (center1) ++ (180:\length) node (Xc) [fill = blue!20, vertex] {$c$};
         \path (center1) ++ (0:\length) node (Xb) [fill = blue!20, vertex] {$b$};
         
         \path (center1) ++ (60:\length * 1) node (Se) [fill = blue!20, vertex, dotted] {$e$};
         \path (center1) ++ (120:\length * 1) node (Sf) [fill = blue!20, vertex, dotted] {$f$} ;
         \path (center1) ++ (0:\length * 0) node (Sg) [fill = blue!20, vertex, dotted] {$g$} ;
         
         \path (center1) ++ (270:\length * 0.5) node {\scriptsize (a) $10$ networks with $\gamma$ = 1, 2, \dots, 10} ;
    \end{pgfonlayer}
               
    \begin{pgfonlayer}{background} 
        \path [-, left, pos = 0.7] (Xa) edge node {$r_X(a, c) = \gamma$} (Xc);
        \path [-, right, pos = 0.3] (Xa) edge node {$r_X(a, b) = \gamma$} (Xb);
        \path [-, below, pos = 0.78] (Xc) edge node {$r_X(b, c) = 11$} (Xb);
        
        \path [-, below, dotted] (Se) edge node {} (Sf);
        \path [-, below, dotted] (Se) edge node {} (Sg);
        \path [-, below, dotted] (Sf) edge node {} (Sg);
    \end{pgfonlayer}
    
\end{tikzpicture}

%% file: sec_5_application.tex
%!TEX root = journal_network_embedding.tex

%%%%%%%%%%%%%%%%%%%%%%%%%%%%%%%%%%%%%%%%%%%%%%%%%%%%%%%%%%%%%%%%%%%%%
%%%   S   E   C   T   I   O   N   %%%%%%%%%%%%%%%%%%%%%%%%%%%%%%%%%%%
%%%%%%%%%%%%%%%%%%%%%%%%%%%%%%%%%%%%%%%%%%%%%%%%%%%%%%%%%%%%%%%%%%%%%
%
\section{Application}\label{sec_application}

We first illustrate the usefulness of considering interiors of networks. We consider 10 networks in the form Figure \ref{fig_application_0} (a) where $\gamma = 1, 2, \dots, 10$. Approximations of network embedding distances are evaluated. Figure \ref{fig_application_0} (b) and (c) illustrate the heat-maps of the distance approximations where the indices in both horizontal and vertical directions denote the value of $\gamma$ in the networks. When interiors are considered by adding midpoints of edges, e.g. nodes $e$, $f$, and $g$ in Figure \ref{fig_application_0} (a), network distance approximations illustrated in Figure \ref{fig_application_0} (b) yield more desired results, as networks with similar $\gamma$ are close to each other with respect to their network distance approximations. This is more apparent for networks with $\gamma \leq 5$, where the relationships in the original networks fail to satisfy triangle inequality. A detailed analysis indicates that adding interior points in the networks (i) preserve the desired property of embedding distance when interiors are not considered (the distance approximations in Figure \ref{fig_application_0} (b) and (c) are very similar for $\gamma > 5$ where triangle inequalities are satisfied) and (ii) fix the undesired issue when the relationships in the original networks fail to satisfy triangle inequality.

We next consider the comparison and classification of three types of synthetic weighted networks. Edge weights in all three types of networks encode proximities. The first type of networks are with weighted Erd\H os-R\' enyi model \cite{erd6s1960}, where the edge weight between any pair of nodes is a random number uniformly selected from the unit interval $[0, 1]$. In the second type of networks, the coordinates of the vertices are generated uniformly and randomly in the unit circle, and the edge weights are evaluated with the Gaussian radial basis function $\exp(-d(i,j)^2 / 2\sigma^2)$ where $d(i,j)$ is the distance between vertices $i$ and $j$ in the unit circle and $\sigma$ is a kernel width parameter. In all simulations, we set $\sigma$ to $0.5$. The edge weight measures the proximity between the pair of vertices and takes value in the unit interval. In the third type of networks, we consider that each vertex $i$ represents an underlying feature $\bbu_i \in \reals^d$ of dimension $d$, and examine the Pearson's linear correlation coefficient $\rho_{ij}$ between the corresponding features $\bbu_i$ and $\bbu_j$ for a given pair of nodes $i$ and $j$. The weight for the edge connecting the pair is then set as $\rho_{ij} / 2 + 0.5$, a proximity measure in the unit interval. The feature space dimension $d$ is set as $5$ in all simulations. We want to see if network comparison tools proposed succeed in distinguishing networks generated from different processes.

We start with networks of equal size $|X| = 25$ and construct $20$ random networks for each aforementioned type. We then use the multi-dimensional scaling methods introduced in \cite{bronstein2006, bronstein2006a} to approximate the embedding network distance $d_\EE$ defined in Definition \ref{dfn_embedding}. To evaluate the effectiveness of considering interiors of networks described in Section \ref{sec_space_induced}, we add midpoints for all edges in a given network; it is apparent that any pair of networks with interiors defined in this way would form a regular sample pair. Approximations of the embedding network distance $d_\EE$ between these networks with midpoints added are then evaluated. Figure \ref{fig_application} (a) and (b) plot the two dimensional Euclidean embeddings \cite{CoxCox08} of the network metric approximations with and without interiors respectively. All embeddings in the paper are constructed with respect to minimizing the sum of squares of the inter-point distances; other common choices to minimize the sum of four power of the inter-point distances yields similar results. Networks constructed with different models form clear separate clusters (1 out of 60 errors with 1.67\%) with respect to approximation of network distances between networks with interior points added, where networks with Erd\H os-R\' enyi model are denoted by red circles, networks with unit circle model are described by blue diamonds, and correlation model represented as black squares. The clustering structure is not that clear (4 out of 60 errors with 6.67\%) in terms of with respect to approximation of network distances between networks without interior points, but networks constructed from different models are in general much more different compared to networks from the same model. 

Next we consider networks with number of nodes ranging between $20$ and $25$. Two networks are randomly generated for each network type and each number of nodes, resulting in $60$ networks in total. Interiors are examined similarly as before by adding midpoints for all edges in a given network. Figure \ref{fig_application} (c) and (d) illustrate the two dimensional Euclidean embeddings of the network metric approximations with and without interiors respectively. Despite the fact that networks with same model have different number of nodes, dissimilarities between network distance approximations are smaller when their underlying networks are from the same process. Similar as in the case with same number of nodes, considering interiors result in a more distinctive clustering pattern. An unsupervised classification with two linear boundaries would yield 1 out of 60 errors ($1.67\%$) for networks with interiors added and 5 errors ($8.33\%$) without interiors.

These results illustrate that (i) comparing networks by using embedding distance succeeds in identifying networks with different properties, and (ii) adding interiors to networks to form regular sample pair as in Section \ref{sec_space_induced} would yield better approximations to the actual network distances. Admittedly, other methods to compare networks may also succeed in distinguishing networks, after some proper treatment towards the issue of different sizes. Nonetheless, interior and embedding method would be more universal, not only for the reason that it establishes an approximation to the actual network metrics, but also since it provides a systematic way to analyze if one network can be well matched to a subset of another network.

%% file: sec_6_conclusion.tex
%!TEX root = journal_network_embedding.tex

%%%%%%%%%%%%%%%%%%%%%%%%%%%%%%%%%%%%%%%%%%%%%%%%%%%%%%%%%%%%%%%%%%%%%
%%%   S   E   C   T   I   O   N   %%%%%%%%%%%%%%%%%%%%%%%%%%%%%%%%%%%
%%%%%%%%%%%%%%%%%%%%%%%%%%%%%%%%%%%%%%%%%%%%%%%%%%%%%%%%%%%%%%%%%%%%%
%
\section{Conclusion}\label{sec_conclusion}

We present a different perspective to consider networks by defining a semimetric space induced from all the relationships in a given network. We demonstrate that comparing the respective induced space between a pair of networks outputs the identical distance as evaluating the discrepancy between the original network by embedding one network into another network, which we prove to be a valid metric in the space of all networks. Therefore, better approximations to the network metric distances can be constructed by examining the respective induced space. We illustrate that such methods succeed in classifying weighted pairwise networks constructed from different processes.

%% file: sec_A_appendix.tex
%!TEX root = journal_network_embedding.tex

\appendices
%%%%%%%%%%%%%%%%%%%%%%%%%%%%%%%%%%%%%%%%%%%%%%%%%%%%%%%%%%%%%%%%%%%%%
%%%   A   P   P   E   N   D   I   X   %%%%%%%%%%%%%%%%%%%%%%%%%%%%%%%
%%%%%%%%%%%%%%%%%%%%%%%%%%%%%%%%%%%%%%%%%%%%%%%%%%%%%%%%%%%%%%%%%%%%%
%
\section{Proofs in Section \ref{sec_embedding_distance}} \label{apx_proof_1}

%%%%%%%%%%%%%%%%%%%%%%%%%%%%%%%%%%%%%%%%%%%%%%%%%%%%%%%%%%%%%%%%%%%%%
%%%   P   R   O   O   F   %%%%%%%%%%%%%%%%%%%%%%%%%%%%%%%
%%%%%%%%%%%%%%%%%%%%%%%%%%%%%%%%%%%%%%%%%%%%%%%%%%%%%%%%%%%%%%%%%%%%%
%
\begin{myproof}[of Proposition \ref{prop_partial_embedding_metric}] To prove that $d_\PE$ is an embedding metric in the space of networks, we prove the (i) nonnegativity, (ii) embedding identity, and (iii) triangle inequality properties in Definition \ref{dfn_embedding_metric}.

%%%%%%%%%%%%%%%%%%%%%%%%%%%%%%%%%%%%%%%%%%%%%%%%%%%%%%%%%%%%%%%%%%%%%
%%%   P   R   O   O   F   %%%%%%%%%%%%%%%%%%%%%%%%%%%%%%%
%%%%%%%%%%%%%%%%%%%%%%%%%%%%%%%%%%%%%%%%%%%%%%%%%%%%%%%%%%%%%%%%%%%%%
%
\begin{myproof}[of nonnegativity property] Since $|r_X(x, x') - r_Y(\phi(x), \phi(x'))|$ is nonnegative, $\Delta_{X,Y}(\phi)$ defined in \eqref{eqn_dfn_partial_embedding_prelim} also is. The partial embedding distance must then satisfy $d_\PE(N_X, N_Y) \geq 0$ because it is a minimum of nonnegative numbers. \end{myproof}

%%%%%%%%%%%%%%%%%%%%%%%%%%%%%%%%%%%%%%%%%%%%%%%%%%%%%%%%%%%%%%%%%%%%%
%%%   P   R   O   O   F   %%%%%%%%%%%%%%%%%%%%%%%%%%%%%%%
%%%%%%%%%%%%%%%%%%%%%%%%%%%%%%%%%%%%%%%%%%%%%%%%%%%%%%%%%%%%%%%%%%%%%
%
\begin{myproof}[of embedding identity property]
First, we need to show that if $N_X$ can be isometrically embedded into $N_Y$, we must have $d_\PE(N_X, N_Y)=0$. To see that this is true recall that for isometric embeddable networks, there exists a mapping $\phi: X \rightarrow Y$ that preserves distance functions \eqref{eqn_network_order_2_isomorphism}. Then, under this mapping, we must have $\Delta_{X,Y}(\phi) = 0$. Since $\phi$ is a particular mapping, taking a minimum over all mappings as in \eqref{eqn_dfn_partial_embedding} yields
\begin{align}\label{eqn_dfn_partial_embedding_proof_identity_1}
    d_\PE(N_X, N_Y) \le \Delta_{X, Y}(\phi) = 0.
\end{align}
Since $d_\PE(N_X, N_Y) \ge 0$, it must be that $d_\PE(N_X, N_Y) = 0$ when $N_X$ can be isometrically embedded into $N_Y$. 

Second, we need to prove $d_\PE(N_X, N_Y) = 0$ must imply that $N_X$ can be isometrically embedded into $N_Y$. If $d_\PE(N_X, N_Y) = 0$, there exists a mapping $\phi: X \rightarrow Y$ such that $r_X(x, x') = r_Y(\phi(x), \phi(x'))$ for any $x, x' \in X$. This implies that $\phi$ is an isometric embedding and therefore $N_X$ can be isometrically embedded into $N_Y$.
\end{myproof}

%%%%%%%%%%%%%%%%%%%%%%%%%%%%%%%%%%%%%%%%%%%%%%%%%%%%%%%%%%%%%%%%%%%%%
%%%   P   R   O   O   F   %%%%%%%%%%%%%%%%%%%%%%%%%%%%%%%
%%%%%%%%%%%%%%%%%%%%%%%%%%%%%%%%%%%%%%%%%%%%%%%%%%%%%%%%%%%%%%%%%%%%%
%
\begin{myproof}[of triangle inequality]
To show that the triangle inequality, let the mapping $\phi$ between $X$ and $Z$ and $\psi$ between $Z$ and $Y$ be the minimizing mappings in (\ref{eqn_dfn_partial_embedding}). We can then write
\begin{align}\label{eqn_dfn_partial_embedding_proof_triangle_1}
      d_\PE(N_X, N_Z) = \Delta_{X,Z}(\phi), ~d_\PE(N_Z, N_Y) = \Delta_{Z,Y}(\psi).
\end{align}
Since both $\phi$ and $\psi$ are mappings, $\psi \circ \phi$ would be a valid mapping from $X$ to $Y$. The mapping $\psi \circ \phi$ may not be the minimizing mapping for the distance $d_\PE(N_X, N_Y)$. However since it is a valid mapping with the definition in \eqref{eqn_dfn_partial_embedding} we can write
\begin{align}\label{eqn_dfn_partial_embedding_proof_triangle_2}
    d_\PE(N_X, N_Y) &\le \Delta_{X,Y}(\psi \circ \phi).
\end{align}
Adding and subtracting $d_Z(\phi(x), \phi(x'))$ in the absolute value of $\Delta_{X,Y}(\psi \circ \phi) = \max_{x, x' \in X} \big| r_X(x, x') - r_Y(\psi(\phi(x)), \psi(\phi(x'))) \big|$ and using the triangle inequality of the absolute value yields
\begin{equation}\begin{aligned}\label{eqn_dfn_partial_embedding_proof_triangle_bound_1}
    \Delta_{X,Y}&(\psi \circ \phi) \leq \max_{x, x' \in X} \Big\{  \big| r_X(x, x') - d_Z(\phi(x), \phi(x'))\big| \\
        & + \! \Big| d_Z(\phi(x), \phi(x')) -r_Y\big(\psi\big(\phi(x)\big), \psi\big(\phi(x')\big)\big) \Big| \Big\}.
\end{aligned}\end{equation}
We can further bound \eqref{eqn_dfn_partial_embedding_proof_triangle_bound_1} by taking maximum over each summand,
\begin{equation}\begin{aligned}\label{eqn_dfn_partial_embedding_proof_triangle_bound_2}
    &\Delta_{X,Y}(\psi \circ \phi) \leq \max_{x, x' \in X} \big| r_X(x, x') - d_Z(\phi(x), \phi(x'))\big| \\
        & \! + \!\! \max_{x, x' \in X} \Big| d_Z(\phi(x), \phi(x')) \!-\!
            r_Y\big(\psi\big(\phi(x)\big), \psi\big(\phi(x')\big)\big) \Big| .
\end{aligned}\end{equation}
The first summand in \eqref{eqn_dfn_partial_embedding_proof_triangle_bound_2} is nothing different from $\Delta_{X, Z}(\phi)$. Since $\phi(x), \phi(x') \in Z$, the second summand in \eqref{eqn_dfn_partial_embedding_proof_triangle_bound_2} can be further bounded by
\begin{equation}\begin{aligned}\label{eqn_dfn_partial_embedding_proof_triangle_bound_3}
    & \max_{x, x' \in X} \Big| d_Z(\phi(x), \phi(x')) - 
        r_Y\big(\psi\big(\phi(x)\big), \psi\big(\phi(x')\big)\big) \Big| \\
            & \leq \max_{z, z' \in Z} \big| d_Z(z, z') - r_Y\big(\psi(z), \psi(z')\big) \big| = \Delta_{Z, Y}(\psi).
\end{aligned}\end{equation}
These two observations implies that
\begin{equation}\begin{aligned}\label{eqn_dfn_partial_embedding_proof_triangle_bound_4}
    &\Delta_{X,Y}(\psi \circ \phi) \leq \Delta_{X, Z}(\phi) + \Delta_{Z, Y}(\psi).
\end{aligned}\end{equation}
Substituting \eqref{eqn_dfn_partial_embedding_proof_triangle_1} and \eqref{eqn_dfn_partial_embedding_proof_triangle_2} into \eqref{eqn_dfn_partial_embedding_proof_triangle_bound_4} yields triangle inequality. \end{myproof}

\noindent Having proven all statements, the global proof completes. \end{myproof}

%%%%%%%%%%%%%%%%%%%%%%%%%%%%%%%%%%%%%%%%%%%%%%%%%%%%%%%%%%%%%%%%%%%%%
%%%   P   R   O   O   F   %%%%%%%%%%%%%%%%%%%%%%%%%%%%%%%
%%%%%%%%%%%%%%%%%%%%%%%%%%%%%%%%%%%%%%%%%%%%%%%%%%%%%%%%%%%%%%%%%%%%%
%
\begin{myproof}[of Theorem \ref{thm_embedding_distance_metric}] To prove that $d_\EE$ is a metric in the space of networks modulo isomorphism, we prove the (i) nonnegativity, (ii) symmetry, (iii) identity, and (iv) triangle inequality properties in Definition \ref{dfn_metric}.

%%%%%%%%%%%%%%%%%%%%%%%%%%%%%%%%%%%%%%%%%%%%%%%%%%%%%%%%%%%%%%%%%%%%%
%%%   P   R   O   O   F   %%%%%%%%%%%%%%%%%%%%%%%%%%%%%%%
%%%%%%%%%%%%%%%%%%%%%%%%%%%%%%%%%%%%%%%%%%%%%%%%%%%%%%%%%%%%%%%%%%%%%
%
\begin{myproof}[of nonnegativity property] Since $d_\PE(N_X, N_Y)$ as well as $d_\PE(N_Y, N_X)$ are both nonnegative, the embedding distance must then satisfy $d_\EE(N_X, N_Y) \geq 0$. \end{myproof}

%%%%%%%%%%%%%%%%%%%%%%%%%%%%%%%%%%%%%%%%%%%%%%%%%%%%%%%%%%%%%%%%%%%%%
%%%   P   R   O   O   F   %%%%%%%%%%%%%%%%%%%%%%%%%%%%%%%
%%%%%%%%%%%%%%%%%%%%%%%%%%%%%%%%%%%%%%%%%%%%%%%%%%%%%%%%%%%%%%%%%%%%%
%
\begin{myproof}[of symmetry property] Since $d_\EE(N_X, N_Y) = d_\EE(N_Y, N_X) = \max\{d_\PE(N_X, N_Y), d_\PE(N_Y, N_X)\}$, the symmetry property follows directly. \end{myproof}

%%%%%%%%%%%%%%%%%%%%%%%%%%%%%%%%%%%%%%%%%%%%%%%%%%%%%%%%%%%%%%%%%%%%%
%%%   P   R   O   O   F   %%%%%%%%%%%%%%%%%%%%%%%%%%%%%%%
%%%%%%%%%%%%%%%%%%%%%%%%%%%%%%%%%%%%%%%%%%%%%%%%%%%%%%%%%%%%%%%%%%%%%
%
\begin{myproof}[of identity property]
First, we need to show that if $N_X$ and $N_Y$ are isomorphic, we must have $d_\EE(N_X, N_Y)=0$. To see that this is true recall that for isomorphic networks there exists a bijective map $\phi: X \rightarrow Y$ that preserves distance functions \eqref{eqn_network_order_2_isomorphism}. This implies $\phi$ is also an injection, and we can find an injection $\psi: Y \rightarrow X$ that preserves distance functions \eqref{eqn_network_order_2_isomorphism}. Then, under the injection $\phi$, we must have $\Delta_{X,Y}(\phi) = 0$. Since $\phi$ is a particular mapping, taking a minimum over all mappings as in \eqref{eqn_dfn_partial_embedding} yields
\begin{align}\label{eqn_dfn_embedding_proof_identity_1}
    d_\PE(N_X, N_Y) \le \Delta_{X, Y}(\phi) = 0.
\end{align}
Since $d_\PE(N_X, N_Y) \ge 0$, as already shown, it must be that $d_\PE(N_X, N_Y) = 0$ when $N_X$ are isomorphic to $N_Y$. By symmetry we have $d_\PE(N_Y, N_X) = 0$, which combines with previous observation implies that $d_\EE(N_X, N_Y) = 0$.

Second, we need to prove $d_\EE(N_X, N_Y) = 0$ must imply that $N_X$ and $N_Y$ are isomorphic. By the definition of embedding distance, $d_\EE(N_X, N_Y) = 0$ means $d_\PE(N_X, N_Y) = 0$ and $d_\PE(N_Y, N_X) = 0$. If $d_\PE(N_X, N_Y) = 0$, there exists a mapping $\phi: X \rightarrow Y$ such that $r_X(x, x') = r_Y(\phi(x), \phi(x'))$ for any $x, x' \in X$. Moreover, this also implies the function $\phi$ must be injective. If it were not, there would be a pair of nodes $x \neq x'$ with $\phi(x) = \phi(x') = y$ for some $y \in Y$. By the definition of networks, we have that
\begin{align}\label{eqn_dfn_embedding_proof_identity_2}
    r_X(x, x') > 0, \quad r_Y(\phi(x), \phi(x')) = r_Y(y, y) = 0,
\end{align}
which contradicts the observation that $r_X(x, x') = r_Y(\phi(x), \phi(x'))$ for any $x, x' \in X$ and shows that $\phi$ must be injective. By symmetry, simultaneously, if $d_\PE(N_Y, N_X) = 0$, there exists an injective mapping $\psi: Y \rightarrow X$ such that $r_Y(y, y') = r_X(\psi(y), \psi(y'))$ for any $y, y' \in Y$. By applying the Cantor-Bernstein-Schroeder Theorem\cite[Section 2.6]{Kolmogorov75} to the reciprocal injections $\phi: X \rightarrow Y$ and $\psi: Y \rightarrow X$, the existence of a bijection between $X$ and $Y$ is guaranteed. This forces $X$ and $Y$ to have same cardinality and $\phi$ and $\psi$ being bijections. Pick the bijection $\phi$ and it follows $r_X(x, x') = r_Y(\phi(x), \phi(x'))$ for all nodes $x, x' \in X$. This shows that $N_X \cong N_Y$ and completes the proof of the identity statement.
\end{myproof}

%%%%%%%%%%%%%%%%%%%%%%%%%%%%%%%%%%%%%%%%%%%%%%%%%%%%%%%%%%%%%%%%%%%%%
%%%   P   R   O   O   F   %%%%%%%%%%%%%%%%%%%%%%%%%%%%%%%
%%%%%%%%%%%%%%%%%%%%%%%%%%%%%%%%%%%%%%%%%%%%%%%%%%%%%%%%%%%%%%%%%%%%%
%
\begin{myproof}[of triangle inequality]
To show that the triangle inequality holds, from the definition of embedding distance, we have that
\begin{align}\label{eqn_dfn_embedding_proof_triangle_1}
    d_\EE(N_X, N_Y) = \max\left\{ d_\PE(N_X, N_Z), d_\PE(N_Z, N_Y) \right\}.
\end{align}
Since partial embedding distance is a valid embedding metric, it satisfies triangle inequality in Definition \ref{dfn_embedding_metric}, therefore, we can bound \eqref{eqn_dfn_embedding_proof_triangle_1} by
\begin{equation}\begin{aligned}\label{eqn_dfn_embedding_proof_triangle_2}
    d_\EE(N_X, N_Y) \! \leq\! \max\big\{ & d_\PE(N_X, N_Z) \!+\! d_\PE(N_Z, N_Y),\\
        &d_\PE(N_Y, N_Z) \!+\! d_\PE(N_Z, N_X) \big\}.
\end{aligned}\end{equation}
To further bound \eqref{eqn_dfn_embedding_proof_triangle_2} we utilize the relationship as next.

%%%%%%%%%%%%%%%%%%%%%%%%%%%%%%%%%%%%%%%%%%%%%%%%%%%%%%%%%%%%%%%%%%%%%
%%%   F   A   C   T   %%%%%%%%%%%%%%%%%%%%%%%
%%%%%%%%%%%%%%%%%%%%%%%%%%%%%%%%%%%%%%%%%%%%%%%%%%%%%%%%%%%%%%%%%%%%%
%
\begin{fact}\label{fact_inequality}
Given real numbers $a,b,c,d$, it holds that
\begin{align}\label{eqn_inequality}
    \max\{a, c\} + \max\{b, d\} \geq \max\{a+b, c+d\}.
\end{align}\end{fact}

%%%%%%%%%%%%%%%%%%%%%%%%%%%%%%%%%%%%%%%%%%%%%%%%%%%%%%%%%%%%%%%%%%%%%
%%%   P   R   O   O   F   %%%%%%%%%%%%%%%%%%%%%%%%%%%%%%%
%%%%%%%%%%%%%%%%%%%%%%%%%%%%%%%%%%%%%%%%%%%%%%%%%%%%%%%%%%%%%%%%%%%%%
%
\begin{myproof}
If $a \geq c$ and $b \geq d$, the inequality holds since the left hand side is $a + b$ and the right hand side is also $a+b$. Similarly, if $c \geq a$ and $d \geq b$, the inequality also holds. What remains to consider are scenarios of $a \geq c, d\geq b$ as well as $c \geq a, b \geq d$. By symmetry, it suffices to consider the first scenario with $a \geq c, d\geq b$. Under this scenario, the statement becomes
\begin{align}\label{eqn_inequality_proof}
    a + d \geq \max\{a+b, c+d\}.
\end{align}
It follows that the state is correct following the assumption. Since we have considered all scenarios, the proof concludes.
\end{myproof}

%%%%%%%%%%%%%%%%%%%%%%%%%%%%%%%%%%%%%%%%%%%%%%%%%%%%%%%%%%%%%%%%%%%%%
%%%   M   A   I   N       M   A   T   T   E   R   %%%%%%%%%%%%%%%%%%%
%%%%%%%%%%%%%%%%%%%%%%%%%%%%%%%%%%%%%%%%%%%%%%%%%%%%%%%%%%%%%%%%%%%%%
%N_Y)
Back to the proof of triangle inequality, applying Fact \ref{fact_inequality} onto \eqref{eqn_dfn_embedding_proof_triangle_2} yields
\begin{equation}\begin{aligned}\label{eqn_dfn_embedding_proof_triangle_3}
    d_\EE(N_X, N_Y)  \leq & \max\big\{ d_\PE(N_X, N_Z), d_\PE(N_Z, N_X) \big\} \\
        + &\max\big\{ d_\PE(N_Y, N_Z), d_\PE(N_Z, N_Y) \big\}.
\end{aligned}\end{equation}
Substituting the definition of $d_\EE(N_X, N_Z)$ and $d_\EE(N_Z, N_Y)$ into \eqref{eqn_dfn_embedding_proof_triangle_3} yields 
\begin{equation}\begin{aligned}\label{eqn_dfn_embedding_proof_triangle_4}
    d_\EE(N_X, N_Y)  \leq d_\EE(N_X, N_Z) + d_\EE(N_Z, N_Y),
\end{aligned}\end{equation}
which is the triangle inequality and completes the proof.
\end{myproof}

%%%%%%%%%%%%%%%%%%%%%%%%%%%%%%%%%%%%%%%%%%%%%%%%%%%%%%%%%%%%%%%%%%%%%
%%%   P   R   O   O   F   %%%%%%%%%%%%%%%%%%%%%%%%%%%%%%%
%%%%%%%%%%%%%%%%%%%%%%%%%%%%%%%%%%%%%%%%%%%%%%%%%%%%%%%%%%%%%%%%%%%%%
%
\noindent Having proven all statements, the global proof completes. \end{myproof}

%%%%%%%%%%%%%%%%%%%%%%%%%%%%%%%%%%%%%%%%%%%%%%%%%%%%%%%%%%%%%%%%%%%%%
%%%   P   R   O   O   F   %%%%%%%%%%%%%%%%%%%%%%%%%%%%%%%
%%%%%%%%%%%%%%%%%%%%%%%%%%%%%%%%%%%%%%%%%%%%%%%%%%%%%%%%%%%%%%%%%%%%%
%
\begin{myproof}[of Lemma \ref{lemma_GH}]
Denote $d'_\C(N_X, N_Y)$ to represent $\min_{\phi:X \rightarrow Y, \psi:Y \rightarrow X} \max \{ \Delta_{X, Y}(\phi), \Delta_{Y, X}(\psi), \delta_{X, Y}(\phi, \psi)\}$. In order to prove prove the statement, we show that given any networks $N_X$ and $N_Y$, we have that (i) $d'_\C(N_X, N_Y) \leq d_\C(N_X, N_Y)$ and that (ii) $d_\C(N_X, N_Y) \leq d'_\C(N_X, N_Y)$.

%%%%%%%%%%%%%%%%%%%%%%%%%%%%%%%%%%%%%%%%%%%%%%%%%%%%%%%%%%%%%%%%%%%%%
%%%   P   R   O   O   F   %%%%%%%%%%%%%%%%%%%%%%%%%%%%%%%
%%%%%%%%%%%%%%%%%%%%%%%%%%%%%%%%%%%%%%%%%%%%%%%%%%%%%%%%%%%%%%%%%%%%%
%
\begin{myproof}[of $d'_\C(N_X, N_Y) \leq d_\C(N_X, N_Y)$]
From the definition of $d_\C(N_X, N_Y)$, there exists a correspondent $C$ such that $|r_X(x, x') - r_Y(y, y')| \leq d_\C(N_X, N_Y)$ for any $(x, y), (x', y') \in C$. Define a function $\phi: X \rightarrow Y$ that associates $x$ with an arbitrary $y$ chosen from the set that form a pair with $x$ in $C$, 
\begin{align}\label{eqn_dfn_embedding_proof_equiv_E_leq_C_1}
    \phi: x \mapsto y_0 \in \{ y \mid (x,y) \in C\}.
\end{align}
Since $C$ is a correspondence the set $\{ y \mid (x,y) \in C\}$ is nonempty for any $x$ implying that $\phi$ is well-defined for any $x \in X$. Hence, 
\begin{align}\label{eqn_dfn_embedding_proof_equiv_E_leq_C_2}
    \big|r_X(x, x') - r_Y(\phi(x), \phi(x'))\big| \leq d_\C(N_X, N_Y),
\end{align}
for any $x, x' \in X$. Since \eqref{eqn_dfn_embedding_proof_equiv_E_leq_C_2} is true for any $x, x' \in X$, it also true for the maximum pair, and therefore
\begin{align}\label{eqn_dfn_embedding_proof_equiv_E_leq_C_3}
    \Delta_{X, Y}\!(\phi) \!=\!\!\!\! \max_{x, x' \in X}\!\!\big|\!r_X(x,\! x') \!-\! r_Y(\phi(x),\! \phi(x'))\!\big| \!\!\leq\! d_\C(N_X,\! N_Y\!)\!.
\end{align}
Define a function $\psi: Y \rightarrow X$ that associates $y$ with an arbitrary $x$ chosen from the set that form a pair with $y$ in $C$, 
\begin{align}\label{eqn_dfn_embedding_proof_equiv_E_leq_Cy_1}
    \psi: y \mapsto x_0 \in \{ x \mid (x,y) \in C\}.
\end{align}
Following the similar argument as above would yield us
\begin{align}\label{eqn_dfn_embedding_proof_equiv_E_leq_Cy_2}
    \Delta_{Y, X}(\psi) \leq d_\C(N_X, N_Y).
\end{align}
Finally, recall that $\delta_{X, Y}(\phi, \psi)$ is defined as $\max_{x \in X, y \in Y} | r_X(x, \psi(y)) - r_Y(\phi(x), y)|$. In the same time, we have $(x, \psi(y)) \in C$ as well as $(\phi(x), y) \in C$, and therefore
\begin{align}\label{eqn_dfn_embedding_proof_equiv_E_leq_Cd}
    \max_{x \in X, y \in Y} \left| r_X(x, \psi(y)) - r_Y(\phi(x), y)\right| \leq d_\C(N_X, N_Y).
\end{align}
Taking a maximum on both sides of inequlities \eqref{eqn_dfn_embedding_proof_equiv_E_leq_C_3}, \eqref{eqn_dfn_embedding_proof_equiv_E_leq_Cy_2}, and \eqref{eqn_dfn_embedding_proof_equiv_E_leq_Cd} yields
\begin{align}\label{eqn_dfn_embedding_proof_equiv_E_leq_final}
   \max \{ \Delta_{X, Y}(\phi), \Delta_{Y, X}(\psi), \delta_{X, Y}(\phi, \psi)\} \! \leq \! d_\C(N_X, N_Y).
\end{align}
The specific $\phi$ and $\psi$ may not be the minimizing mappings for the left hand side of \eqref{eqn_dfn_embedding_proof_equiv_E_leq_final}. Nonetheless, they are valid mappings and therefore taking a minimum over all mappings yields the desired inequality $d'_\C(N_X, N_Y) \leq d_\C(N_X, N_Y)$.
\end{myproof}

%%%%%%%%%%%%%%%%%%%%%%%%%%%%%%%%%%%%%%%%%%%%%%%%%%%%%%%%%%%%%%%%%%%%%
%%%   P   R   O   O   F   %%%%%%%%%%%%%%%%%%%%%%%%%%%%%%%
%%%%%%%%%%%%%%%%%%%%%%%%%%%%%%%%%%%%%%%%%%%%%%%%%%%%%%%%%%%%%%%%%%%%%
%
\begin{myproof}[of $d_\C(N_X, N_Y) \leq d'_\C(N_X, N_Y)$]
From the definition of $d'_\C(N_X, N_Y)$, there exists a pair of mappings $\phi: X \rightarrow Y$ and $\psi: Y \rightarrow X$ such that 
\begin{align}
\label{eqn_dfn_embedding_proof_2_start_1}
    |r_X(x, x') - r_Y( \phi(x), \phi(x'))| &\leq d'_\C(N_X, N_Y), \\
\label{eqn_dfn_embedding_proof_2_start_2}
    |r_X(\psi(y), \psi(y')) - r_Y( y, y' )| &\leq d'_\C(N_X, N_Y), \\
\label{eqn_dfn_embedding_proof_2_start_3}
    |r_X(x, \psi(y)) - r_Y( \phi(x), y)| &\leq d'_\C(N_X, N_Y),
\end{align}
for any $x, x' \in X$ and $y, y' \in Y$. Define a correspondence by taking the union of the pairs associated by $\phi$ and $\psi$ such that
\begin{align}\label{eqn_dfn_embedding_proof_2_corespondence}
    C = \left\{ (x, \phi(x)) \mid x \in X \right\} \cup \left\{ (\psi(y), y) \mid y \in Y \right\}.
\end{align}
Since $\phi(x)$ is defined for any $x$ and $\psi(y)$ is defined for any $y$, $C$ is a well-defined correspondence. Notice that any pair $(x, y) \in C$ in the correspondence would be one of the following two forms: $(x, \phi(x))$ or $(\psi(y), y)$. Therefore, for any pairs $(x, y), (x', y') \in C$, they must be from one of the following three forms (i) $(x, \phi(x)), (x', \phi(x'))$, (ii) $(\psi(y), y), (\psi(y'), y')$, or (iii) $(x, \phi(x)), (\psi(y), y)$. If they are in the form (i), from \eqref{eqn_dfn_embedding_proof_2_start_1}, we can bound the difference between the respective relationship as
\begin{align}\label{eqn_dfn_embedding_proof_2_bound}
    \left| r_X(x, x') - r_Y(y, y') \right| \leq d'_\C(N_X, N_Y).
\end{align}
If the pairs are in the form (ii), \eqref{eqn_dfn_embedding_proof_2_start_2} also implies the correctness of \eqref{eqn_dfn_embedding_proof_2_bound}. Finally, if the pairs are in the form (iii), \eqref{eqn_dfn_embedding_proof_2_bound} would be established from \eqref{eqn_dfn_embedding_proof_2_start_3}. Consequently, \eqref{eqn_dfn_embedding_proof_2_bound} holds for any $(x, y), (x', y') \in C$. Therefore, they must also hold true for the bottleneck pairs achieving the maximum $\Gamma_{X, Y}(C)$ in \eqref{eqn_conventional_network_distance_prelim} which implies that $\Gamma_{X, Y}(C) \leq d'_\C(N_X, N_Y)$. The specific correspondence $C$ may not be the minimizing one in defining $d_\C(N_X, N_Y)$. Nonetheless, they are valid mappings and therefore taking a minimum over all mappings yields the desired inequality $d_\C(N_X, N_Y) \leq d'_\C(N_X, N_Y)$.
\end{myproof}

%%%%%%%%%%%%%%%%%%%%%%%%%%%%%%%%%%%%%%%%%%%%%%%%%%%%%%%%%%%%%%%%%%%%%
%%%   P   R   O   O   F   %%%%%%%%%%%%%%%%%%%%%%%%%%%%%%%
%%%%%%%%%%%%%%%%%%%%%%%%%%%%%%%%%%%%%%%%%%%%%%%%%%%%%%%%%%%%%%%%%%%%%
%
Since we have proven the two inequalities, it follows that $d_\C \equiv d'_\C$ and this completes the proof of the statement.\end{myproof}

%%%%%%%%%%%%%%%%%%%%%%%%%%%%%%%%%%%%%%%%%%%%%%%%%%%%%%%%%%%%%%%%%%%%%
%%%   P   R   O   O   F   %%%%%%%%%%%%%%%%%%%%%%%%%%%%%%%
%%%%%%%%%%%%%%%%%%%%%%%%%%%%%%%%%%%%%%%%%%%%%%%%%%%%%%%%%%%%%%%%%%%%%
%
\begin{myproof}[of Corollary \ref{coro_lower_bound}]
The network distance $d_\C(N_X, N_Y)$ would be no smaller than the right hand side of \eqref{eqn_lemma_GH}, if we remove the term $\delta_{X, Y}(\phi, \psi)$ in the maximum, i.e.
\begin{align}\label{eqn_coro_lower_bound_1}
    d_\C(N_X, \!N_Y) \!\geq\!\!\! \min_{\phi:X \rightarrow Y, \psi:Y \rightarrow X}\!\!
        \max\!\left\{\! \Delta_{X, Y}(\phi),\! \Delta_{Y, X}(\psi) \!\right\}\!.
\end{align}
The right hand side of \eqref{eqn_coro_lower_bound_1} would become smaller if we take the respective minimum for mappings $\phi$ and $\psi$ before taking the maximum, yielding us
\begin{align}\label{eqn_coro_lower_bound_2}
    d_\C(N_X, \!N_Y) \geq \max\left\{ \!\min_{\phi: X \rightarrow Y} \Delta_{X, Y}(\phi), \!\!
        \min_{\phi: Y \rightarrow X} \Delta_{Y, X}(\psi) \!\right\}\!.
\end{align}
From \eqref{eqn_dfn_partial_embedding} and \eqref{eqn_dfn_embedding} in Definitions \ref{dfn_embedding} and \ref{dfn_partial_embedding}, it is not hard to observe that the right hand side of \eqref{eqn_coro_lower_bound_2} is $\max\{d_\PE(N_X, N_Y), d_\PE(N_Y, N_X)\} =: d_\EE(N_X, N_Y)$, yielding the desired result $d_\EE(N_X, N_Y) \leq d_\C(N_X, N_Y)$.
\end{myproof}

%%%%%%%%%%%%%%%%%%%%%%%%%%%%%%%%%%%%%%%%%%%%%%%%%%%%%%%%%%%%%%%%%%%%%
%%%   A   P   P   E   N   D   I   X   %%%%%%%%%%%%%%%%%%%%%%%%%%%%%%%
%%%%%%%%%%%%%%%%%%%%%%%%%%%%%%%%%%%%%%%%%%%%%%%%%%%%%%%%%%%%%%%%%%%%%
%
\section{Proofs in Section \ref{sec_space_induced}} \label{apx_proof_2}

%%%%%%%%%%%%%%%%%%%%%%%%%%%%%%%%%%%%%%%%%%%%%%%%%%%%%%%%%%%%%%%%%%%%%
%%%   P   R   O   O   F   %%%%%%%%%%%%%%%%%%%%%%%%%%%%%%%
%%%%%%%%%%%%%%%%%%%%%%%%%%%%%%%%%%%%%%%%%%%%%%%%%%%%%%%%%%%%%%%%%%%%%
%
\begin{myproof}[of Proposition \ref{prop_semimetric}] To prove that space $(S_X, s_X)$ induced from $N_X = (X, r_X)$ is a semimetric space, we prove the (i) nonnegativity, (ii) symmetry, (iii) identity properties in Definition \ref{dfn_metric} and (iv) $s_X(p, m) = r_X(p, m)$ when $p, m \in X$.

%%%%%%%%%%%%%%%%%%%%%%%%%%%%%%%%%%%%%%%%%%%%%%%%%%%%%%%%%%%%%%%%%%%%%
%%%   P   R   O   O   F   %%%%%%%%%%%%%%%%%%%%%%%%%%%%%%%
%%%%%%%%%%%%%%%%%%%%%%%%%%%%%%%%%%%%%%%%%%%%%%%%%%%%%%%%%%%%%%%%%%%%%
%
\begin{myproof}[of nonnegativity property] Since $r_X(i, j) > 0$ for any different nodes in the original networks $i, j \in X, i \neq j$, $|w_{ij}^\star|r_X(i, j) \geq 0$ in \eqref{eqn_arbitrary_pmDistance}. Therefore, the induced distance $s_X(p, m) = \sum_{i \neq j} |w_{ij}^\star|r_X(i, j) \geq 0$.\end{myproof}

%%%%%%%%%%%%%%%%%%%%%%%%%%%%%%%%%%%%%%%%%%%%%%%%%%%%%%%%%%%%%%%%%%%%%
%%%   P   R   O   O   F   %%%%%%%%%%%%%%%%%%%%%%%%%%%%%%%
%%%%%%%%%%%%%%%%%%%%%%%%%%%%%%%%%%%%%%%%%%%%%%%%%%%%%%%%%%%%%%%%%%%%%
%
\begin{myproof}[of symmetry property] Given a pair of nodes $p, m \in S_X$, we would like to demonstrate that $s_X(p, m) = s_X(m, p)$. Denote $\{w_{ij}^\star\}$ as the collection of units of transformation along the direction from $i$ to $j$ in the original network. These vectors together make up the path from $p$ to $m$ with smallest amount of transformation. By definition, $\{w_{ij}^\star\}$ is the optimal solution to \eqref{eqn_constraint_arbitraryNodes}. Denote $\{v_{ij}^\star\}$ as the collection of units of transformation along the direction from $i$ to $j$ which makes up the path from $m$ to $p$ with the smallest amount of transformation. By definition, $\{v_{ij}^\star\}$ is the optimal solution to the following problem
\begin{equation}\begin{aligned}\label{eqn_proof_semimetric_symmetry}
    \left\{ v_{ij}^\star \right\} = \argmin ~ & \sum_{i, j \in X, i < j}\left| v_{ij} \right| \\
    \st ~ & p_i = m_i - \sum_{j \in X, j > i} v_{ij} + \sum_{j \in X, j < i} v_{ji}, ~ \forall i
\end{aligned}\end{equation}
Comparing \eqref{eqn_constraint_arbitraryNodes} with \eqref{eqn_proof_semimetric_symmetry}, it is easy to observe that if we take $v_{ij} = -w_{ij}$ for any $i < j$, the two problems becomes identical. Therefore, for the optimal solutions, we have the relationship $v_{ij}^\star = - w_{ij}^\star$ for any $i < j$. By definition in \eqref{eqn_arbitrary_pmDistance}, this implies the two relationships are the same 
\begin{equation}\begin{aligned}\label{eqn_proof_semimetric_same}
    s_X(p, m) \!=\! \sum_{\substack{i, j \in X,\\ i< j}}\left|w_{ij}^\star\right| r_X(i, j) \!=\! \sum_{\substack{i, j \in X, \\i< j}}\left|v_{ij}^\star\right| r_X(i, j) \!=\! s_X(m, p),
\end{aligned}\end{equation}
%
%\begin{equation}\begin{aligned}\label{eqn_proof_semimetric_same}
%    s_X(p, m) &= \sum_{i, j \in X, i< j}\left|w_{ij}^\star\right| r_X(i, j) \\
%        &= \sum_{i, j \in X, i< j}\left|v_{ij}^\star\right| r_X(i, j) = s_X(m, p),
%\end{aligned}\end{equation}
%
and completes the proof.
\end{myproof}

%%%%%%%%%%%%%%%%%%%%%%%%%%%%%%%%%%%%%%%%%%%%%%%%%%%%%%%%%%%%%%%%%%%%%
%%%   P   R   O   O   F   %%%%%%%%%%%%%%%%%%%%%%%%%%%%%%%
%%%%%%%%%%%%%%%%%%%%%%%%%%%%%%%%%%%%%%%%%%%%%%%%%%%%%%%%%%%%%%%%%%%%%
%
\begin{myproof}[of identity property] First we want to show that if $m$ and $p$ are identical points, their induced relationship $s_X(p, m) = 0$. In such scenario, $m$ and $p$ must have same tuple representation $(m_1, \dots, m_n)$ and $(p_1, \dots, p_n)$ with $m_i = p_i$ for any $i \in X$. In this case, it is apparent that the optimal solution $\{w_{ij}^\star\}$ in \eqref{eqn_constraint_arbitraryNodes} is $w_{ij}^\star = 0$ for any $i \neq j$. Therefore, $s_X(p, m) = 0$ shows the first part of the proof for identity property.

Second, we need to prove $s_X(p, m) = 0$ must imply that $p$ and $m$ are the same. By definition in \eqref{eqn_arbitrary_pmDistance}, the induced relationship can be written $s_X(p, m) = \sum_{i \neq j}|w_{ij}^\star|r_X(i,j)$, where the original relationship is always positive with $r_X(i,j) > 0$ for any $i \neq j$. Therefore, $s_X(p, m) = 0$ must imply that $|w_{ij}^\star|$ given any $i \neq j$. Combining this observation with the constraints in \eqref{eqn_constraint_arbitraryNodes} imply that $p_i = m_i$ for any $i \in X$. Therefore, $p$ and $m$ are identical point in the induced space, and this completes the proof.
\end{myproof}

%%%%%%%%%%%%%%%%%%%%%%%%%%%%%%%%%%%%%%%%%%%%%%%%%%%%%%%%%%%%%%%%%%%%%
%%%   F   I   G   U   R   E   %%%%%%%%%%%%%%%%%%%%%%%%%%%%%%%%%%%%%%%
%%%%%%%%%%%%%%%%%%%%%%%%%%%%%%%%%%%%%%%%%%%%%%%%%%%%%%%%%%%%%%%%%%%%%
%
\begin{figure}[t]
\centerline{\input{figures/simplex.tex}\vspace{-3mm} 
}
\caption{Examples of $(n-1)$-simplex used in the proof of Proposition \ref{prop_semimetric}. The induced space is the simplex with interior. Points in the original spaces are vertices of the simplex. The shortest path connecting two vertices is the edge joining them.\vspace{-4mm}
}
\label{fig_simplex}
\end{figure}
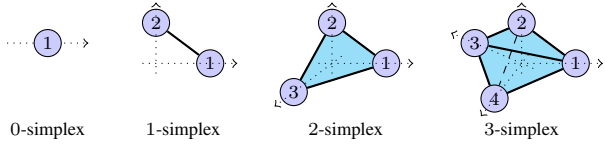

%%%%%%%%%%%%%%%%%%%%%%%%%%%%%%%%%%%%%%%%%%%%%%%%%%%%%%%%%%%%%%%%%%%%%
%%%   P   R   O   O   F   %%%%%%%%%%%%%%%%%%%%%%%%%%%%%%%
%%%%%%%%%%%%%%%%%%%%%%%%%%%%%%%%%%%%%%%%%%%%%%%%%%%%%%%%%%%%%%%%%%%%%
%
\begin{myproof}[of the property that $s_X(p, m) = r_X(p, m)$ when $p, m \in X$] When both $p, m \in X$, the respective tuple representation in the space is $p = (p_1, \dots, p_n)$ with $p_i = 1$ if $i = p$ and $p_i = 0$ otherwise, and $m = (m_1, \dots, m_n)$ with $m_i = 1$ if $i = m$ and $m_i = 0$. It is apparent that the path with the smallest amount of transformation from $p$ into $m$ is the exact vector from $p$ to $m$. Here we give a geometric proof using Figure \ref{fig_simplex}. The induced space is the $(n-1)$-simplex with interior defined. Nodes $p$ and $m$ correspond to the vertices in the simplex with their coordinates given by the tuple representations $p = (p_1, \dots, p_n)$ and $m = (m_1, \dots, m_n)$. The problem in \eqref{eqn_constraint_arbitraryNodes} searches for the shortest path in the simplex joining $p$ to $m$. It is then apparent that the shortest path should be the edge joining then; consequently $w_{pm}^\star = 1$ for the edge and $w_{ij}^\star = 0$ for any other edges $ij$. Taking this observation into \eqref{eqn_arbitrary_pmDistance} implies that $s_X(p, m) = r_X(p, m)$ and concludes the proof.
\end{myproof}

%%%%%%%%%%%%%%%%%%%%%%%%%%%%%%%%%%%%%%%%%%%%%%%%%%%%%%%%%%%%%%%%%%%%%
%%%   P   R   O   O   F   %%%%%%%%%%%%%%%%%%%%%%%%%%%%%%%
%%%%%%%%%%%%%%%%%%%%%%%%%%%%%%%%%%%%%%%%%%%%%%%%%%%%%%%%%%%%%%%%%%%%%
%
\noindent Having proven all statements, the global proof completes.\end{myproof}

%%%%%%%%%%%%%%%%%%%%%%%%%%%%%%%%%%%%%%%%%%%%%%%%%%%%%%%%%%%%%%%%%%%%%
%%%   P   R   O   O   F   %%%%%%%%%%%%%%%%%%%%%%%%%%%%%%%
%%%%%%%%%%%%%%%%%%%%%%%%%%%%%%%%%%%%%%%%%%%%%%%%%%%%%%%%%%%%%%%%%%%%%
%
\begin{myproof}[of Theorem \ref{thm_induced_space_distance_same}] To prove the statements, it suffices to show $d_\PES \equiv d_\PE$. Then the fact that $d_\PES$ is an embedding metric in the space $\ccalN$ follows since $d_\PE$ is an embedding metric in $\ccalN$. To prove the equivalence of the pair of distances, we show that given any networks $N_X$ and $N_Y$, we have that (i) $d_\PE(N_X, N_Y) \leq d_\PES(N_X, N_Y)$ and (ii) $d_\PES(N_X, N_Y) \leq d_\PE(N_X, N_Y)$.

%%%%%%%%%%%%%%%%%%%%%%%%%%%%%%%%%%%%%%%%%%%%%%%%%%%%%%%%%%%%%%%%%%%%%
%%%   P   R   O   O   F   %%%%%%%%%%%%%%%%%%%%%%%%%%%%%%%
%%%%%%%%%%%%%%%%%%%%%%%%%%%%%%%%%%%%%%%%%%%%%%%%%%%%%%%%%%%%%%%%%%%%%
%
\begin{myproof}[of $d_\PE(N_X, N_Y) \leq d_\PES(N_X, N_Y)$] 
From the definition of $d_\PES(N_X, N_Y)$, there exists a mapping $\phi: S_X \rightarrow S_Y$ in the induced space such that 
\begin{align}\label{eqn_proof_thm_induced_space_distance_same_first}
    \left|s_X(x, x') - s_Y(\phi(x), \phi(x'))\right| \leq d_\PES(N_X, N_Y),
\end{align}
for any $x, x' \in S_X$. Define a map $\hat \phi : X \rightarrow Y$ with $\hhatphi: i \mapsto \phi(i)$ for all $i \in X$. The map $\hhatphi$ is well defined since $\phi(x) \in Y$ for any $x \in X$. Proposition \ref{prop_semimetric} guarantees that $r_X(i, j) = s_X(i, j)$ when $i \in X$ and $j \in X$. Therefore, we also have $r_Y(\hhatphi(i), \hhatphi(j)) = s_Y(\hhatphi(i), \hhatphi(j))$. Substituting these two observations in \eqref{eqn_proof_thm_induced_space_distance_same_first} and searching for the maximum over all nodes $i, j \in X$ yields
\begin{align}\label{eqn_proof_thm_induced_space_distance_same_first_2}
    \Delta\!_{X, Y}\!(\hhatphi) 
        \!\!=\!\! \max_{i, j \in X}\left|r_X(i,\! j)\!\! -\!\! r_Y(\hhatphi(i),\! \hhatphi(j))\right| 
            \!\!\leq\!\! d_\PES(N_X, \!N_Y).
\end{align}
The specific $\hhatphi$ may not be the minimizing mapping for the left hand side of \eqref{eqn_proof_thm_induced_space_distance_same_first_2}. Nonetheless, it is a valid mapping and therefore taking a minimum over all mappings $\phi: X \rightarrow Y$ yields the desired inequality $d_\PE(N_X, N_Y) \leq d_\PES(N_X, N_Y)$.
\end{myproof}

%%%%%%%%%%%%%%%%%%%%%%%%%%%%%%%%%%%%%%%%%%%%%%%%%%%%%%%%%%%%%%%%%%%%%
%%%   P   R   O   O   F   %%%%%%%%%%%%%%%%%%%%%%%%%%%%%%%
%%%%%%%%%%%%%%%%%%%%%%%%%%%%%%%%%%%%%%%%%%%%%%%%%%%%%%%%%%%%%%%%%%%%%
%
\begin{myproof}[of $d_\PES(N_X, N_Y) \leq d_\PE(N_X, N_Y)$] From the definition of $d_\PE(N_X, N_Y)$, there exists a mapping $\hhatphi: X \rightarrow Y$ in the original node space such that
\begin{align}\label{eqn_proof_thm_induced_space_distance_same_second}
    \left|r_X(i, j) - r_Y(\hhatphi(i), \hhatphi(j))\right| \leq d_\PE(N_X, N_Y),
\end{align}
for any $i, j \in X$. Denote the cardinality of the node sets as $|X| = n$ and $|Y| = n'$. For any $x \in S_X$, it has the tuple representation of $(x_1, \dots, x_n)$, and any $y \in S_Y$ possess the tuple representation of $(y_1, \dots, y_{n'})$. Define a map $\phi: S_X \rightarrow S_Y$ as in \eqref{eqn_dfn_sample_same_rule_phi_mapping}. We showed in \eqref{eqn_dfn_sample_same_rule_phi_mapping_proof} that $\phi$ is a well-defined mapping. For a given pair of points $x, x' \in S_X$, denote $\{w_{ij}^\star\}$ as the collection of weights consisting the shortest path from $x$ to $x'$ solving the problem in \eqref{eqn_constraint_arbitraryNodes}. By \eqref{eqn_arbitrary_pmDistance}, the distance between $x$ and $x'$ in the induced space $S_X$ is then given by $s_X(x, x') = \sum_{i < j}|w_{ij}^\star|r_X(i, j)$. For the mapped pair $\phi(x)$ and $\phi(x')$ in the induced space $S_Y$, we show in the next fact that the optimal path between $\phi(x)$ and $\phi(x')$ is the optimal path between $x$ and $x'$ mapped under $\phi$.

%%%%%%%%%%%%%%%%%%%%%%%%%%%%%%%%%%%%%%%%%%%%%%%%%%%%%%%%%%%%%%%%%%%%%
%%%   F   A   C   T   %%%%%%%%%%%%%%%%%%%%%%%
%%%%%%%%%%%%%%%%%%%%%%%%%%%%%%%%%%%%%%%%%%%%%%%%%%%%%%%%%%%%%%%%%%%%%
%
\begin{fact}\label{fact_mapped_path}
Given a pair of networks $(X, r_X)$ and $(Y, r_Y)$ with their respective induced space $(S_X, s_X)$ and $(S_Y, s_Y)$, for any pair of nodes $x, x'$ in the induced space $S_X$, if $\{w_{ij}^\star\}$ is the collection of weights consisting the optimal path from $x$ to $x'$ solving the problem in \eqref{eqn_constraint_arbitraryNodes}, then $\{w_{\phi(i)\phi(j)}^\star\}$ is the collection of weights consisting the optimal path from $\phi(x)$ to $\phi(x')$ solving the problem in \eqref{eqn_constraint_arbitraryNodes}, where $w_{\phi(i)\phi(j)}^\star$ is the length traversed along the path parallel to the direction from $\phi(i)$ to $\phi(j)$.
\end{fact}

%%%%%%%%%%%%%%%%%%%%%%%%%%%%%%%%%%%%%%%%%%%%%%%%%%%%%%%%%%%%%%%%%%%%%
%%%   P   R   O   O   F   %%%%%%%%%%%%%%%%%%%%%%%%%%%%%%%
%%%%%%%%%%%%%%%%%%%%%%%%%%%%%%%%%%%%%%%%%%%%%%%%%%%%%%%%%%%%%%%%%%%%%
%
\begin{myproof}
Notice that the map $\phi$ may not be surjective; in other words, if we define $\tdS_Y = \{\phi(x) \mid x \in S_X\}$ as the image under $\phi$, then it is likely that $\tdS_Y \subsetneq S_Y$. We first show that the entire path of the shortest path from $\phi(x)$ to $\phi(x')$ must lie entirely inside the space $\tdS_Y$. To see this, define $n'' = |\{\hhatphi(i)\mid i \in X\}| \subseteq Y$ as the number of unique elements in $Y$ hit by nodes in $X$ under the map $\hhatphi$. It then follows that $\tdS_Y$ is the $(n''-1)$-simplex (convex hull) defined by vertices of $\hhatphi(i)$ for all $i \in X$. It follows naturally that $n'' \leq n'$. Therefore, in the language of topology, the $(n''-1)$ simplex describing $\tdS_Y$ would be a face of the larger $(n'-1)$-simplex $S_Y$. As an example, if $S_Y$ is the 3-simplex on the right of Figure \ref{fig_simplex} given by vertices $[0, 1, 2, 3]$, then $\tdS_Y$ would be the face of this 3-simplex: it may be one of the 2-simplices $[0,1,2]$, $[0, 1, 3]$, $[0, 2, 3]$, or $[1, 2, 3]$, or one of the 1-simplices $[0, 1]$, $[0, 2]$, $[0, 3]$, $[1, 2]$, $[1, 3]$ or $[2, 3]$, or one of the 0-simplices $[0]$, $[1]$, $[2]$, or $[3]$. Both $\phi(x)$ and $\phi(x')$ are on this $(n''-1)$-simplex. It then follows geometrically that the optimal path transforming $\phi(x)$ into $\phi(x')$ would also be on this $(n''-1)$-simplex. This implies that the entire path of the optimal path from $\phi(x)$ to $\phi(x')$ must lie entirely inside the space $\tdS_Y$.

Now, suppose the statement in Fact \ref{fact_mapped_path} is false, that the optimal path transforming $\phi(x)$ into $\phi(x')$ is not given by $\{w_{\phi(i)\phi(j)}^\star\}$ as the collection of weights. Since this path is entirely inside the space $\tdS_Y$, it needs to be of the form $\phi(x) \rightarrow \phi(k_0) \rightarrow \phi(k_1) \cdots \rightarrow \phi(k_L) \rightarrow \phi(x')$ for some $\phi(k_0), \dots, \phi(k_L) \in \tdS_Y$. Denote the collection of weights for this path as $\{v_{\phi(i)\phi(j)}^\star\}$, where $v_{\phi(i)\phi(j)}^\star$ is the length traversed along the path parallel to the direction from $\phi(i) \in \tdS_Y$ to $\phi(j) \in \tdS_Y$. The assumption that Fact \ref{fact_mapped_path} is false implies that
\begin{align}\label{eqn_proof_fact_mapped_path_1}
    \sum_{i,j \in X, i < j} \left| v_{\phi(i)\phi(j)}^\star \right| 
        < \sum_{i,j \in X, i < j} \left| w_{\phi(i)\phi(j)}^\star \right|.
\end{align}
Since $\phi(k_0), \dots, \phi(k_L) \in \tdS_Y$, we have that $k_0, \dots, k_L \in S_X$, and therefore the path of the form $x \rightarrow k_0 \rightarrow k_1 \cdots \rightarrow k_L \rightarrow x'$ would be a path in $S_X$ from $x$ to $x'$. This path has the collection of weights as $\{v_{ij}^\star\}$, where $v_{ij}^\star$ is the length traversed along the path parallel to the direction from $i \in X$ to $j \in X$. Moreover, utilizing the relationship in \eqref{eqn_proof_fact_mapped_path_1} yields
\begin{align}\label{eqn_proof_fact_mapped_path_2}
    \sum_{i,j \in X, i < j} \left| v_{ij}^\star \right| 
        < \sum_{i,j \in X, i < j} \left| w_{ij}^\star \right|,
\end{align}
which contradicts the assumption beforehand that the optimal path from $x$ to $x'$ is given by the collection of weights $\{w_{ij}^\star\}$. Therefore, it must be that the statement in Fact \ref{fact_mapped_path} is true and that $\{w_{\phi(i)\phi(j)}^\star\}$ is the collection of weights consisting the optimal path from $\phi(x)$ to $\phi(x')$ solving the problem in \eqref{eqn_constraint_arbitraryNodes}.
\end{myproof}

%%%%%%%%%%%%%%%%%%%%%%%%%%%%%%%%%%%%%%%%%%%%%%%%%%%%%%%%%%%%%%%%%%%%%
%%%   M   A   I   N       M   A   T   T   E   R   %%%%%%%%%%%%%%%%%%%
%%%%%%%%%%%%%%%%%%%%%%%%%%%%%%%%%%%%%%%%%%%%%%%%%%%%%%%%%%%%%%%%%%%%%
%
Back to the proof of $d_\PES(N_X, N_Y) \leq d_\PE(N_X, N_Y)$, leveraging the relationship established in Fact \ref{fact_mapped_path}, for any pair $x, x' \in X$, the distance between their mapped points $\phi(x)$ and $\phi(x')$ in the induced space is given by $s_Y(\phi(x), \phi(x')) = \sum_{i < j}|w_{\phi(i)\phi(j)}^\star|r_Y(\phi(i), \phi(j))$. Therefore,
\begin{equation}\begin{aligned}\label{eqn_proof_thm_induced_space_distance_same_second_1}
    & \left| s_X(x, x') - s_Y(\phi(x), \phi(x')) \right| \\
        &\!= \!\!\left| \sum_{i, j \in X, i < j} \!\!\!\!\left| w_{ij}^\star \right| r_X(i, j) \!- \!\!\!\!\!\!\!
            \sum_{i, j \in X, i < j} \!\!\!\!\left| w_{\phi(i)\phi(j)}^\star \right| r_Y(\phi(i), \phi(j)) \!\right|\!.
\end{aligned}\end{equation}

\noindent $w_{\phi(i)\phi(j)}^\star$ is for notation purposes and its value is no different from $w_{ij}^\star$. Therefore, combining terms for each $|w_{ij}^\star|$ on the right hand side of in \eqref{eqn_proof_thm_induced_space_distance_same_second_1} yields
\begin{align}\label{eqn_proof_thm_induced_space_distance_same_second_2}
    \left| \!s_X(x, x') \!-\! s_Y(\phi(x), \phi(x'))\! \right| 
        \!= \!\!\! \sum_{ i < j} \!\!\left| w_{ij}^\star \right| \left|r_X(i, j) \!- \! r_Y(\phi(i), \phi(j)) \!\right|\!.
\end{align}

Note that $\phi(i) = \hhatphi(i)$ and $\phi(j) = \hhatphi(j)$ by construction. Substituting \eqref{eqn_proof_thm_induced_space_distance_same_second} into \eqref{eqn_proof_thm_induced_space_distance_same_second_2} yields an inequality as 
\begin{align}\label{eqn_proof_thm_induced_space_distance_same_second_3}
    \left| s_X(x, x') - s_Y(\phi(x), \phi(x')) \right| 
        \leq \sum_{ i < j} \left| w_{ij}^\star \right| d_\PE(N_X, N_Y).
\end{align}
To finish the proof we further bound $\sum_{i < j}|w_{ij}^\star| \leq 1$ by utilizing the following important observation for the induced space.

%%%%%%%%%%%%%%%%%%%%%%%%%%%%%%%%%%%%%%%%%%%%%%%%%%%%%%%%%%%%%%%%%%%%%
%%%   F   I   G   U   R   E   %%%%%%%%%%%%%%%%%%%%%%%%%%%%%%%%%%%%%%%
%%%%%%%%%%%%%%%%%%%%%%%%%%%%%%%%%%%%%%%%%%%%%%%%%%%%%%%%%%%%%%%%%%%%%
%
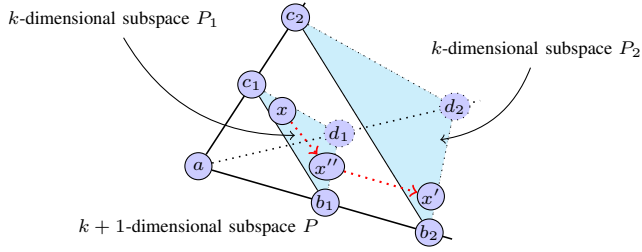
\begin{figure}[t]
\centerline{\input{figures/proof_bound_one}\vspace{-3mm} }
\caption{Illustration for the proof of Fact \ref{fact_bound_of_one}. In order to prove the statement for $k+1$ from $k$, we consider $x$ and $x'$ living in some $(k+1)$-dimensional subspace $P$. We can then find a point $x''$ sharing $n-k$ tuple coefficients with $x$ and therefore residing in some $k$-dimensional subspace $P_1$ with $x$ such that the length of the path from $x$ to $x''$ is upper bounded by $1 - \sum_{i=k+1}^n x_i$ and the length of the path from $x''$ to $x'$ is upper bounded by $x_{k+1}$. The existence of the path from $x$ to $x'$ via $x''$ shows the validity of the statement for $k+1$.\vspace{-4mm}
}
\label{fig_proof_bound_one}
\end{figure}

%%%%%%%%%%%%%%%%%%%%%%%%%%%%%%%%%%%%%%%%%%%%%%%%%%%%%%%%%%%%%%%%%%%%%
%%%   F   A   C   T   %%%%%%%%%%%%%%%%%%%%%%%
%%%%%%%%%%%%%%%%%%%%%%%%%%%%%%%%%%%%%%%%%%%%%%%%%%%%%%%%%%%%%%%%%%%%%
%
\begin{fact}\label{fact_bound_of_one}
Consider a network $(X, r_X)$ with its induced space $(S_X, s_X)$ and $|X| = n$, for any pair of nodes $x, x' \in S_X$ in the induced space with their respective tuple representation $x = (x_1, \dots, x_n)$ and $x' = (x_1', \dots, x_n')$. Suppose that $x$ and $x'$ share at least $n - k$ of their tuple coefficient such that for any $0 \leq k \leq n$, we have $x_{l_1} = x'_{l_1}$ for $1 \leq l_i \leq n - k$, then if $\sum_{1 \leq i \leq n - k}x_{l_k} = \alpha$, the optimal path $\{w_{ij}^\star\}$ from $x$ to $x'$ solving the problem in \eqref{eqn_constraint_arbitraryNodes} satisfies that
\begin{align}\label{eqn_fact_bound_of_one}
    \sum_{i < j} \left| w_{ij}^\star \right| \leq 1 - \alpha.
\end{align}\end{fact}

%%%%%%%%%%%%%%%%%%%%%%%%%%%%%%%%%%%%%%%%%%%%%%%%%%%%%%%%%%%%%%%%%%%%%
%%%   P   R   O   O   F   %%%%%%%%%%%%%%%%%%%%%%%%%%%%%%%
%%%%%%%%%%%%%%%%%%%%%%%%%%%%%%%%%%%%%%%%%%%%%%%%%%%%%%%%%%%%%%%%%%%%%
%
\begin{myproof}
We prove the statement by induction. The base case with $k = 0$ is trivial since $x$ and $x'$ would be the identical points. The case with $k = 1$ do not exist, since $\sum_i x_i = \sum_i x'_i = 1$ and therefore we cannot have a single tuple coefficient being different. For the case with $k = 2$, without loss of generality, suppose the two tuple coefficients being different for $x$ and $x'$ are the first and second, i.e. $x = (x_1, x_2, x_3, \dots, x_n)$ and $x' = (x_1', x_2', x_3, \dots, x_n)$, then it is apparent $x_1' = x_2$ and $x_2' = x_1$ and that the optimal path from $x$ to $x'$ only involves a path from node $1$ to node $2$ with length $w_{12} = x_1 - x_2 \leq 1 - \sum_{i = 3}^n x_i$ from requirement $\sum_{i = 1}^n x_i = 1$.

For the induction, suppose the statement is true for $k$, we would like to show the validity of the statement for $k + 1$. Without loss of generality, suppose the last $n-(k+1)$ coefficients in the tuple representations of $x$ and $x'$ are the same. Hence, they can be represented as $x = (x_1, \dots, x_k, x_{k+1}, x_{k+2}, \dots, x_n)$ and $x' = (x_1', \dots, x_k', x_{k+1}', x_{k+2}, \dots, x_n)$. $x$ and $x'$ can therefore be considered as points in some $k+1$-dimensional subspace $P$, see Figure \ref{fig_proof_bound_one}. Without loss of generality, we further assume that $x_{k+1} \geq x'_{k+1}$. For any node $x'' \in S_X$ with tuple representation $x'' = (x_1'', \dots, x_k'', x_{k+1}, \dots, x_n)$, it shares the last $n - k$ coefficients with $x$ and therefore $x$ and $x''$ can be considered as some point in the $k$-dimensional subspace $P_1$. The statement holds true for $k$; consequently, there exists a path with collection of weights $\{v_{ij}^\star\}$ from $x$ to $x''$ solving the problem in \eqref{eqn_constraint_arbitraryNodes} such that 
\begin{align}\label{eqn_proof_fact_bound_of_one_1}
    \sum_{i < j} \left|v_{ij}^\star\right| \leq 1 - \sum_{i = k+1}^n x_k.
\end{align}
We argue that we can find one such $x''$ sharing $n-k$ coefficients with $x$ in the subspace $P_1$ such that there exists a path from $x''$ to $x'$ that only consists of one vector parallel with the direction from the $(k+1)$-th node in $X$ to some node $i \in X$. We give a proof by geometry here: for the subspace $P_1$, from any points at its vertices (e.g. $b_1$, $c_1$, or $d_1$ in Figure \ref{fig_proof_bound_one}), we could reach a corresponding point at the vertices (e.g. $b_2$, $c_2$, or $d_2$) of the subspace $P_2$ via a link parallel to the direction from node the $(k+1)$-th node in $X$ to some node $i \in X$. For example, from $b_1$, we can reach $b_2$ simply via the vector from $b_1$ to $b_2$. This means the subspace $P_1$ can cover the subspace $P_2$ by moving along the direction only consists of one vector parallel to the direction from node the $(k+1)$-th node in $X$ to some node $i \in X$. Hence, we can find one such $x''$ sharing $n-k$ coefficients with $x$ in $P_1$ which can reach $x'$ via a path only consisting of one vector with length $x_{k+1} - x'_{k+1}$ (recall we assume $x_{k+1} \geq x'_{k+1}$). Therefore, we can construct a path with collection of weights $\{w_{ij}\}$ from $x$ to $x'$ by transversing the path from $x$ to $x''$ with collection of weights $\{v_{ij}^\star\}$ followed by the line segment from $x''$ to $x'$. This collection of path would have the property that
\begin{equation}\begin{aligned}\label{eqn_proof_fact_bound_of_one_2}
    \sum_{i < j} \left|w_{ij}\right| &= \sum_{i < j} \left|v_{ij}^\star\right| + \left| x_{k+1} - x'_{k+1}\right|
        \leq \sum_{i < j} \left|v_{ij}^\star\right| + x_{k+1} \\ 
        &\leq \left( 1 - \sum_{i = k+1}^n x_i \right)+ x_{k+1} = 1 - \sum_{i = k+2}^n x_i.
\end{aligned}\end{equation}
The path from $x$ to $x'$ via $x''$ with collection of weights $\{w_{ij}\}$ may not be the optimal path from $x$ to $x'$, but since it is a valid path, we can safely bound $\sum_{i < j} |w_{ij}^\star| \leq \sum_{i < j}|w_{ij}| \leq 1 - \sum_{i = k+2}^n x_k$, which shows the statement for $k+1$. This completes the induction step and therefore concludes the proof.
\end{myproof}

%%%%%%%%%%%%%%%%%%%%%%%%%%%%%%%%%%%%%%%%%%%%%%%%%%%%%%%%%%%%%%%%%%%%%
%%%   P   R   O   O   F   %%%%%%%%%%%%%%%%%%%%%%%%%%%%%%%
%%%%%%%%%%%%%%%%%%%%%%%%%%%%%%%%%%%%%%%%%%%%%%%%%%%%%%%%%%%%%%%%%%%%%
%
Back to the proof of $d_\PES(N_X, N_Y) \leq d_\PE(N_X, N_Y)$, considering the case with $k = n$ and $\alpha = 0$ in Fact \ref{fact_bound_of_one} yields $\sum_{ij}|w_{ij}^\star| \leq 1$ for any pair of nodes $x,  x' \in S_X$. Substituting this relationship into \eqref{eqn_proof_thm_induced_space_distance_same_second_3} yields
\begin{align}\label{eqn_proof_thm_induced_space_distance_same_second_final}
    \left| s_X(x, x') - s_Y(\phi(x), \phi(x')) \right| 
        \leq d_\PE(N_X, N_Y).
\end{align}
Since \eqref{eqn_proof_thm_induced_space_distance_same_second_final} holds true for any $x, x' \in S_X$, it must also be for the pair of nodes yielding the maximum discrepancy between $s_X(x, x')$ and $s_Y(\phi(x), \phi(x'))$; consequently, this implies $d_\PES(N_X, N_Y) \leq d_\PE(N_X, N_Y)$ and concludes the proof.
\end{myproof}

%%%%%%%%%%%%%%%%%%%%%%%%%%%%%%%%%%%%%%%%%%%%%%%%%%%%%%%%%%%%%%%%%%%%%
%%%   P   R   O   O   F   %%%%%%%%%%%%%%%%%%%%%%%%%%%%%%%
%%%%%%%%%%%%%%%%%%%%%%%%%%%%%%%%%%%%%%%%%%%%%%%%%%%%%%%%%%%%%%%%%%%%%
%
Since we have proven the two inequalities, it follows that $d_\PES \equiv d_\PE$ and this completes the proof of the statements.
\end{myproof}

%%%%%%%%%%%%%%%%%%%%%%%%%%%%%%%%%%%%%%%%%%%%%%%%%%%%%%%%%%%%%%%%%%%%%
%%%   P   R   O   O   F   %%%%%%%%%%%%%%%%%%%%%%%%%%%%%%%
%%%%%%%%%%%%%%%%%%%%%%%%%%%%%%%%%%%%%%%%%%%%%%%%%%%%%%%%%%%%%%%%%%%%%
%
\begin{myproof}[of Theorem \ref{thm_sampled_induced_space}]
To prove the statements, it suffices to show $d_\PEQ \equiv d_\PE$. Then the fact that $d_\PEQ$ is an embedding metric in the space $\ccalN$ follows since $d_\PE$ is an embedding metric in $\ccalN$. To prove the equivalence of the pair of distances, we show that given any networks $N_X$ and $N_Y$ with the sample spaces form a regular sample pair, we have that (i) $d_\PE(N_X, N_Y) \leq d_\PEQ(N_X, N_Y)$ and (ii) $d_\PEQ(N_X, N_Y) \leq d_\PE(N_X, N_Y)$.

%%%%%%%%%%%%%%%%%%%%%%%%%%%%%%%%%%%%%%%%%%%%%%%%%%%%%%%%%%%%%%%%%%%%%
%%%   P   R   O   O   F   %%%%%%%%%%%%%%%%%%%%%%%%%%%%%%%
%%%%%%%%%%%%%%%%%%%%%%%%%%%%%%%%%%%%%%%%%%%%%%%%%%%%%%%%%%%%%%%%%%%%%
%
\begin{myproof}[of $d_\PE(N_X, N_Y) \leq d_\PEQ(N_X, N_Y)$] 
Utilizing $\hhatr_X(i, j) = s_X(i, j) = r_X(i, j)$ for $i, j \in X$ by the definition the sampled space, the proof follows from the proof of $d_\PE(N_X, N_Y) \leq d_\PES(N_ X, N_Y)$ in the first part of proof for Theorem \ref{thm_induced_space_distance_same} in Appendix \ref{apx_proof_2}.
\end{myproof}

%%%%%%%%%%%%%%%%%%%%%%%%%%%%%%%%%%%%%%%%%%%%%%%%%%%%%%%%%%%%%%%%%%%%%
%%%   P   R   O   O   F   %%%%%%%%%%%%%%%%%%%%%%%%%%%%%%%
%%%%%%%%%%%%%%%%%%%%%%%%%%%%%%%%%%%%%%%%%%%%%%%%%%%%%%%%%%%%%%%%%%%%%
%
\begin{myproof}[of $d_\PEQ(N_X, N_Y) \leq d_\PE(N_X, N_Y)$]  
From the definition of $d_\PE(N_X, N_Y)$, there exists a mapping $\hhatphi: X \rightarrow Y$ in the original node space such that
\begin{align}\label{eqn_proof_thm_sampled_induced_space_initial}
    \left|r_X(i, j) - r_Y(\hhatphi(i), \hhatphi(j))\right| \leq d_\PE(N_X, N_Y),
\end{align}
for any $i, j \in X$. Define a mapping $\tilde \phi: S_X \rightarrow S_Y$ induced from $\hhatphi$ as in \eqref{eqn_dfn_sample_same_rule_phi_mapping}. The fact that $Q_X$ and $Q_Y$ form a regular sample pair implies that $\phi'(x) \in Q_Y$ for any $x \in Q_X$ and any mapping $\phi': Q_X \rightarrow Q_Y$. Therefore, the specific mapping $\phi: Q_X \rightarrow Q_Y$ formed by restricting $\tilde \phi: S_X \rightarrow S_Y$ onto $Q_X \subset S_X$ is also well-defined. In the second part of proof for Theorem \ref{thm_induced_space_distance_same} in Appendix \ref{apx_proof_2}, we have demonstrated that 
\begin{align}\label{eqn_proof_thm_sampled_induced_space_2}
    \left| s_X(x, x') - s_Y(\tilde \phi(x), \tilde \phi(x')) \right| 
        \leq d_\PE(N_X, N_Y).
\end{align}
for any pair of nodes $x, x' \in S_X$ and $\tilde \phi: S_X \rightarrow S_Y$. Restricting \eqref{eqn_proof_thm_sampled_induced_space_2} on pair of nodes $x, x' \in Q_X \subset S_X$ and the mapping $\phi = \tilde \phi |_{Q_X}$ and utilizing the fact that 
\begin{align}\label{eqn_proof_thm_sampled_induced_space_r_s}
    \hhatr_X(x, x') \!=\! s_X(x, x'), ~\hhatr_Y(\phi(x), \phi(x')) \!=\! s_Y(\phi(x), \phi(x'))\!,
\end{align}
yields the following relationship
\begin{align}\label{eqn_proof_thm_sampled_induced_space_3}
    \left| \hhatr_X(x, x') - \hhatr_Y(\phi(x), \phi(x')) \right| 
        \leq d_\PE(N_X, N_Y).
\end{align}
Since \eqref{eqn_proof_thm_sampled_induced_space_3} holds true for any $x, x' \in Q_X$, it must also be the case for the pairs yielding the maximum discrepancy between $\hhatr_X(x, x')$ and $\hhatr_Y(\phi(x), \phi(x'))$; hence, this implies $d_\PEQ(N_X, N_Y) \leq d_\PE(N_X, N_Y)$ and completes the proof.
\end{myproof}

%%%%%%%%%%%%%%%%%%%%%%%%%%%%%%%%%%%%%%%%%%%%%%%%%%%%%%%%%%%%%%%%%%%%%
%%%   P   R   O   O   F   %%%%%%%%%%%%%%%%%%%%%%%%%%%%%%%
%%%%%%%%%%%%%%%%%%%%%%%%%%%%%%%%%%%%%%%%%%%%%%%%%%%%%%%%%%%%%%%%%%%%%
%
Since we have proven the two inequalities, it follows that $d_\PEQ \equiv d_\PE$ and this concludes the proof of the statements.
\end{myproof}

%% file: figures/simplex.tex
%!TEX root = ../journal_network_embedding.tex

\def \thisplotscale {0.45}
\def \unit {\thisplotscale cm}

\pgfdeclarelayer{back}
\pgfdeclarelayer{fore}
\pgfdeclarelayer{mid}
\pgfsetlayers{back,mid,fore}

\begin{tikzpicture}[-stealth,  shorten >=0, x = 0.8*\unit, y=0.6*\unit, font=\scriptsize]

%%%%%%%%%%%%%%%%%%%%%%%%
    %%%%%% N  E  T  W  O  R  K     X   %%%%%%
    %%%%%%%%%%%%%%%%%%%%%%%%

    \begin{pgfonlayer}{fore}
    % 0-simplx
    \draw[dotted,->] (-1.5,0) -- (1.5,0);
    \node [blue vertex] at (0,0) (v) {$1$};
    \node at (0, -4.3) {$0$-simplex};	
    % 1-simplex
    \node [blue vertex] at (6,-1) (e1) {$1$};
    \node [blue vertex] at (4,1) (e2) {$2$}; 
    \draw[dotted,->] (3.5,-1) -- (7,-1);
    \draw[dotted,->] (4, -1.5) -- (4, 2);
    \node at (5, -4.3) {$1$-simplex};	 
    % 2-simplex
    \node [blue vertex] at (12.5, -1) (tr1) {$1$};
    \node [blue vertex] at (10.5, 1) (tr2) {$2$}; 
    \node [blue vertex] at (10.5 - 1.414, -1 - 1.414) (tr3) {$3$}; 
    \draw[dotted,->] (10, -1) -- (13.5, -1);
    \draw[dotted,->] (10.5, -1.5) -- (10.5, 2);
    \draw[dotted,->] (10.5 + 2.12 / 6, -1 + 2.12 / 6) -- (10.5 - 2.12, -1 - 2.12);
    \node at (11, -4.3) {$2$-simplex};	  
    % 3-simplex
    \node [blue vertex] at (19.5, -1) (ti1) {$1$};
    \node [blue vertex] at (17.5, 1) (ti2) {$2$}; 
    \node [blue vertex] at (17.5 - 5.2 / 3, -1 + 1) (ti3) {$3$}; 
    \node [blue vertex] at (17.5 - 1, -1 - 5.2 / 3) (ti4) {$4$}; 
    \draw[dotted,->] (17.5 - 0.5, -1) -- (20.5, -1);
    \draw[dotted,->] (17.5, -1 - 0.5) -- (17.5, 2);
    \draw[dotted,->] (17.5 + 2.6 / 6, -1 - 1.5 / 6) -- (17.5 - 2.6, -1 + 1.5);
    \draw[dotted,->] (17.5 + 1.5 / 6, -1 + 2.6 / 6) -- (17.5 - 1.5, -1 -2.6);
    \node at (17.5, -4.3) {$3$-simplex};		
    \end{pgfonlayer}
    
    \begin{pgfonlayer}{mid}
    % 1-simplex
    \path[-, thick] (e1) edge (e2);	
    % 2-simplex
    \path[-, thick] (tr1) edge (tr2);	
    \path[-, thick] (tr1) edge (tr3);	
    \path[-, thick] (tr2) edge (tr3);	
    % 3-simplex
    \path[-, thick] (ti1) edge (ti2);	
    \path[-, thick] (ti1) edge (ti3);	
    \path[-, thick] (ti2) edge (ti3);	
    \path[-, thick] (ti1) edge (ti4);	
    \path[-, thick] (ti3) edge (ti4);	
    \path[dashed, -] (ti2) edge (ti4);	
    \end{pgfonlayer}
    
    \begin{pgfonlayer}{back}
    % 2-simplex
    \filldraw[ultra thin, fill = cyan, fill opacity = 0.4] (tr1.center) --(tr2.center) --(tr3.center) -- cycle;
    % 3-simplex
    \filldraw[ultra thin, fill = cyan, fill opacity = 0.4] (ti1.center) --(ti2.center) --(ti3.center) -- (ti4.center) -- cycle;
    \end{pgfonlayer}
     
\end{tikzpicture}

%% file: figures/proof_bound_one.tex
%!TEX root = ../journal_network_embedding.tex

\def \thisplotscale {0.45}
\def \unit {\thisplotscale cm}

\def \zdivisor {2.2}
\def \length {1.2}

\pgfdeclarelayer{backback}
\pgfdeclarelayer{back}
\pgfdeclarelayer{middle}
\pgfdeclarelayer{fore}
\pgfsetlayers{backback,back,middle,fore}

\tdplotsetmaincoords{75}{48}
\begin{tikzpicture}[tdplot_main_coords, scale = 0.42, font = \scriptsize]

%%%%%%%%%%%%%%%%%%%%%%%%%%%%%%%%%%%%%%%%%%%%%%%%%%%%%%%%%%%%%%%%%%%%%
%%%   P   E   R   T   U   R   B   E   D   %%%%%%%%%%%%%%%%%%%%%%%
%%%%%%%%%%%%%%%%%%%%%%%%%%%%%%%%%%%%%%%%%%%%%%%%%%%%%%%%%%%%%%%%%%%%%
%
\begin{pgfonlayer}{fore} 
    % nodes    
    \node at (0, 0, 0) (center1) {};
    \path (center1) ++ (0, 0, 0) node (a) [fill = blue!20, vertex] {$a$};
    \path (center1) ++ (5*\length, 0, 0) node (b1) [fill = blue!20, vertex] {$b_1$};
    \path (center1) ++ (9.1*\length, 0, 0) node (b2) [fill = blue!20, vertex] {$b_2$};
    \path (center1) ++ (0, 5*\length, 0) node (d1) [fill = blue!20, vertex, dotted] {$d_1$};
    \path (center1) ++ (0, 9.1*\length, 0) node (d2) [fill = blue!20, vertex, dotted] {$d_2$};
    \path (center1) ++ (0, 5/\zdivisor, 5/\zdivisor) node (c1) [fill = blue!20, vertex] {$c_1$};
    \path (center1) ++ (0, 9.1/\zdivisor, 9.1/\zdivisor) node (c2) [fill = blue!20, vertex] {$c_2$};
    \path (center1) ++ (0, 0, -2) node {$k+1$-dimensional subspace $P$};
    \path (center1) ++ (-4, 0, 4) node (k) {$k$-dimensional subspace $P_1$};
    \path (k) ++ (9, 0, -2) node (k2) {};
    \path (center1) ++ (16, 0, 7) node (l) {$k$-dimensional subspace $P_2$};
    \path (l) ++ (-5, 0, -4) node (l2) {};
    
    \path (center1) ++ (2.3, 1.5, 2) node (x) [fill = blue!20, vertex] {$x$};
    \path (x) ++ (1, 1, -1.8) node (x2) [fill = blue!20, vertex] {$x''$};
    \path (x2) ++ (4.1*\length, 0, 0) node (x1) [fill = blue!20, vertex] {$x'$};
\end{pgfonlayer}

\begin{pgfonlayer}{backback}  
    % subspace
     \filldraw [ultra thin, fill = cyan, fill opacity = 0.2, dotted] 
                (b1.center) --(c1.center) --(d1.center) -- cycle;
     \filldraw [ultra thin, fill = cyan, fill opacity = 0.2, dotted] 
                (b2.center) --(c2.center) --(d2.center) -- cycle;
\end{pgfonlayer}

\begin{pgfonlayer}{back}  
    % graphs
    \path [draw, semithick] (a) -- ++ (10 * \length, 0, 0);
    \path [draw, semithick, dotted] (a) -- ++ (0, 10 * \length, 0);
    \path [draw, semithick] (a) -- ++ (0, 10/\zdivisor, 10/\zdivisor);
    
    \path [draw] (b1) -- (c1);
    \path [draw, dotted] (c1) -- (d1);
    \path [draw, dotted] (d1) -- (b1);
    \path [draw] (b2) -- (c2);
    \path [draw, dotted] (c2) -- (d2);
    \path [draw, dotted] (d2) -- (b2);
    
    \path [->, draw, thin, bend right = 30] (k) edge (k2);	
    \path [->, draw, thin, bend left = 30] (l) edge (l2);	
\end{pgfonlayer}

\begin{pgfonlayer}{middle}  
    % signals
    \path [->, draw, thick, red, dotted] (x) -- (x2);
    \path [->, draw, thick, red, dotted] (x2) -- (x1);
\end{pgfonlayer}    

\end{tikzpicture}